\documentclass[notitlepage,floatfix,nofootinbib,british,balancelastpage,hideallsubsections,preprintnumbers]{revtex4-1}
\usepackage{amsmath,mathrsfs}
\usepackage{amssymb}
\usepackage{appendix}
\usepackage{balance}
\usepackage{booktabs}
\usepackage{capt-of}
\usepackage{hyperref}
\usepackage{cleveref}
\usepackage{fancyvrb}
\usepackage[pdftex]{graphicx}
\usepackage{multirow}
\usepackage{relsize}
\usepackage{rotate}
\usepackage{rotating}
\usepackage{slashed}
\usepackage{subcaption}
\usepackage{tabularx}

\usepackage{units}
\usepackage{url}
\usepackage[usenames,dvipsnames]{xcolor}

\hypersetup{
    colorlinks=false,
    pdfborder={0 0 0},
}

\newcommand{\TeV}{\text{Te\hspace{-0.05cm}V}}
\newcommand{\GeV}{\text{Ge\hspace{-0.05cm}V}}
\newcommand{\fb}{\text{fb}}

\newcommand{\invfb}{fb${}^{-1}$}
\newcommand{\etmiss}{$\slashed{E}_T$}
\newcommand{\Root}{\textsc{Root}}
\newcommand{\Checkmate}{Check\textsc{mate}}
\newcommand{\CLs}{CL$_\text{S}$\ }
\newcommand{\pT}{p_\text{T}}

\usepackage{soul}

\crefname{appsec}{Appendix}{Appendices}

\makeatletter
\g@addto@macro\@verbatim\scriptsize
\makeatother

\begin{document}
\fvset{samepage=true, fontsize=\scriptsize}
\title{CheckMATE: Confronting your Favourite New Physics Model with LHC Data}%

\author{Manuel Drees }%
\email{drees@th.physik.uni-bonn.de}
\affiliation{Physikalisches
  Institut and Bethe
  Center for Theoretical Physics, University of Bonn, Bonn, Germany}

\author{Herbi Dreiner }%
\email{dreiner@th.physik.uni-bonn.de}
\affiliation{Physikalisches
  Institut and Bethe
  Center for Theoretical Physics, University of Bonn, Bonn, Germany}

\author{Daniel Schmeier }%
\email{daschm@th.physik.uni-bonn.de}
\affiliation{Physikalisches
  Institut and Bethe
  Center for Theoretical Physics, University of Bonn, Bonn, Germany}

\author{Jamie Tattersall }%
\email{j.tattersall@thphys.uni-heidelberg.de}
\affiliation{Institut f\"ur Theoretische Physik, University of Heidelberg, Heidelberg, Germany}
\affiliation{Physikalisches
  Institut and Bethe
  Center for Theoretical Physics, University of Bonn, Bonn, Germany}

\author{Jong Soo Kim}%
\email{jong.kim@csic.es}
\affiliation{Instituto de Fisica Teorica UAM/CSIC, Madrid, Spain and ARC Centre of Excellence for Particle Physics at the Terascale, School of   Chemistry and Physics, University of Adelaide, Australia}

\preprint{ADP-13-29/T849}
\preprint{IFT-UAM/CSIC-13-133}

\begin{abstract}
In the first three years of running, the LHC has 
delivered a wealth of new data that is now being analysed. 
With over 20~fb$^{-1}$
of integrated luminosity, both ATLAS and CMS have performed many searches for new physics that 
theorists are eager to test their model against. However, tuning the detector simulations, 
understanding the particular analysis details and interpreting the results can be a tedious task.

\Checkmate{} (Check Models At Terascale Energies) is a program package which accepts simulated event files
in many formats for 
any model. The program then determines whether the model is excluded or not at 95$\%$ C.L. 
by comparing to many recent
experimental analyses. Furthermore 
the program can calculate confidence limits and provide detailed information about 
signal regions of interest. It is simple to use and the program structure allows for 
easy extensions to upcoming LHC results in the future. \newline
\\ \Checkmate\; can be found at: \verb@http://checkmate.hepforge.org@ 
\end{abstract}


\setcounter{tocdepth}{1}
\setcounter{secnumdepth}{2}
\date{\today}%
\begin{center}
\includegraphics[width=0.5\textwidth]{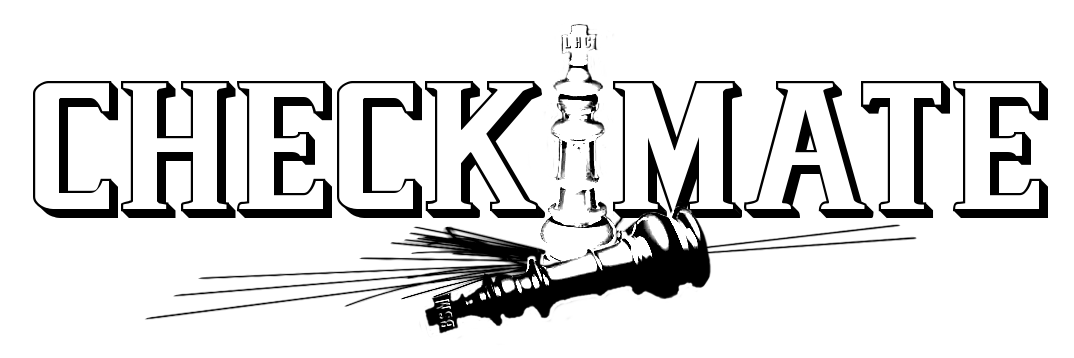}
\end{center}
\maketitle
\makeatletter
  \def\l@subsubsection#1#2{}%
\makeatother 

\pagebreak
\tableofcontents
\pagebreak


\subsection*{Important Note}

\begin{itemize}
 \item \Checkmate{} is built upon the tools and hard work of many people. If \Checkmate{} is used in your 
    publication it is extremely important that all of the following citations are included,
\begin{itemize}
  \item Delphes 3 \cite{deFavereau:2013fsa}.
  \item FastJet \cite{Cacciari:2011ma,Cacciari:2005hq}.
  \item Anti-kt jet algorithm \cite{Cacciari:2008gp}.
  \item \CLs prescription \cite{0954-3899-28-10-313}.
  \item In analyses that use the MT2 kinematical discriminant we use the Oxbridge Kinetics 
  Library \cite{Lester:1999tx,Barr:2003rg} and the algorithm developed by Cheng and Han \cite{Cheng:2008hk}.
  \item All experimental analyses that were used to set limits in the study.
  \item The Monte Carlo event generator that was used.
\end{itemize}  


\end{itemize}  

\section{Introduction}

\subsection{Motivation}
In the three years that the Large Hadron Collider (LHC) has been running, the Beyond 
the Standard Model (BSM) landscape has changed enormously. To take just a few 
examples, the possible parameter spaces of 
Supersymmetry (SUSY) \cite{Haber:1984rc,Nilles:1983ge}, Technicolour \cite{Weinberg:1975gm,Susskind:1978ms}
and Extra Dimensional models \cite{ArkaniHamed:1998rs,Randall:1999ee} are 
now significantly more limited and constrained than in the pre--LHC era.

LHC experimentalists typically classify analyses in terms of a 
specific final state configuration. They interpret the results in essentially 
two different approaches:
The first strategy is to take a constrained version of a complete model, with a small number of free parameters
and investigate exactly what set of 
model parameters have been excluded by the latest data and which are still allowed.
The advantage of such a complete model is its predictivity and the small number of input parameters. However, 
as soon as the model is 
slightly modified, it can be very difficult, if not impossible, to set a new appropriate 
limit. Therefore, only a small number of models tested by the experimental 
collaborations can be reliably checked.

The second approach employed to great effect at the LHC uses simplified 
models \cite{Alves:2011wf}. The idea here is that an effective Lagrangian with only a small number of particles and interactions 
is considered. Limits on the mass of BSM particles, in particular production 
and decay topologies, are presented. The production and decay modes chosen are those that often occur in a 
particular class of models and consequently 
can be applied widely. 

However, the particular interpretation of these results can lead to very different 
results. For example, the headline limits are often presented in terms of 100\% branching ratios 
and in models with a dark matter candidate \cite{Bertone:2004pz}, this is often taken as being massless. In most \textquoteleft realistic' 
models, neither of these assumptions is true which can lead to a far more conservative bound.

One may nevertheless attempt to map the particular model of interest onto the simplified model
by considering the rate of events that have exactly the same production and decay topologies. However, 
 rarely are branching ratios 100\% and many different production modes may 
contribute. These different topologies may not be identical to those given in the simplified
model and so cannot be safely used to set a limit. Despite this, it is inevitable that some events
from these processes will contribute to signal regions. Consequently, the limit set may be far
weaker than if the full model had been properly analysed by the particular experimental collaboration.

The above two issues lead to the obvious question of how to reinterpret the data 
to place bounds on the BSM models that have not previously been analysed by the experiments. Unavoidably, the 
only way to do this accurately is to generate simulated events and then run these 
through an implementation of the experimental analysis, including all detector effects. 
\Checkmate{} is meant to provide the best possible reproduction of both the analysis and detector effects. In 
addition we aim to simplify and improve confidence in the second step so that any user can investigate the 
model of their choice.

The idea behind \Checkmate{} is that the user only has to provide event file(s) 
and the corresponding cross section(s). The user also selects the analyses\footnote{If the particular analysis of interest is missing from the 
current \Checkmate{} list, we welcome requests for new analyses. Indeed, the analysis under 
question may be almost ready for public release and is only awaiting final validation.} against 
which the event files should be tested.
\Checkmate{} then informs the user whether the particular 
model is excluded or not using the 95\% \CLs \cite{0954-3899-28-10-313} method. In addition, the 
actual exclusion likelihood can be given if desired.

\subsection{Overview}
\label{sec:overview}
\begin{figure}
\centering
\includegraphics[width=0.95\textwidth]{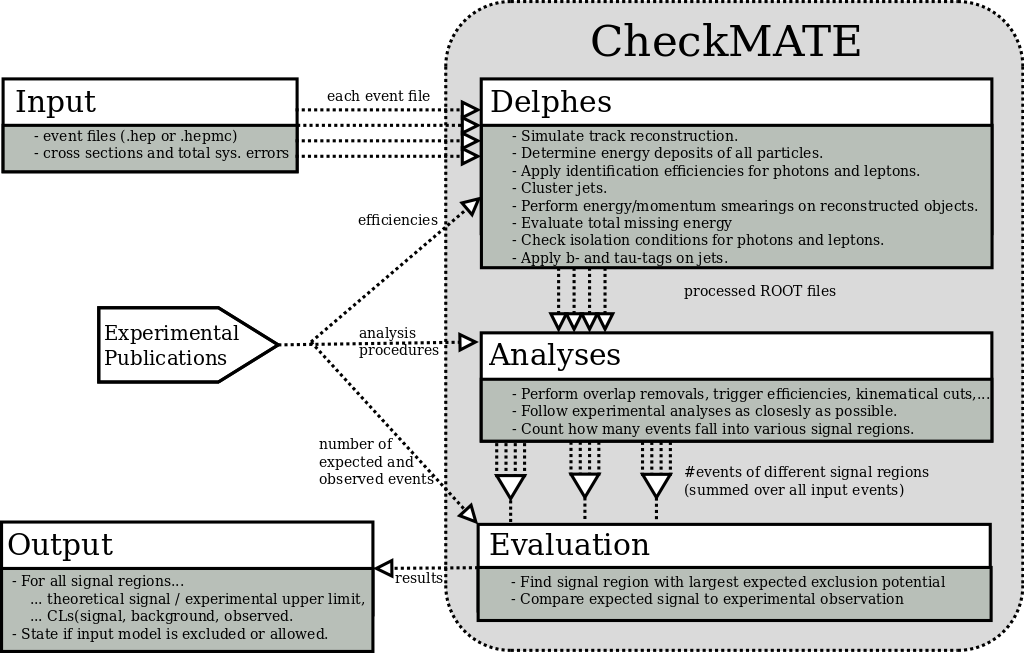}
\caption{Flow chart to demonstrate the chain of data processing within \Checkmate. }
\label{fig:flowdiagram}
\end{figure}
To run \Checkmate, the user has to provide simulated event files (in HepMC \cite{Dobbs:2001ck} or 
HepEvt \cite{Knowles:1995kj}) from an event generator of choice (Pythia \cite{Sjostrand:2006za, 
Sjostrand:2007gs} , Herwig++ \cite{Bahr:2008pv}, 
CalcHEP\cite{Belyaev:2012qa}, MadGraph \cite{Alwall:2011uj}, Sherpa \cite{Gleisberg:2008ta}, ...) plus
the effective cross sections that correspond to these events, e.\/g.\ taking into account potential phase space cuts or 
branching ratios. The total systematic $1 \sigma$ error on the 
cross section has to be provided which enters the calculation of the total uncertainty. Input events can be weighted and these weights are properly taken into account by \Checkmate{}.

This information is then consecutively processed by three independent 
modules (see \Cref{fig:flowdiagram}). 
As a first step, the events are run through a fast \emph{detector simulation}. We have chosen to use Delphes \cite{deFavereau:2013fsa},
which we have heavily modified to accurately model the ATLAS detector and the final state
reconstructions that are performed. This includes the smearing of the visible particles' momenta according to the 
respective experimental resolution, efficiency--based object identification, b--/$\tau$--tagging 
for jets and reconstruction of the missing transverse momentum vector. The program has
been optimised to check for all the final state information needed by the given selection of analyses and 
to minimise the computational effort by removing redundant and/or superfluous parts from the simulation, see 
\Cref{sec:internal:delphes} for more details. The results of this simulation are stored in \Root{} 
trees, which can optionally be read and/or manipulated by the user externally \cite{Brun:1997pa}.

These files are automatically processed by those of the supplied \emph{analyses} 
the user selects when starting \Checkmate{}. A list of currently implemented analyses can 
be found in \Cref{table:analyses} and will be significantly extended in the 
coming months. They are internally coded in a well--structured framework that 
allows for an easy extension to new upcoming experimental 
results (see \Cref{sec:app:addanalyses} for more details). It furthermore 
allows users to easily update given analyses or implement their own.

The input data is processed event by event by checking isolation criteria, removing overlapping objects, and implementing the cuts that define the signal regions. The analysis program determines 
how many events in total satisfy certain signal region criteria and stores this 
information in human--readable output for each separate input event file. Both 
the true number of Monte Carlo events as well as the number of events normalised 
to the given cross section and luminosity are stored.

The final step of the program consists of a statistical \emph{evaluation} of the result. For 
each individual signal region of the chosen analyses, the total number of expected 
signal events $S$ is determined by summing up the results from each input event 
file. The total $1 \sigma$ uncertainty $\Delta S$ on this number 
is determined from both the statistical uncertainty, given by the number of 
Monte Carlo events, and the systematic uncertainty, which is estimated from the 
total uncertainty on the signal cross section given by the user. These numbers are compared to the 
results from the respective experimental search, listed in the 
references shown in \Cref{table:analyses}. There are two possible ways of comparison:
\begin{enumerate}
\item Many experimental searches translate their results into model 
      independent 95\% upper confidence limits on the number of signal 
      events coming from new physics. A quick--and--easy way of comparison 
      is given by computing the parameter,
      \begin{align}
	  r \equiv \frac{S - 1.96 \cdot \Delta S}{S^{95}_\text{Exp.}} = \frac{\text{95\% lower limit on the number of signal events, determined by \Checkmate{}}}{\text{Experimentally measured 95\% confidence limit on signal events}}.
	  \label{eqn:rlimit}
      \end{align}
      In that case, a model can be considered as excluded to the 95\% confidence level, if 
      $r \geq 1$.
\item The user can ask for the explicit confidence level for the given 
      signal. Following common use of experimental searches, \Checkmate{} uses a 
      profile log--likelihood ratio test paired with the \CLs 
      prescription \cite{0954-3899-28-10-313} to determine the confidence 
      level that corresponds to the determined number of signal events and 
      the expected and observed number of events at the experiment. Both the uncertainty 
      on the signal and on the background are taken into account as Gaussian probability distribution functions around 
      the nominal values. The calculation of the \CLs allows for \Checkmate{} to be easily
      included into a fitting routine for the chosen model.
\end{enumerate} 
The aforementioned parameters are determined for all signal regions of 
all selected analyses. If available from the experimental papers, not only 
the observed but also the expected limits are used. The signal region with 
the strongest expected limit is determined and the corresponding 
observed limit is used to state whether the input can be considered excluded or not.

\begin{table}
\centering
\renewcommand{\tabcolsep}{0.3cm}
\begin{tabularx}{\textwidth}{l r | X r r r r}
\hline \hline
Internal name & \#SR & Description & $\sqrt{s}$ & $\mathcal{L}$ \quad \  & Ref & CR? \\
\hline
atlas\_1308\_2631 & 2 & 2 b-jets + \etmiss  & 8 \TeV & 20.1 \invfb & \cite{Aad:2013ija} & $\checkmark$  \\
atlas\_conf\_2012\_104 & 2 &  1 lepton + $\geq$ 4 jets + \etmiss & 8 \TeV & 5.8 \invfb &\cite{ATLAS-CONF-2012-104}&  \\
atlas\_conf\_2012\_147 & 4 & Monojet + \etmiss & 8 \TeV & 10.5 \invfb &\cite{ATLAS-CONF-2012-147}& $\checkmark$ \\
atlas\_conf\_2013\_024 & 3 & 0 lepton + 6 (2 b-)jets + \etmiss & 8 \TeV & 20.5 \invfb &\cite{ATLAS-CONF-2013-024}& $\checkmark$ \\
atlas\_conf\_2013\_035 & 6 & 3 leptons + \etmiss & 8 \TeV & 20.7 \invfb &\cite{ATLAS-CONF-2013-035}& \\
atlas\_conf\_2013\_047 & 9 & 0 leptons + 2-6 jets + \etmiss & 8 \TeV & 20.3 \invfb &\cite{ATLAS-CONF-2013-047}& $\checkmark$ \\
atlas\_conf\_2013\_049 & 9 &2 leptons + \etmiss & 8 \TeV & 20.3 \invfb &\cite{ATLAS-CONF-2013-049}& \\
atlas\_conf\_2013\_061 & 9 &0-1 leptons + $\geq$ 3 b-jets + \etmiss & 8 \TeV & 20.1 \invfb &\cite{ATLAS-CONF-2013-061}& $\checkmark$ \\
atlas\_conf\_2013\_062 & 19 & 1-2 leptons + 3-6 jets + \etmiss & 8 \TeV & 20.3  \invfb& \cite{ATLAS-CONF-2013-062}&  \\
atlas\_conf\_2013\_089 & 12 &  2 leptons + jets + \etmiss  & 8 \TeV &  20.3 \invfb&\cite{ATLAS-CONF-2013-089} & $\checkmark$ \\
cms\_pas\_exo\_12\_048 & 7 & Monojet & 8 \TeV & 19.5 \invfb & \cite{CMS-PAS-EXO-12-048} &  \\
cms\_1303\_2985 & 59 & $\alpha_T$ + b jet multiplicity & 8 \TeV & 11.7 \invfb & \cite{Chatrchyan:2013lya} & $\checkmark$ \\
\hline \hline
\end{tabularx}
\caption{List of currently available analyses in \Checkmate. `\#SR' shows the number 
of signal regions the respective analysis provides.  The rightmost column 
shows whether control regions for that particular analysis have been implemented or not.}
\label{table:analyses}
\end{table}

\section{Using \Checkmate{}}
\label{sec:tutorial}
This section introduces the basic setup of the program. It should give the users enough insight into the program to run 
the code for their own purposes, which does not require knowledge of the 
internal details. These will be introduced in \Cref{sec:internal:delphes} and the Appendices. An even simpler version showing the minimum required information to run \Checkmate{} and 
extract the relevant results is given on the \Checkmate{} website.
\Checkmate{} can easily be downloaded and installed\footnote{If you encounter problems during installation, please contact us via the webpage and we are happy to
help.} by the following set of commands:
\begin{Verbatim}
   wget http://www.hepforge.org/archive/checkmate/CheckMATE-Current.tar.gz
   tar -xvf CheckMATE-Current.tar.gz
   cd CheckMATE-X.Y.Z
   ./configure
   make
\end{Verbatim}
More information on \Root{} prerequisites and installation parameters can be 
found in \Cref{sec:app:installation}. A successful installation should 
have created a \verb@CheckMATE@ binary within the \verb@bin/@ folder of the 
installation directory.


To run \Checkmate{}, the user has to provide at least the following information:
\begin{description}
\item[Name of the run]
A unique name has to be set to define the current run and to set up the output 
folder accordingly.
\item[Analyses to be tested]
The user has to choose at least one of the provided analyses to be used 
by \Checkmate{} for testing the input data against experiment. Any combination 
of the available analyses can be chosen. Note that a single analysis usually defines 
several signal regions, all of which will be investigated by \Checkmate{}.
\item[Events]
Any number of event files can be given as long as they are either in .hep, .hepmc
or .lhe\footnote{The use of lhe event files is strongly discouraged. If they have been
created from a parton level Monte-Carlo program, the lack of a parton shower can significantly alter the analysis
acceptances and lead to incorrect limits. Even if the events have been showered, the 
choice of jet clustering before a detector simulation has been performed can lead to
large differences. If lhe files are required care must be taken when employing analysis that make use of 
flavour tagging. For flavour tagging to be performed, the parton level particles must be 
stored in the event file along with the final state particles.} format, which allows linkage to all modern event generators.
Weighted event files are allowed 
and properly considered within the detector simulation, the analyses and the evaluation. 
\item[Processes]
In case there are multiple input event files, they have to be sorted into processes, since \Checkmate{} needs to know whether they 
correspond to the same or to different production modes. This will change the final summation over all individual 
results; see \Cref{sec:example:evaluation} for more details

\item[Total cross section and error]
In order to properly normalise the final number of events in each signal 
region to the corresponding luminosity of the respective experimental data, the 
total cross section that corresponds to a given process has to be provided. Furthermore, 
an absolute error on that value is necessary to quantify the effect of this 
systematic uncertainty on the final exclusion. Note that all event files for a given process (see above) 
are considered to have identical cross sections, i\/.e.\ they must have been created with the same event generator settings.
\end{description}
In addition to the aforementioned list, several optional parameters are 
available. They are listed and explained in \Cref{tbl:optional_parameters}. \Checkmate{} provides two possibilities to let the user define the options of a certain run:
\begin{description}
\item[Parameter File]
A convenient method to provide the parameters of choice by the user is 
the usage of a separate text-file that stores all information in 
human readable form. A minimal example looks as follows:
\begin{Verbatim}
## General Options
[Mandatory Parameters]
Name: My_New_Run
Analyses: atlas_conf_2013_047

[Optional Parameters]

## Process Information (Each new process 'X' must start with [X])
[gluinogluino]
XSect: 3.53*FB
XSectErr: 1e-5*PB
Events: testfile.hep
\end{Verbatim}
Parameter names and corresponding values have to be separated 
by a colon. Most optional parameters need a Boolean answer; they 
can be either \verb@True@/\verb@False@, \verb@Yes@/\verb@No@, \verb@1@/\verb@0@ 
or \verb@on@/\verb@off@. The items \verb@Analyses@ and \verb@Events@ can consist 
of more than one entry. In that case, they have to be separated by 
commas (line breaks are allowed as long as the new lines start with a whitespace). Every 
separate process has to be introduced with a new \verb@[PROCESSNAME]@ block and 
has to contain the three entries \verb@XSect, XSectErr@ and \verb@Events@ as shown 
in the above example.

In order to load \Checkmate{} via such an input parameter file, the 
user has to run \verb@./CheckMATE PARAMETERFILE@.

\item[Command Line Input]
For an automatised embedding of \Checkmate{} into a program chain, it can be useful if one is not obliged 
to save the parameters in a separate file. For that purpose, all parameters 
mentioned before and in \Cref{tbl:optional_parameters} can also be 
directly entered into \Checkmate{} in the form of command line parameters. The 
above example would look as follows:
\begin{Verbatim}
bin/: ./CheckMATE -n My_New_Run -a atlas_conf_2013_047 -p "gluinogluino" -xs "3.53*FB" -xse "1e-5*PB" testfile.hep
\end{Verbatim}
Note that the list of event files comes last without any prefix indicator and separated 
by spaces only. In case of $N$ event files, there have to be either 
exactly $1$ or exactly $N$ entries in the options 
for \verb@-p, -xs@ and \verb@-xse@ which have to be separated 
by semicolons. If there is only one entry, \Checkmate{} will assume that all 
given event files belong to the same process with identical cross sections 
and errors. Otherwise it will consider event file $i$ to belong to the process 
at the $i$th position etc.
\end{description}
Running either of the aforementioned examples with the event file \verb@testfile.hep@ provided by the \Checkmate{} installation should, after confirmation and some 
intermediate output, produce the following result (which should not be surprising as the 
test events effectively do not contain any signal events):
\begin{Verbatim}
[...]
Test: Calculation of r = signal/(95%CL limit on signal)
Result: Allowed
Result for r: r_max = 0.0
SR: atlas_conf_2013_047 - AL
\end{Verbatim}
The most important result is given by the \verb@Result:@ line, which will either state \verb@Allowed@ or \verb@Excluded@. This statement alone allows the user to quickly perform simple tests on his model of 
interest. However, if they should be interested in the intermediate details that lead to this result, they can find further information in \Cref{sec:app:investigation}.
\begin{table}
\begin{tabularx}{\textwidth}{llX}
\hline \hline
Input File & Command Line & Definition \\
\hline
\verb@FullCL@ & \verb@-cl, --full-cl@ & Calculates \CLs explicitly for each signal region and uses it for the exclusion statement. \\
\verb@RandomSeed@ & \verb@-rs, --random-seed@ & Defines a fixed seed for the random number generator. \\
\verb@OutputDirectory@ & \verb@-od, --outdir@ & This defines the path to the directory into which all \Checkmate{} output should be saved. Both absolute and relative paths are allowed. If not set, \Checkmate{} will create and use a \verb@results/@ directory within its main folder. \\
\verb@OutputExists@ & \verb@-oe, --output-exists@ & If there are already output files with the same run name in the same output directory as the current run, \Checkmate{} can deal with this in different ways: \verb@overwrite@ will delete the old output and restart with a new run. \verb@add@ will consider the current run as an addendum to the previous run and will add the current results to the old ones.  \verb@overwrite@ will always ask the user via prompt. \\
\verb@SkipParamCheck@ & \verb@-sp, --skip-paramcheck@ & Skip startup parameter check. \\
\verb@SkipDelphes@ & \verb@-sd, --skip-delphes@ & Only works if the output directory already exists and has Delphes \Root{} files. Delphes won't run and the given \Root{} files in the output directory are re--processed by the given analyses. This is only useful in case the user changed parts of the analysis and wants to test with the same input. \\
\verb@SkipEvaluation@ & \verb@-se, --skip-evaluation@ & The input files are only processed by the detector simulation and the chosen analyses, but the analysis result will not be further evaluated and compared to experimental data. It should be used when doing a control region analysis or for debugging purposes. \\
\verb@QuietMode@ & \verb@-q, --quiet@ & No output will be print on screen. \\
\verb@VerboseMode@ & \verb@-v, --verbose@ & All Delphes and analysis output will be printed to the standard output. \\
\verb@TempMode@ & \verb@-t, --temporary@ & All Delphes \Root{} files will be deleted after the analysis step to save hard disk space. \\
\hline \hline
\end{tabularx} 
\fvset{samepage=true, fontsize=\normalsize}
\SaveVerb{v1}|[Optional Parameters]|
\SaveVerb{v2}|PARAMETER = VALUE|
\SaveVerb{v3}|True/False, Yes/No, 1/0|
\SaveVerb{v32}|No|
\SaveVerb{v4}|on/off|
\SaveVerb{v5}|-PARAMETER=VALUE|
\SaveVerb{v6}|-PARAMETER|
\SaveVerb{v7}|True|
\caption{List of all currently available optional parameters  for \Checkmate{}. The first column 
denotes the string to be used in the parameter card; it has to be added as a separate 
line in the \protect\UseVerb{v1} block as \protect\UseVerb{v2}. Boolean values can be given 
as any of the following: \protect\UseVerb{v3}
or \protect\UseVerb{v4}. The standard value for all these is \protect\UseVerb{v32}. The second 
column names the corresponding parameter to be used in the command line input in both a short 
and an extended format. They can be used by adding \protect\UseVerb{v5} or \protect\UseVerb{v6} to set 
Boolean parameters to \protect\UseVerb{v7}.}
\label{tbl:optional_parameters}
\fvset{samepage=true, fontsize=\scriptsize}
\end{table}
\vfill
\pagebreak
\section{Internal Details}
\subsection{Investigating the Results}
\label{sec:app:investigation}
In order to demonstrate the structure of \Checkmate's output, we show and discuss 
the results for a reasonably complex example. We will describe the steps that \Checkmate{} performs in detail and discuss the information that can be extracted from the results. Knowing these details allows use of \Checkmate{} for more detailed studies and/or to adapt the code to the user's needs.
\makeatletter
\renewcommand*{\toclevel@subsection}{-2} 
\subsubsection{Description of the Example Run}
\makeatother
\begin{figure}[b]
\scriptsize
\begin{SaveVerbatim}{input}   
## General Options
[Mandatory Parameters]
Name: cMSSM
Analyses: atlas_conf_2013_047

[Optional Parameters]
FullCL: yes
OutputDirectory: /hdd/results
RandomSeed: 42

## Process Information (Each new process 'X' must start with [X])
[gluino_pair]
XSect: 5.52*FB
XSectErr: 0.68*FB
Events: /hdd/CheckMATE_example/gg_1.hep, 
        /hdd/CheckMATE_example/gg_2.hep
        
[gluino_squark]
XSect: 0.926*FB
XSectErr: 0.173*FB
Events: /hdd/CheckMATE_example/gs_1.hep, 
        /hdd/CheckMATE_example/gs_2.hep
\end{SaveVerbatim} 
\fbox{\BUseVerbatim{input}} 
\caption{Parameter input file that corresponds to the example scenario described in the text.}
\label{fig:verb:input}
\end{figure}
Input events have 
been produced with MadGraph using the CMSSM model 
with $m_0 = 2000 ~ \GeV, m_{1/2} = 450 ~ \GeV, \tan \beta = 30, A_0 = -4000 ~ \GeV$ 
and $\mu > 0$. The total cross sections for $\tilde{g}\tilde{g}$ 
and $\tilde{g}\tilde{q}$ production at $pp$ collisions at $\sqrt{s} = \unit[8]{\TeV}$ have been estimated with NLL-Fast \cite{Beenakker:1997ut,Kulesza:2008jb,Kulesza:2009kq,Beenakker:2009ha,Beenakker:2010nq,Beenakker:2011fu}, together 
with the respective theoretical uncertainties from scale 
variations and PDF dependences, to 
be $\sigma_{\tilde{g}\tilde{g}} = (5.52 \pm 0.68) ~ \fb$ 
and $\sigma_{\tilde{g}\tilde{q}} = (0.926 \pm 0.173) ~ \fb$. Squark pair 
production has been neglected\footnote{$m_{\tilde{q}} = \unit[2.17]{\TeV},  \sigma_{\tilde{q}\tilde{q}} \approx 0.02  ~ \fb$}. For each 
production process, we 
created two event files with $10000$ events each, which we desire to be tested 
against \verb@atlas_conf_2013_047@, i.e.\ a zero lepton search with 
missing energy and two to six jets. The \CLs{} value should be used for the final exclusion statement. The corresponding input parameter 
file is shown in \Cref{fig:verb:input}.

\subsubsection{Structure of the Results Folder}
A successful run with the parameter file shown in \Cref{fig:verb:input} should have 
created a results folder at the desired destination. It consists of the 
following files and directories:
\begin{Verbatim}
/hdd/results/cMSSM: ls
analysis  delphes  evaluation  progress.txt  result.txt
\end{Verbatim}

To keep a general overview on all processed data files, \Checkmate{} produces 
a file \verb@progress.txt@ at the top level of the results directory. It lists 
the cross section and process information of all events files that have been 
fully analysed by \Checkmate{}, together with the time of production and a unique 
prefix that identifies the output files which correspond to a given input file. 
In the aforementioned example, the progress file would look as follows:
\begin{Verbatim}
/CheckMATE_example: cat progress.txt
#Prefix  EventFile                        Sigma     dSigma    Process        Date and Time of Procession  
000      /hdd/CheckMATE_example/gg_1.hep  5.52 FB   0.68 FB   gluino_pair    2013-11-28 13:29:46.981908   
001      /hdd/CheckMATE_example/gg_2.hep  5.52 FB   0.68 FB   gluino_pair    2013-11-28 13:30:43.669558   
002      /hdd/CheckMATE_example/gs_1.hep  0.926 FB  0.173 FB  gluino_squark  2013-11-28 13:31:52.322075   
003      /hdd/CheckMATE_example/gs_2.hep  0.926 FB  0.173 FB  gluino_squark  2013-11-28 13:32:59.544565 
\end{Verbatim}
This file is particularly important when the user uses the \verb@add@ 
(see \Cref{tbl:optional_parameters} for more details) feature 
of \Checkmate{}, i.e.\ when they add data to an already existing output directory, 
as all necessary information from previous runs is stored in the above shown progress file.
%
%
\subsubsection{Delphes}
As a first step, the input event file has to be processed by the 
detector simulation Delphes, which will produce the following output:
\begin{Verbatim}
/hdd/results/cMSSM: ls delphes
000_delphes.root  001_delphes.root  002_delphes.root  003_delphes.root  log_delphes.txt  merged.tcl
\end{Verbatim}
First, the simulation has to be set up according to the required information 
the later analyses need (see \Cref{sec:delphes_sim_set}). This so-called \emph{detector--card}, which 
saves the settings in a specific syntax the Delphes code is able to parse 
internally, is generated at the beginning of the \Checkmate{} run and is 
saved in \verb@/delphes/merged.tcl@ for later inspection by the user. All 
output that is generated by Delphes will be stored in a single log 
file, \verb@log_delphes.txt@ and can be examined later; this can e.g.\ help to understand unexpected 
behaviour of the simulation.

Furthermore, \Checkmate{} will keep all \Root{} trees Delphes generates and 
stores them in the same folder. They can be investigated directly by the 
user, e.g.\ by using the \Root{} internal TBrowser object 
 or indirectly by the subsequent analysis 
step. The naming of the individual files uses the prefix mapping as listed in the previously mentioned \verb@progress.dat@. There will always be 
one \Root{} file per input event file, irrespective of the number of chosen analyses.

It should be noted that, depending on the number and complexity of the 
chosen analyses and the size of the input event files, the 
resulting \Root{} trees can be very large and the corresponding 
file might consume large amounts of disk space\footnote{The size of the \Root{} file depends
upon the type of signal events produced (e.g. events with lots of hard jets give 
larger files), the particular Monte-Carlo program used to generate the events and the particular 
analysis being run. As a rough estimate however, 1000 events 
should produce a \Root{} file size below 20 Mb.}.
To avoid this, the user 
can declare that these files should be immediately deleted after 
the analysis step by using the \verb@tempMode@ setting 
in \Checkmate{} (see \Cref{tbl:optional_parameters}). Still, the 
complex \Root{} structure needs the output of \emph{one} Delphes run to 
be saved as \emph{one complete file}, i.e.\ it is impossible to pipe the 
data on the fly from Delphes to the analysis code. The user therefore 
has to make sure that enough disk space is available to save at least 
one full \Root{} file, even when \verb@tempMode@ is activated. Should 
disk space be very limited, the user might consider splitting the 
input event file into multiple subfiles. 
%

\subsubsection{Analysis}
\begin{figure}[b]
\begin{SaveVerbatim}{Cutflow}
# ATLAS
# ATLAS-CONF-2013-047
# 0 leptons, 2-6 jets, etmiss
# sqrt(s) = 8 TeV
# int(L) = 20.3 fb^-1

Inputfile:       /hdd/results/cMSSM/delphes/000_delphes.root
XSect:           5.52 fb
 Error:          0.68 fb
MCEvents:        10000
 SumOfWeights:   10000
 SumOfWeights2:  10000
 NormEvents:     111.608

Cut  Sum_W  Sum_W2  Acc     N_Norm   
0    10000  10000   1       111.608  
1    4439   4439    0.4439  49.5427  
2    3694   3694    0.3694  41.2279  
3    3680   3680    0.368   41.0717  
3J1  3676   3676    0.3676  41.027   
3J2  3048   3048    0.3048  34.0181  
4    3680   3680    0.368   41.0717  
4J1  3622   3622    0.3622  40.4243  
4J2  3002   3002    0.3002  33.5047  
[...]
\end{SaveVerbatim}

\begin{SaveVerbatim}{Signal}
# ATLAS
# ATLAS-CONF-2013-047
# 0 leptons, 2-6 jets, etmiss
# sqrt(s) = 8 TeV
# int(L) = 20.3 fb^-1

Inputfile:       /hdd/results/cMSSM/delphes/000_delphes.root
XSect:           5.52 fb
 Error:          0.68 fb
MCEvents:        10000
 SumOfWeights:   10000
 SumOfWeights2:  10000
 NormEvents:     111.608

SR  Sum_W  Sum_W2  Acc     N_Norm    
AL  2866   2866    0.2866  31.9868   
AM  353    353     0.0353  3.93975   
BM  478    478     0.0478  5.33485   
BT  35     35      0.0035  0.390627  
CM  1416   1416    0.1416  15.8037   
CT  89     89      0.0089  0.993309  
D   870    870     0.087   9.70988   
EL  1655   1655    0.1655  18.4711   
EM  1255   1255    0.1255  14.0068   
ET  609    609     0.0609  6.79691  
\end{SaveVerbatim}
\fbox{\BUseVerbatim{Cutflow} \quad } 
\qquad
\fbox{\BUseVerbatim{Signal} \quad }
\caption{Example analysis output for one particular input event file. The left 
figure shows the first lines of the cutflow file \texttt{analysis/000\_atlas\_conf\_2013\_047\_cutflow.dat}. It first shows general information for the whole event file, like the total number of events, the sum of their weights, the sum of their squared weights and the number of physical events given by cross section $\times$ integrated luminosity. It follows a list of all cutflow levels with the sum of all weights (and squared weights) of the events that passed the corresponding cuts, the acceptance relative to the total sum of weights of the file and the number of physical events given by acceptance $\times$ cross section $\times$ integrated luminosity. The right figure shows the full signal result file \texttt{analysis/000\_atlas\_conf\_2013\_047\_signal.dat}, which gives the the same 
information in the individual signal regions.}
\label{fig:verb:cutflow_signal}
\end{figure}
Looking into the \verb@analysis@ folder reveals the following files:
\begin{Verbatim}
/hdd/results/cMSSM: ls analysis
000_atlas_conf_2013_047_cutflow.dat  002_atlas_conf_2013_047_cutflow.dat
000_atlas_conf_2013_047_signal.dat   002_atlas_conf_2013_047_signal.dat
001_atlas_conf_2013_047_cutflow.dat  003_atlas_conf_2013_047_cutflow.dat
001_atlas_conf_2013_047_signal.dat   003_atlas_conf_2013_047_signal.dat
log_analysis.txt
\end{Verbatim}
The \Root{} files produced by Delphes will automatically be processed by all the 
chosen analyses. Each input file will create its individual outputs, identified by the prefix defined 
in the progress file. Most analyses that are included in \Checkmate{} will produce 
only two types of output: \verb@cutflow@-files show the absolute and relative 
amounts of data that pass the individual event selection cuts of the 
corresponding analysis, whereas \verb@signal@-files give the amount of 
events that fall into each signal region defined by the analysis. The latter file is more 
important for \Checkmate{} as its content is used for the later evaluation step.  
The cutflow information can be used by the user for various purposes (e.\/g.\ validating 
analyses by comparing with the cutflows produced by the collaborations), but it is not 
needed by \Checkmate{}. 

Examples for both types of output are shown in \Cref{fig:verb:cutflow_signal}. After 
some general information on the analysis\footnote{This header information is printed in 
all output files, but has been removed from all further example output files in 
this section.} and the processed event file, a list of all individual 
cutflow milestones / signal regions follows. For each of these, \Checkmate{} lists 
the sum of weights (and squared weights) of all events that passed the corresponding cut(s) (\verb@Sum_W, Sum_W2@), the 
relative number compared to the total sum of weights (\verb@Acc@) as well as the 
 physically expected number of events (\verb@N_Norm@),  after normalising 
to the given total cross section of the data and the luminosity of the respective analysis. In case of unweighted events, \verb@Sum_W@ and \verb@Sum_W2@ corresponds to the number of Monte Carlo events in the respective region. However, in case of weighted events, both these columns are needed by \Checkmate{} to properly calculate the statistical error in the upcoming evaluation step.

We note that usually the analyses do not produce any output on the screen but 
just write into the above described output files. However, in case of 
errors\footnote{Warnings \textit{'** WARNING: cannot access branch 
'Jet2', return NULL pointer'} might appear in the analysis log file. These are 
inevitably produced internally by \Root{} but do not hint to any problems and 
should be ignored by the user.} and/or if the user adds extra print 
commands to the analysis code, the output will be stored in \verb@log_analysis.txt@.

\subsubsection{Evaluation}
\label{sec:example:evaluation}
\begin{figure}
\centering
\scriptsize
\begin{SaveVerbatim}{Counting_Events}
Prefix  N_TotMC   AL     stat  sys   AM    stat  sys   BM    stat  sys   BT    stat  sys   CM     stat  sys   CT    stat  sys   ..

Process: gluino_pair
000     10000.00  31.99  0.60  3.94  3.94  0.21  0.49  5.33  0.24  0.66  0.39  0.07  0.05  15.80  0.42  1.95  0.99  0.11  0.12  ..
001     10000.00  32.84  0.61  4.04  4.04  0.21  0.50  5.49  0.25  0.68  0.41  0.07  0.05  16.91  0.43  2.08  1.08  0.11  0.13  ..
----------------------------------------------------------------------------------------------------------------------------------
Tot     20000.00  32.41  0.43  3.99  3.99  0.15  0.49  5.41  0.17  0.67  0.40  0.05  0.05  16.36  0.30  2.01  1.04  0.08  0.13  ..

Process: gluino_squark
002     10000.00  4.99   0.10  0.93  1.34  0.05  0.25  1.92  0.06  0.36  0.47  0.03  0.09  1.89   0.06  0.35  1.31  0.05  0.24  ..
003     10000.00  5.00   0.10  0.93  1.43  0.05  0.27  1.92  0.06  0.36  0.42  0.03  0.08  1.84   0.06  0.34  1.27  0.05  0.24  ..
----------------------------------------------------------------------------------------------------------------------------------
Tot     20000.00  4.99   0.07  0.93  1.38  0.04  0.26  1.92  0.04  0.36  0.44  0.02  0.08  1.86   0.04  0.35  1.29  0.03  0.24  ..

==================================================================================================================================
Tot     40000.00  37.41  0.43  4.10  5.37  0.15  0.56  7.33  0.18  0.76  0.84  0.05  0.10  18.22  0.30  2.04  2.32  0.08  0.27  ..
\end{SaveVerbatim}

\fbox{\BUseVerbatim{Counting_Events}} 
\caption{Example result of the event summation routine during \Checkmate's evaluation 
step from \texttt{evaluation/atlas\_conf\_2013\_047\_event\_numbers.txt}. For each signal region, the final event numbers are averaged for events from the 
same and added for events in different processes. The errors are combined accordingly.}
\label{fig:verv:eval}
\end{figure}

We discuss the last step, namely examining the \verb@evaluation@ folder:
\begin{Verbatim}
/hdd/results/cMSSM: ls evaluation
atlas_conf_2013_047_cl_limits.txt  atlas_conf_2013_047_event_numbers.txt  atlas_conf_2013_047_r_limits.txt  best_signal_regions.txt
\end{Verbatim}
In the evaluation step, the results of all individual analyses have to be combined 
to compare the total number of events to the experimental limits. The result of this summation 
is stored in \texttt{evaluation/ANALYSISNAME} \texttt{\_event\_numbers.txt} (as an example, see 
\Cref{fig:verv:eval}) and utilises the following logic. Using the information 
from the \verb@progress@ file, \Checkmate{} will determine which \verb@*_signal.dat@ files 
correspond to the same and which to different input processes. Events that correspond 
to the same process will be considered as if they originated from one large 
input file, i.e.\ their results will be \textit{averaged} by taking the corresponding weights properly into account. The statistical error 
is calculated from the sum of weights and sum of squared weights in the given signal region, whereas the systematic error is given as $\delta \sigma / \sigma$ times the 
number of signal events after cuts, where $\sigma$ and $\delta \sigma$ are the cross section 
and error given as input by the user for the process the events correspond to (see \Cref{sec:tutorial}). Subsequently, results 
from different processes will contribute individually and therefore will be \textit{added} to determine the total expected number of signal events. The 
errors are considered independent and hence added in quadrature. This 
is done for each signal region in each selected analysis separately.

These results are then compared to the experimental limits, separately for each 
signal region in each selected analysis. The conservative $r$--limit as 
defined in \Cref{eqn:rlimit} is always evaluated and the results 
are stored in \texttt{evaluation/} \texttt{ANALYSISNAME\_r\_limits.txt}, together with 
the total number of signal events after cuts, the errors and the experimental 
upper limits. An example is shown in \Cref{fig:verb:rlimits} for the above example run. 

If the \verb@FullCLs@ option has been activated in the input parameters, \Checkmate{} will 
furthermore calculate the \CLs value, using the total number and uncertainty 
of signal events determined by \Checkmate, the total number and uncertainty of 
expected background events and the number of observed events, all taken from the 
respective experimental source. The expected limit is determined by 
setting the `observed' number of events in the signal region equal
to the expected background (\verb@O = B@).  Since \CLs is determined 
numerically, the result has a computational uncertainty that is also printed 
in the output to state the numerical stability of the result. All these numbers 
are stored in \verb@evaluation/ANALYSISNAME_cl_limits.txt@.

\begin{figure}[b]
\begin{SaveVerbatim}{r_limits}
SR  S      dS_stat  dS_sys  dS_tot  S95_obs  S95_exp  r^c_obs  r^c_exp  
AL  37.41  0.43     4.10    4.12    1341.20  1135.00  0.02     0.03     
AM  5.37   0.15     0.56    0.58    51.30    42.70    0.08     0.10     
BM  7.33   0.18     0.76    0.78    14.90    17.00    0.39     0.34     
BT  0.84   0.05     0.10    0.11    6.70     5.80     0.09     0.11     
CM  18.22  0.30     2.04    2.07    81.20    72.90    0.17     0.19     
CT  2.32   0.08     0.27    0.28    2.40     3.30     0.74     0.54     
D   12.15  0.24     1.29    1.32    15.50    13.60    0.62     0.70     
EL  21.56  0.33     2.38    2.41    92.40    57.30    0.18     0.29     
EM  16.08  0.29     1.79    1.81    28.60    21.40    0.44     0.59     
ET  8.10   0.20     0.89    0.91    8.30     6.50     0.76     0.97     
\end{SaveVerbatim}

\fbox{\BUseVerbatim{r_limits}} 
\caption{Example result of the conservative $r$ test defined in \Cref{eqn:rlimit}, written into \texttt{evaluation/atlas\_conf\_2013\_047\_r\_limits.txt}. The 
observed and expected upper 95\% confidence 
limits on the number of signal events is taken from the respective experimental source.}
\label{fig:verb:rlimits}
\end{figure}

\begin{figure}
\begin{SaveVerbatim}{cl_limits}
SR  S      dS_stat  dS_sys  dS_tot  B        dB      O        CL_obs  dCL_obs  CL_exp  dCL_exp  
AL  37.41  0.43     4.10    4.12    4700.00  500.00  5333.00  0.9754  0.0147   0.9204  0.0224   
AM  5.37   0.15     0.56    0.58    122.00   18.00   135.00   0.9269  0.0137   0.8307  0.0210   
BM  7.33   0.18     0.76    0.78    33.00    7.00    29.00    0.3453  0.0137   0.4566  0.0125   
BT  0.84   0.05     0.10    0.11    2.40     1.40    4.00     0.8518  0.0095   0.6890  0.0158   
CM  18.22  0.30     2.04    2.07    210.00   40.00   228.00   0.7395  0.0136   0.6691  0.0171   
CT  2.32   0.08     0.27    0.28    1.60     1.40    0.00     0.1063  0.0077   0.2225  0.0082   
D   12.15  0.24     1.29    1.32    15.00    5.00    18.00    0.1590  0.0049   0.0824  0.0041   
EL  21.56  0.33     2.38    2.41    113.00   21.00   166.00   0.7574  0.0072   0.3704  0.0109   
EM  16.08  0.29     1.79    1.81    30.00    8.00    41.00    0.3927  0.0064   0.1296  0.0055   
ET  8.10   0.20     0.89    0.91    2.90     1.80    5.00     0.0609  0.0027   0.0077  0.0013   
\end{SaveVerbatim}

\fbox{{\BUseVerbatim{cl_limits}}}
\caption{Example output of the \CLs test written into \texttt{evaluation/atlas\_conf\_2013\_047\_cls\_limits.txt}. B (expected background), dB (error on the 
expected background) and O (observed events) are taken from the 
experimental collaborations.}
\label{fig:verb:cllimits}
\end{figure}

\begin{figure}[b]
\begin{SaveVerbatim}{summeduplimits}
analysis             bestSR  r_obs^c  r_exp^c  CLs_obs  dCLs_obs  CLs_exp  dCLs_exp  [...]
atlas_conf_2013_047  ET      0.76     0.97     0.0609   0.0027    0.0077   0.0013    [...]
\end{SaveVerbatim}

\fbox{{\BUseVerbatim{summeduplimits}}}
\caption{Example output written into \texttt{evaluation/best\_signal\_regions.txt}. For each 
analysis that has been considered in the given run, this file sums up the results of the signal region 
with the largest expected sensitivity.}
\label{fig:verb:summeduplimits}
\end{figure}

As the last step, \Checkmate{} will search for the signal region with 
the largest expected sensitivity; for the $r$-limits this corresponds to 
the signal region with the largest \verb@r^c_exp@, whereas for 
the \CLs test the one with the smallest \verb@CL_exp@ is chosen. The results of the most 
sensitive signal region 
of each analysis separately is written in the file \verb@best_signal_regions.txt@ and an example is shown in
\Cref{fig:verb:summeduplimits}. \Checkmate{} then 
chooses the most sensitive region among these and the corresponding 
observed result will be used to finally conclude whether the input can be 
considered to be excluded or not, i.e.\ in the case of the $r$-limit if \verb@r^c_obs@ is 
larger than one or for the \CLs test if the corresponding \verb@CL_obs@ is 
smaller than 0.05. This statement is printed on the screen during 
the \Checkmate{} run and can be found in the file \verb@result.txt@ at the 
top of the output directory. In the above example, it reads as follows:
\begin{Verbatim}
Test: Calculation of CLs from determined signal
Result: Allowed
Result for CLs: cls_min = 0.0695756126718
Result for r: r_max = 0.757777916596
SR: atlas_conf_2013_047 - ET
\end{Verbatim}

It should be noted that the $r$-limit is usually weaker than 
the \CLs{} result, as the first uses the total uncertainty on the 
signal in a more conservative manner than the latter. It is therefore not 
impossible that \CLs{} $\lesssim 0.05$ and $r \lesssim 1$, i.e.\ \CLs{} excludes 
but $r$ allows the input model. In those cases, the model will still be 
considered excluded, since if \CLs{} is evaluated, it will exclusively be used 
to determine the final exclusion statement.

\subsection{Delphes Tunings}
\label{sec:internal:delphes}
Each of the given event files will be processed with version 3.0.10 of the fast 
multipurpose detector simulation Delphes \cite{deFavereau:2013fsa}. This tool includes the experimental resolutions and efficiencies of the two LHC multipurpose detectors, \textsc{Atlas} 
and \textsc{Cms}, by parametrising a large list of detector 
effects (see \Cref{fig:flowdiagram} or details in Ref.~\cite{deFavereau:2013fsa}). However 
exhaustive this list seems, many analyses \Checkmate{} uses need extra functionalities, 
which lie beyond the published version of Delphes. We have added these extra functionalities
as well as re-parametrised many of the pre-existing functions of Delphes in order to
better model the LHC detectors. These changes were motivated by detailed 
comparisons to the experimental results of individual simulation components  
which showed that the standard Delphes configuration needed to be improved. In 
the remainder of this section we qualitatively describe these changes in detail. Quantitative results are listed in 
\Cref{sec:app:detector_tune}.

\subsubsection{Improved Description of Lepton Reconstruction}
In order to properly estimate the measurement of leptons inside a detector, there are 
two main effects which have to be taken into account:
\begin{enumerate}
\item Inaccuracies in both the tracker and the calorimeter or muon spectrometer 
lead to an uncertainty in the kinematical properties of the lepton candidate. These 
can be estimated by applying a Gaussian smearing on every candidate, depending on its 
energy and position in the detector. 
\item Algorithms to reconstruct electrons inside the calorimeter and to identify muons 
by associating tracks to hits in the muon chamber might fail. Hence a given truth lepton 
should only appear in the list of reconstructed lepton objects with a probability $\epsilon_\ell < 1$ describing the reconstruction/identification efficiency.
\end{enumerate}
There exist sophisticated experimental studies which provide quantitative 
statements for these effects, e.g.\ in the form of probability functions
. These can 
be used in detector simulations like Delphes to reproduce the experimental effects 
without having to reproduce all the internal details. We provide references to all the studies we used in this section and \Cref{sec:app:detector_tune}.

Delphes uses text files, so--called \emph{detector--cards}, which provide all 
necessary functions for both ATLAS and CMS, such that the users do not need to parametrise these effects themselves but can use the simulation 
straight away. However, on further investigation it turns out that the implemented 
standard functions are rather simplified versions of the usually complex distributions 
the experiments measure. Depending on the kinematics of the model and the considered 
analysis, the results from the simulation can deviate significantly from the experimental measurement.

\Checkmate{} has therefore replaced most of these functions, using 
publicly available data from the experiments\footnote{For now, these improvements 
have only been implemented in \Checkmate{} for the ATLAS experiment. They are planned for CMS at a 
later point.}. These include:
\begin{itemize}
\item Detailed $\pT$ and $\eta$ dependent reconstruction and identification 
efficiency functions for `medium' and `tight' electrons\footnote{`loose' electrons are considered in a special manner, see  \Cref{sec:app:detector_tune}.} \cite{Aad:2011mk, ATL-COM-PHYS-2013-1287}.
\item Muon reconstruction efficiency with respect to the type of detector that will 
measure the candidate according to its position, using a $\phi$--$\eta$ map of the muon 
spectrometer \cite{ATLAS-CONF-2011-063}.
\item Gaussian smearing of the momenta of the electron and muon candidates with $E_T$-- and $\eta$--dependent 
width \cite{2012EPJWC..2812039S, ATLAS-CONF-2011-063, Aad:2011mk, Aad:2009wy}.
\end{itemize}

\subsubsection{Improved Description of Jet Tagging}
Jets that originate from $b$-quarks or hadronically decaying $\tau$-leptons can in some cases 
be distinguished from other hadronic jets, e.g.\ by measuring displaced vertices or reconstructing 
a distinctive track signature. Sophisticated algorithms are used by the experimental 
collaborations  on every reconstructed jet object and tag them as a $b$-quark or a $\tau$ 
if they pass certain criteria. These algorithms usually have various working points 
depending on how pure a sample is demanded and what background rejection is required.
Similar to the lepton efficiencies, Delphes already provides rough efficiency functions 
for these tagging algorithms, which have been replaced by more detailed functions in \Checkmate{}:
\begin{itemize}
\item $\pT$ dependent $b$-quark identification efficiency and mis-tagging probability for jets with and 
without a $c$-quark component \cite{ATLAS-CONF-2011-102, ATLAS-CONF-2012-039, ATLAS-CONF-2012-040, 
ATLAS-CONF-2012-097}.
\item $\pT$ dependent $\tau$ tagging efficiencies for signal and background, distinguishing 
between 1-track and 3-track candidates and deducing the candidate's charge from the combined 
charge of the track particles \cite{ATLAS-CONF-2011-152}.
\end{itemize}

\subsubsection{Flag Members for Efficiency, Isolation and Tagging}
Delphes provides an isolation module, which checks the calorimeter and/or tracking environment 
around a particular candidate and removes that candidate if there is too much activity in its 
vicinity. This functionality is sufficient if only one analysis is being performed with Delphes
and the analysis in question only requires a single isolation condition. However, many analyses require
at least two different isolation conditions per final state object. Examples are many of the 
exclusive ATLAS lepton studies that use loosely isolated leptons in veto conditions but require
tightly isolated leptons for signal regions.

The same problem appears for efficiencies, which remove objects with a 
particular  $E_T$- and $\eta$-dependent probability to parametrise losses during 
reconstruction and identification. These efficiencies depend on the precise experimental definition of the object in question. Many analyses use  a loose definition --- corresponding to a relatively large efficiency --- to 
perform event vetoes or overlap removals, and then use stronger requirements with smaller efficiency for 
the actual signal objects. For jet tagging, i.e.\ for labelling a jet as having most 
likely originated from a $b$ quark or a hadronically decaying $\tau$ lepton, the original version 
of Delphes only provides binary member variables to state whether a particular tagging test 
succeeded or not. Therefore problems arise
as soon as one needs to store results of multiple tagging algorithms,
e.g.\ with different working points for the signal efficiency.

For these purposes, charged lepton\footnote{Throughout this article, the word `lepton' excludes 
$\tau$ leptons, which are reconstructed as jets with a $\tau$ tag.}, photon and jet objects 
have been extended by one additional member, called \emph{flag}, for each of the above 
mentioned checks, with the tagging being used exclusively for jets: If the $n$th flag condition succeeds, 
the member's flag value is increased by $2^n$. The final flag value of a candidate therefore 
represents, using binary decomposition, which tests have been passed or failed, and as such allows 
different test outcomes to be checked independently. The current implementation allows for the 
check of at most 32 independent conditions per flag. 

\subsubsection{Detector Simulation Settings}
\label{sec:delphes_sim_set}
As described above, Delphes has been modified to be able to 
produce all necessary information needed by the various analyses. To do so, the following 
information is required for each analysis:
\begin{itemize}
\item  The experiment (ATLAS or CMS) the analysis corresponds to,
\item  the dR cone parameter and $\pT^\text{min}$ for the FastJet program (all current analyses in \Checkmate{} use the 
anti-kt jet clustering algorithm \cite{Cacciari:2008gp}),
\item  potentially, the dR parameter of a second type of jet,
\item  the working signal efficiencies of all required b-/$\tau$-tagging algorithms and
\item  which isolation criteria for electrons/muons/photons have to be tested, including
\begin{itemize}
  \item  the size $\Delta R$ of the cone around the candidate used to define the isolation criterion,
  \item  whether tracks or calorimeter entries should be used for the isolation\footnote{The distance parameter $\Delta R$ is defined 
  as $\sqrt{\Delta \eta^2 + \Delta \phi^2}$, where $\Delta \eta$ denotes the pseudorapidity 
  difference and $\Delta \phi$ the relative angle projected onto the plane perpendicular to the 
  beam axis.},
  \item  the minimum $\pT$ of objects within the $\Delta R$ cone to be taken into account and
  \item  the maximum $\sum_{\Delta R} \pT$ (absolute or relative to $\pT^\text{cand}$) that is allowed within the cone of a candidate.
\end{itemize}
\end{itemize}

\subsubsection{Merge of Settings}
Many of the aforementioned settings are identical for different analyses, e.g.\ signal electrons 
with the same identification efficiency, jet tags corresponding to common working points or 
identical isolation conditions for final state objects. Before each run, \Checkmate{} collects the 
individual requirements of each chosen analysis, checks for common settings and  merges them to a detector 
card with the smallest number of necessary modules. This both saves computing time and
significantly reduces the size of the Delphes result files.

\section{Analyses}
\Checkmate{} provides a variety of different analyses against which the users can test their models. In 
this section we give an overview of what defines a particular analysis. 
See \Cref{sec:app:addanalyses} for details on where 
this information can be found within \Checkmate{} and what users who would 
like to create their own analysis need to provide.


\subsection{Analysis Code}
The analysis part of \Checkmate{} is based on the program package \Root{} which takes as 
input the \Root file output of the modified Delphes detector simulation.  The final output from 
any analysis contains the numbers of events that pass each recorded cut and
the final number of events in all the signal regions of interest, as can be seen for an example in \Cref{fig:verb:cutflow_signal}. All counters for these
cutflows and the various signal regions are defined in the initialization process. In 
addition, kinematical variables relevant for the analysis can be given here.

The analysis begins with the definition of reconstructed 
final state objects. Potential isolation and tagging conditions must be defined in the relevant Delphes detector card, see \Cref{sec:delphes_sim_set}. 
If relevant for the analysis, trigger cuts on the events have to be imposed, such as cuts on 
the transverse momentum of  leptons, jets or transverse missing momentum.
Many of the experimental analyses are based on triggers that are not 100\% efficient for 
all signal regions.
Unfortunately, as of now, the experimental collaborations do not provide 
detailed efficiency maps for the employed triggers.
In the absence of this data, \Checkmate{} analyses uses tuned triggers according
to the supplied cutflows and parameter scans to match the experimental result as closely as possible.


Many analyses apply various cleaning cuts to remove beam and/or cosmic background events.
However, since \Checkmate{} is not running with real experimental data, no event cleaning 
needs to take place. In order to account for 
the loss of experimental luminosity due to these effects however, the reduction in 
the number of events used in the 
experimental analysis is applied as a flat efficiency factor. The reduction depends on the particular analysis but is never more than a few percent.

The analysis code tries to replicate as closely as possible the analysis structure of 
the corresponding experimental study.
Cuts are placed on various final states and kinematical variables\footnote{More 
advanced observables such as the stransverse mass are 
included by integrating the Oxbridge Kinetics Library into \Checkmate{} \cite{Lester:1999tx,Barr:2003rg}.} and corresponding 
cutflow counters keep track of the number of events that survive each cut. 

The analyses in \Checkmate{} all follow the philosophy that whenever a cutflow table is presented in 
the original study, the analysis follows the cutflow
exactly. Following the corresponding cutflows has the consequence that the
analyses are not optimised for computing time performance.
However, we believe that this is a sacrifice worth making since the running 
time of the analysis is usually small in 
comparison to the event generation steps and detector simulation. The big 
advantage of our approach is that the same
analysis code that is used in validation is also used for setting limits. In 
addition, users can check the particular
cutflow themselves. At the end of the analysis, events are counted in each of the 
signal regions used to set limits. 
For a detailed overview on how to extend \Checkmate{} with your own analysis, please 
see \Cref{sec:app:addanalyses}.

\subsection{Reference Data}

The reference data that \Checkmate{} uses to set limits on models all come from the corresponding experimental 
analysis. Since different analyses provide different levels of detail regarding errors, backgrounds and experimental limits, 
the format is flexible, but a certain minimum set of information is required.

As already described in \Cref{sec:overview}, \Checkmate{} can set limits on the model that is being tested in two
different ways. One method is to compare the number of events in the most sensitive signal region with the 95\% 
exclusion given by the experiment, \Cref{eqn:rlimit}. Note that, in order to determine which signal 
region should be used to set the limit, the {\em expected} limit for each signal region is also 
required\footnote{For experimental analyses that do not contain this information, the 95\% \CLs has been
pre-calculated by \Checkmate{} so that this method is always available.}. The uncertainty of the expected 
limit, determined by the experimental collaboration, is also required and is split into positive and negative fields if this information is provided.

The second method is to calculate the likelihood for the model 
using the \CLs prescription, which requires the following information:
 \begin{itemize}
  \item The number of background events expected for a particular signal region along with the associated error.
  If given, the error is split into statistical and systematic components, each of which have positive and 
  negative fields.
  \item The number of observed events in the signal region.
 \end{itemize}

\subsection{Currently Validated Analyses}
The following analyses have been implemented in \Checkmate{} and the results have been 
cross-checked against the published results. The current analyses were chosen to give a broad overview of the
current ATLAS searches for final states containing a significant amount of missing transverse momentum, plus 
various combinations of jets, leptons and b-tags. In particular, we included the most sensitive ATLAS New Physics 
searches for final states containing an 
invisible particle. 

In the following section we briefly 
discuss the corresponding signatures and example cases to which the analysis has been optimised.

\subsubsection{atlas\_conf\_2012\_104: 1 lepton + $\geq$ 4 jets + \etmiss, \cite{ATLAS-CONF-2012-104}}
The one lepton search requires one isolated electron or muon with at least four jets and 
large missing transverse momentum 
in the final state. The study has two independent signal regions corresponding to an 
electron as well as a muon channel.

The study focuses on the pair production of new strongly interacting particles 
which then decay primarily into jets and a 
particle that evades detection, e.g.\ the Lightest Supersymmetric Particle (LSP) in SUSY. In addition
at least one decay chain must contain a detected charged lepton. The search is motivated
by the commonly studied mSUGRA (minimal supergravity) model where a left handed squark dominantly decays to a
light chargino which subsequently produces a lepton pair and the LSP.

\subsubsection{atlas\_conf\_2012\_147: Monojet search + \etmiss, \cite{ATLAS-CONF-2012-147}}
The monojet search is optimised for a signal consisting of a single hard jet recoiling against missing energy. A second 
jet is allowed as long as the $\Delta\phi$ between the second jet and the missing $p_T$ is greater than 0.5,
but events with further jets are vetoed. Events 
containing reconstructed electrons, muons or hard photons are vetoed.

The signal regions have been optimised for a variety of signal models including graviton production in 
large extra dimension models \cite{ArkaniHamed:1998rs}, SUSY gravitino production in the context of gauge 
mediated SUSY breaking models \cite{Giudice:1998bp} and a model independent search for dark 
matter \cite{Goodman:2010ku}. In addition, if the model of new physics has a very compressed spectrum, 
this search can also become relevant \cite{Dreiner:2012gx,Dreiner:2012sh}.

\subsubsection{atlas\_conf\_2013\_024: 0 lepton + 6 (2 b-)jets + \etmiss\; (All-hadronic stop search), \cite{ATLAS-CONF-2013-024}}

The all-hadronic stop search requires at least 6 jets including 2 $b$-jets to fully reconstruct two final state top quarks. In addition, 
a significant amount of  \etmiss{} is required and a lepton--, including tau--, veto is applied. The search is targeted at
SUSY top squark pair production where the top squark decays into a top quark and the LSP. However, the search can equally be applied
to any model producing a pair of top quarks and significant missing energy in the final state.

\subsubsection{atlas\_conf\_2013\_035: 3 leptons + \etmiss, \cite{ATLAS-CONF-2013-035}}

The trilepton search considers final states containing three charged leptons (i.e.\ electrons or muons), with at least 2 being from a 
same-flavour opposite-sign (SFOS) pair, and missing transverse 
energy in the final state. It has six signal regions that are either enriched or
depleted in $Z$ boson decays (the SFOS leptons lie within the $Z$ mass window). For 
each enriched and depleted $Z$ region, loose, medium and tight 
signal regions are defined. These correspond to tighter
missing transverse momentum and transverse mass requirements.

The search focuses on the electroweak production of charginos and neutralinos with subsequent decays into final states with leptons and missing
transverse momentum via slepton or electroweak gauge bosons.

\subsubsection{atlas\_conf\_2013\_047: 2-6 jets + \etmiss, \cite{ATLAS-CONF-2013-047}}

The multijet search is optimised for models with large production cross sections for heavy particles that decay to 
hard jets and a significant component 
of missing energy. For SUSY models this corresponds to squark and gluino production with direct 
decays into first or second generation quarks 
and the LSP. Signal regions with various numbers of final state jets and different proportions 
of missing energy are adapted to the different 
combinations of squark and gluino production.

\subsubsection{atlas\_conf\_2013\_049: 2 leptons + \etmiss, \cite{ATLAS-CONF-2013-049}}

The search for two opposite--sign leptons, missing transverse momentum and no jets in the final state focuses on electroweak production of charginos and sleptons. 
The study has five signal regions in total with two signal regions imposing a $Z$ veto and a low missing transverse momentum cut. The three remaining signal regions 
are sensitive to on--shell $W$ bosons and require stricter cuts on the transverse momentum of the leptons.

Both signal regions with a $Z$ veto are optimised for direct slepton pair production or provide sensitivity to sleptons from chargino decays. The three remaining signal regions are adapted 
to direct chargino pair production with charginos subsequently decaying into a W boson and the LSP.

\subsubsection{atlas\_conf\_2013\_061: 0-1 leptons + $\geq$ 3 b-jets + \etmiss, \cite{ATLAS-CONF-2013-061}}

The 3 $b$-jets study concentrates on the pair production of a heavy coloured particle which then decays with at least 2 $b$-jets 
in the final state, giving 4 $b$-jets in total. The reason only 3 $b$-jets are used is that the tagging algorithm has an efficiency of approximately \unit[70]{\%}, hence consequently requiring all 4 $b$-jets to be tagged leads to a significant reduction in acceptance. In addition, a large \etmiss{} cut is
used in all signal regions.

The primary motivation for the search is to find gluino pair production followed by the decay into the LSP and either a pair of bottom quarks 
or top quarks. Another possible application is the production of Higgs bosons in decay chains, since the dominant decay mode of the Higgs is
into 2 $b$-jets.


\subsubsection{atlas\_conf\_2013\_089: 2 leptons + jets + \etmiss (razor), \cite{ATLAS-CONF-2013-089}}

This 2-leptons + jets study focuses on the strong production of new states followed by decays that include 
leptons. A specific emphasis 
is placed on looking for soft leptons and the study allows for both electrons and muons to
have transverse momenta as low as 10~\GeV. To gain a better discrimination between the signal and
SM backgrounds, the `Razor' \cite{Rogan:2010kb} variables are used. These variables exploit 
the symmetry that can be expected in the pair production and decay of new states with sizable missing momentum. 

\subsubsection{cms\_1303\_2985: At least 2 jets + b jet multiplicity + \etmiss ($\alpha_T$), \cite{Chatrchyan:2013lya}}

This all hadronic search uses the $\alpha_T$ variable to search for strongly produced high mass states that decay to hard
jets and a significant component of missing energy. The strength of the $\alpha_T$ variable is that it is robust against 
jet energy mis-measurements of \etmiss\, and consequently can search for new physics at lower jet $\pT$ than more traditional
`jet + \etmiss searches'. There are also signal regions containing different $b$-jet multiplicities to search for bottom
and top quark partners. 

We remind the reader that the CMS detector is currently only modelled by the default Delphes settings for final states and 
has not been tuned to the latest experimental data. However, in order to simulate realistic $b$-tagging functionality, the 
ATLAS performance efficiency is used.

\section{Performance Studies}

\subsection{Validation}

\begin{figure}
 
\begin{subfigure}{0.49\textwidth}
 \includegraphics[width=1.\textwidth]{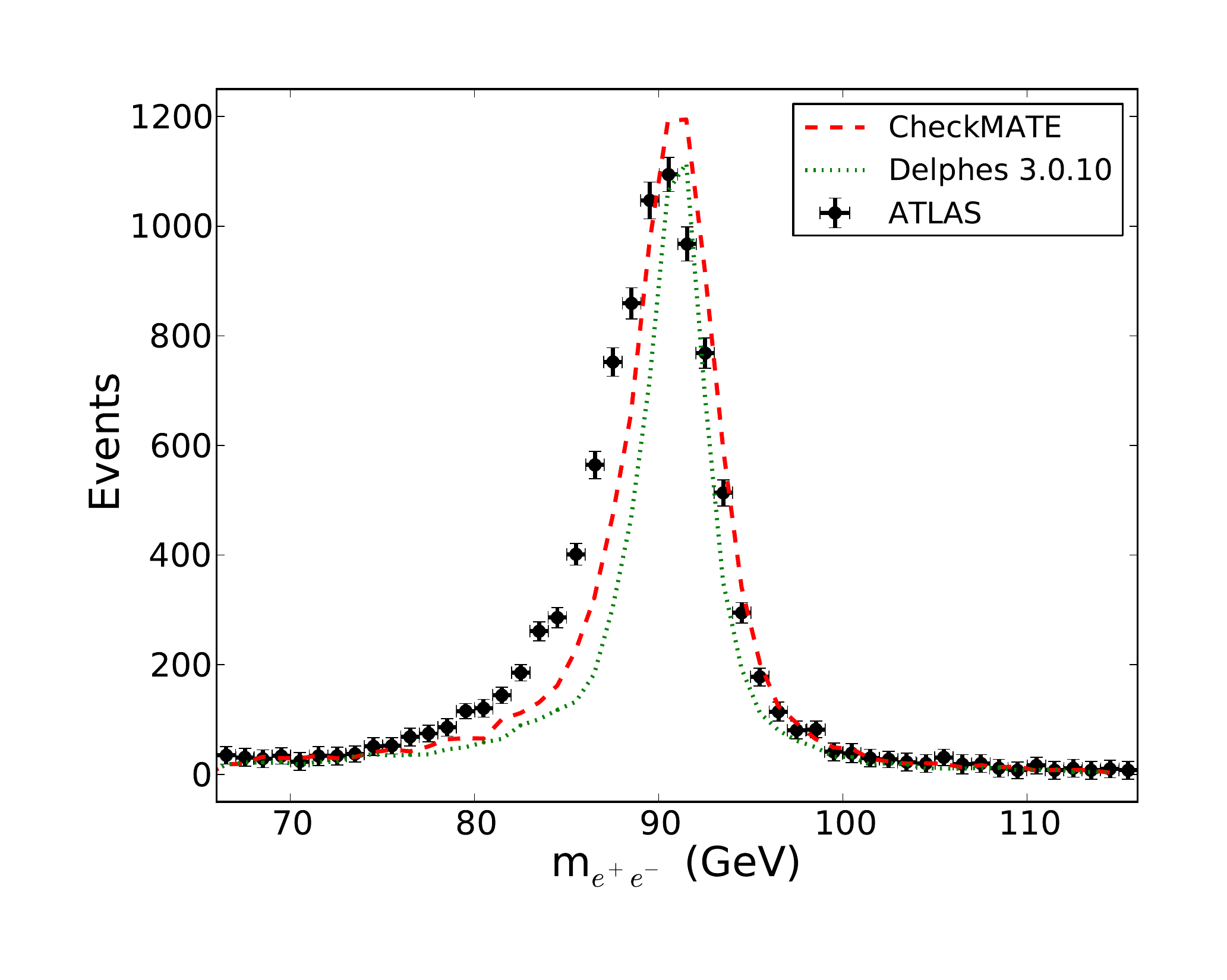}
 \caption{Invariant mass of dielectron pairs in $Z \rightarrow e^+ e^-$.}
\end{subfigure}
\begin{subfigure}{0.49\textwidth}
 \includegraphics[width=1.\textwidth]{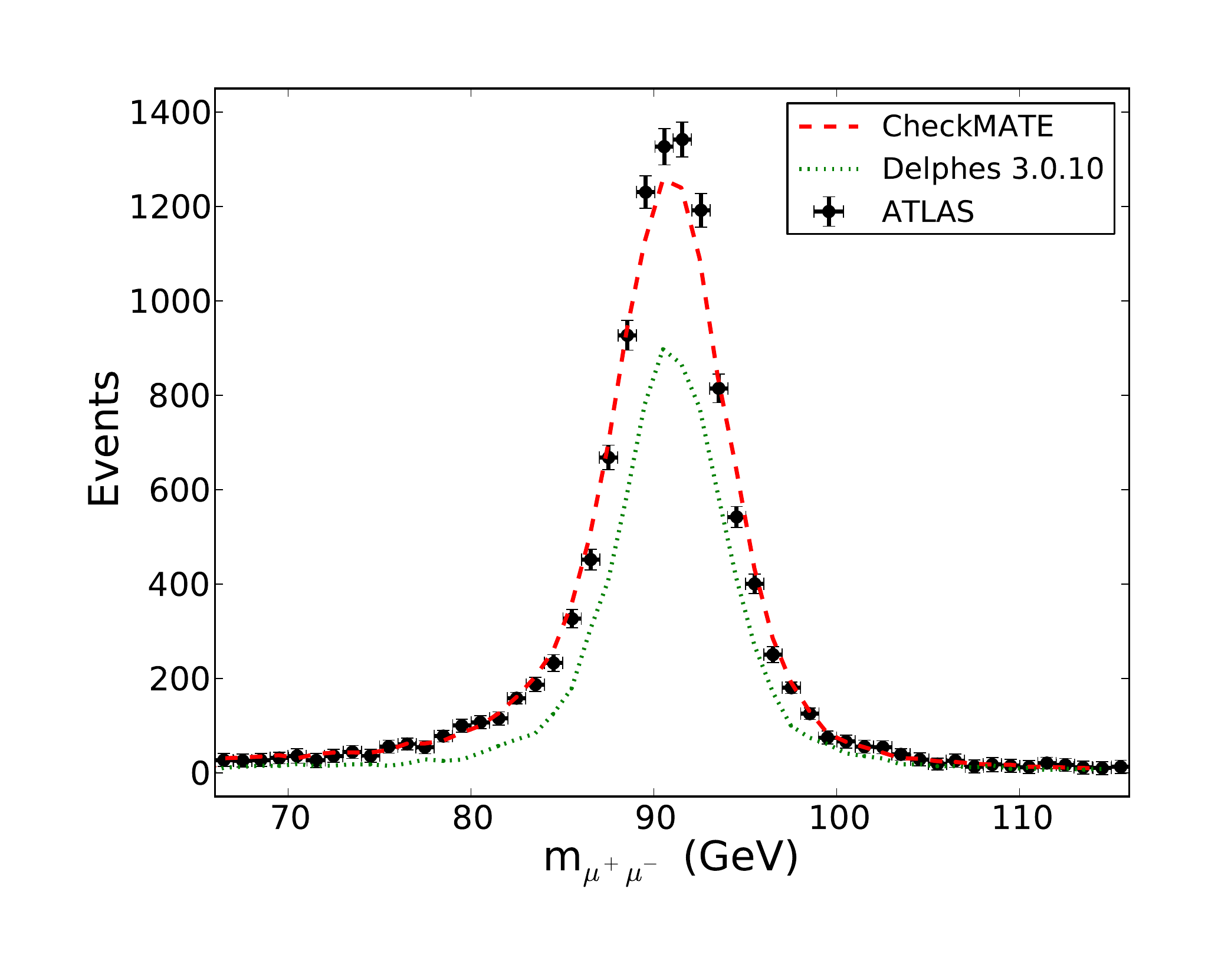}
 \caption{Invariant mass of dimuon pairs in $Z \rightarrow \mu^+ \mu^-$.}
\end{subfigure}
\begin{subfigure}{0.49\textwidth}
 \includegraphics[width=1.\textwidth]{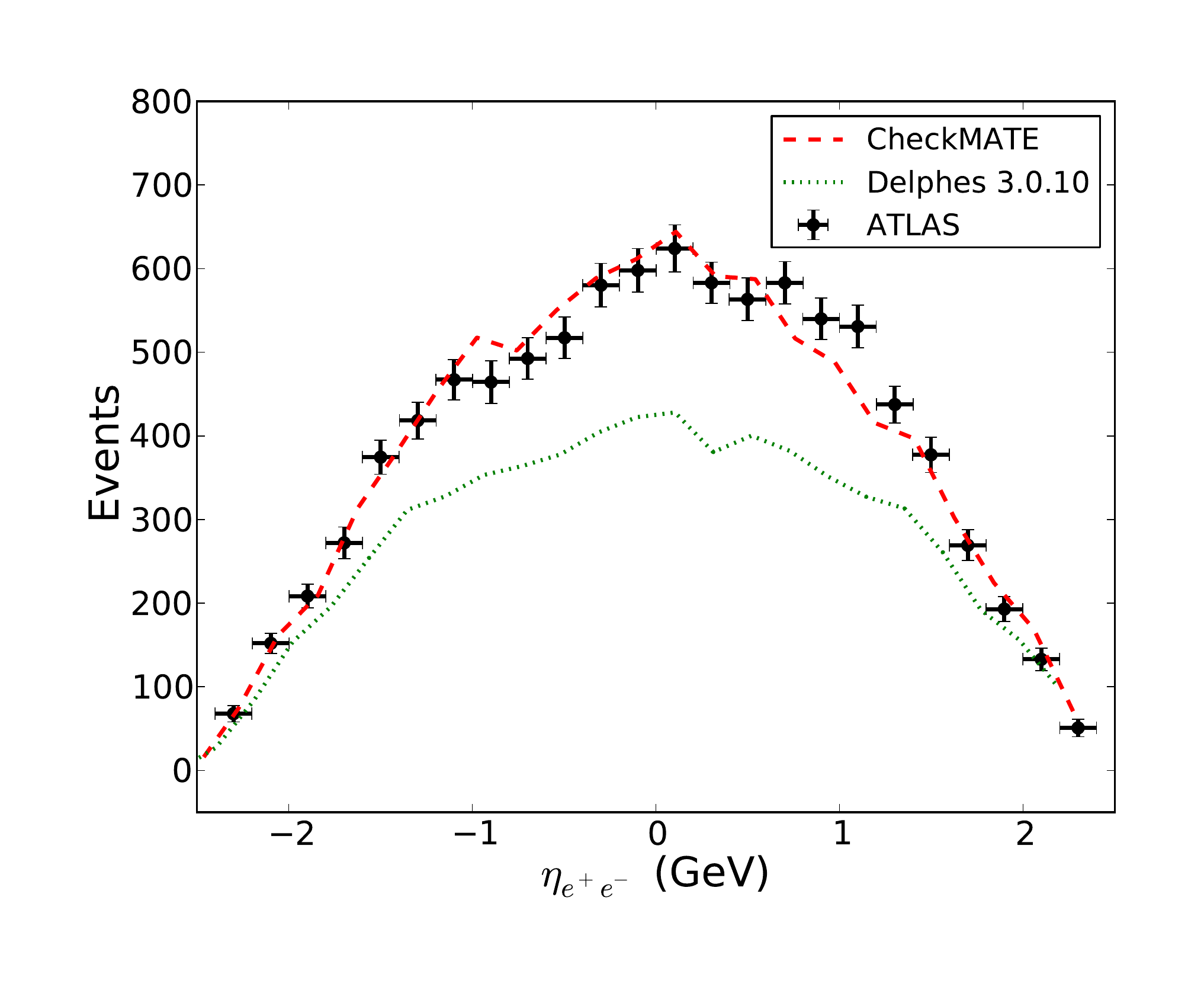}
 \caption{Total pseudorapidity of dielectron pairs in $Z \rightarrow e^+ e^-$. }
\end{subfigure}
\begin{subfigure}{0.49\textwidth}
 \includegraphics[width=1.\textwidth]{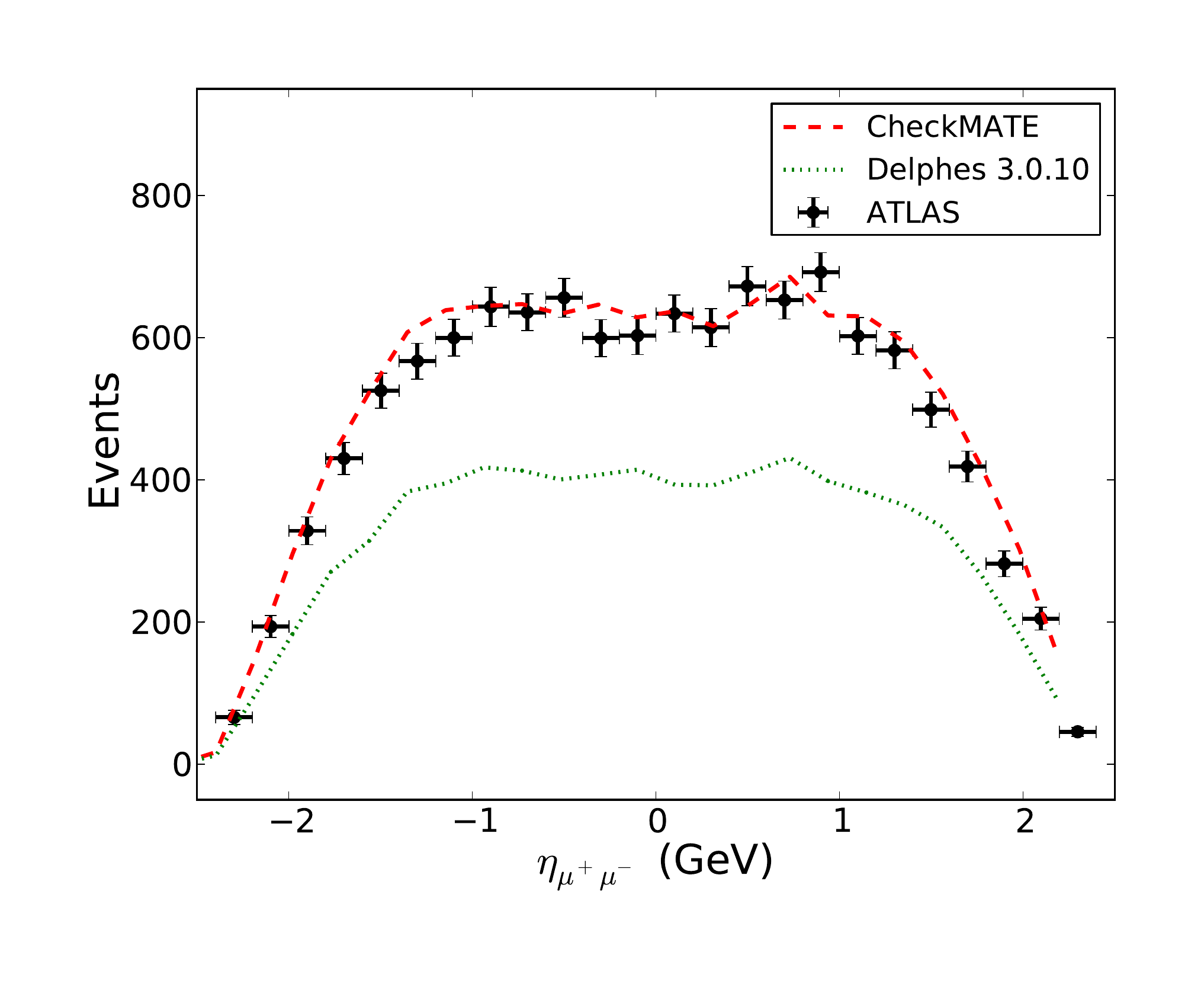}
 \caption{Total pseudorapidity of dimuon pairs in $Z \rightarrow \mu^+ \mu^-$.}
\end{subfigure}
 \caption{Comparison of results in dilepton channels of ATLAS $Z$--boson searches \cite{Aad:2011dm}. We use the same set of input events, produced with Pythia8, and process them separately with Delphes 3.0.10 using the standard ATLAS detector card and the Delphes version implemented in \Checkmate{} using the described detector tunings. The results of both Delphes runs are then analysed by identical analysis codes. Discrepancies in the dielectron width are most likely caused by problems in the full consideration of ISR/FSR effects within the simulation.} 
 \label{fig:Z_muon}
 \end{figure}

As stated in \Cref{sec:internal:delphes}, the default detector settings of Delphes have been significantly 
improved to match the latest ATLAS performance. In particular, all final state objects have been matched within
the experimental uncertainties to the latest
data available and the detailed results are given in \Cref{sec:app:detector_tune}. These tunings have been validated by comparing with 
some ATLAS analyses of Standard Model processes.
As an example we show the reconstruction of SM $Z$ production\footnote{Note that this analysis has been 
performed at $\sqrt{s} = \unit[7]{\TeV}$ and the efficiencies \Checkmate{} used for the results 
shown in \Cref{fig:Z_muon} correspond to the detector performance during this phase. However, 
recently ATLAS improved their electron reconstruction and identification algorithms with better overall 
efficiency factors. \Checkmate{} uses these updated results for its public version.} versus ATLAS data for both electrons
and muons in \Cref{fig:Z_muon}. More detailed results are given in \Cref{sec:app:detector_tune}. 

Before any analysis is accepted into the the latest version of \Checkmate{} it must be validated against 
the various data given in the published result. Whenever cutflows are given, these are checked at each step 
to make sure that all experimental cuts perform as expected, c.\/f.\ \Cref{sec:app:analysis_val}. The majority of the cutflows we have evaluated with \Checkmate{} have an 
acceptance that is within 10\% of the published value for the signal regions. If this accuracy is not
achieved, the analysis has been checked thoroughly to ensure no bugs remain and possible (or confirmed)
reasons for the discrepancy are clearly stated, see \Cref{sec:app:analysis_val}. 

 \begin{figure}[h]
 \centering \vspace{-0.4cm}
 \includegraphics[width=0.49\textwidth]{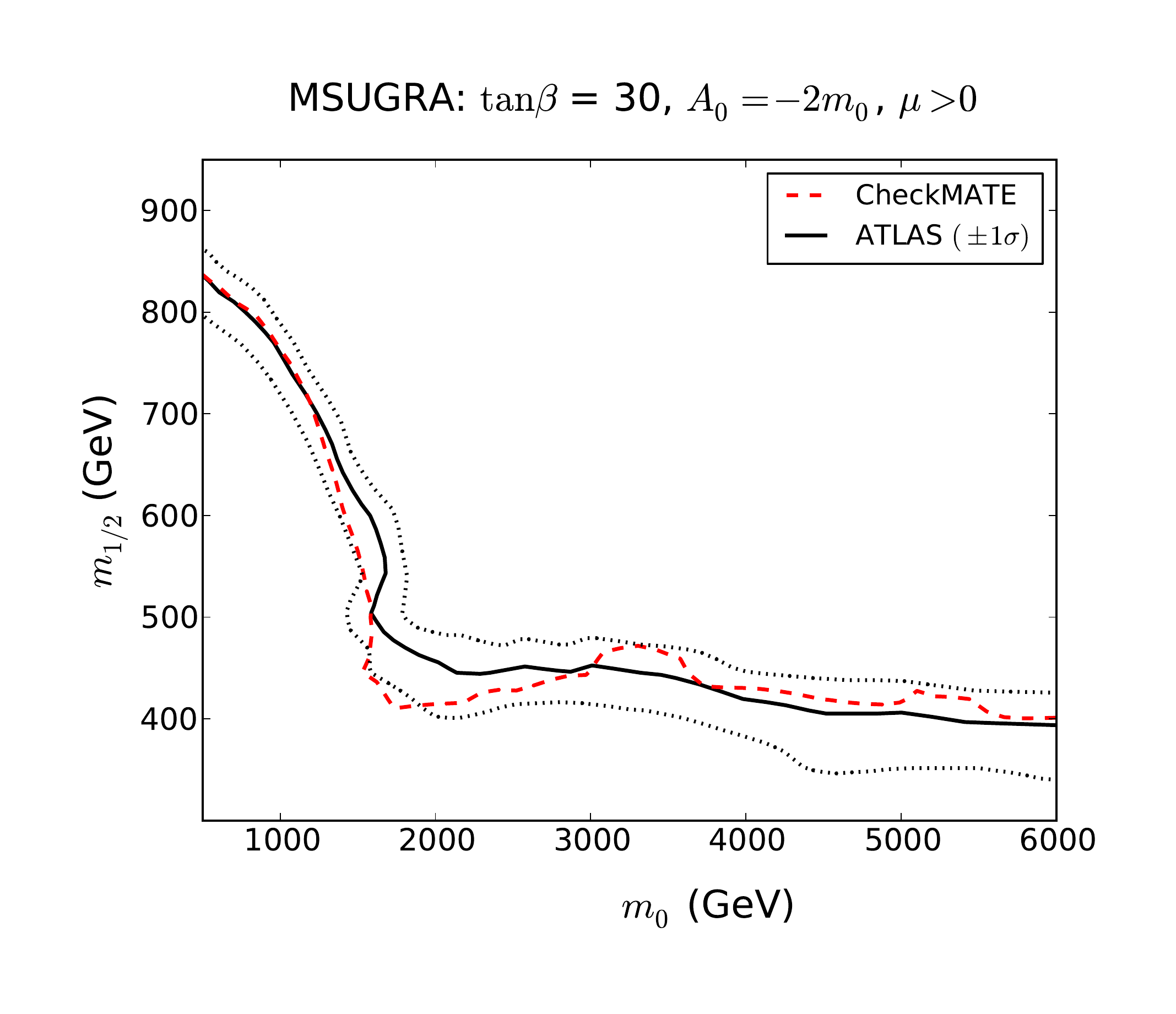}
 \includegraphics[width=0.49\textwidth]{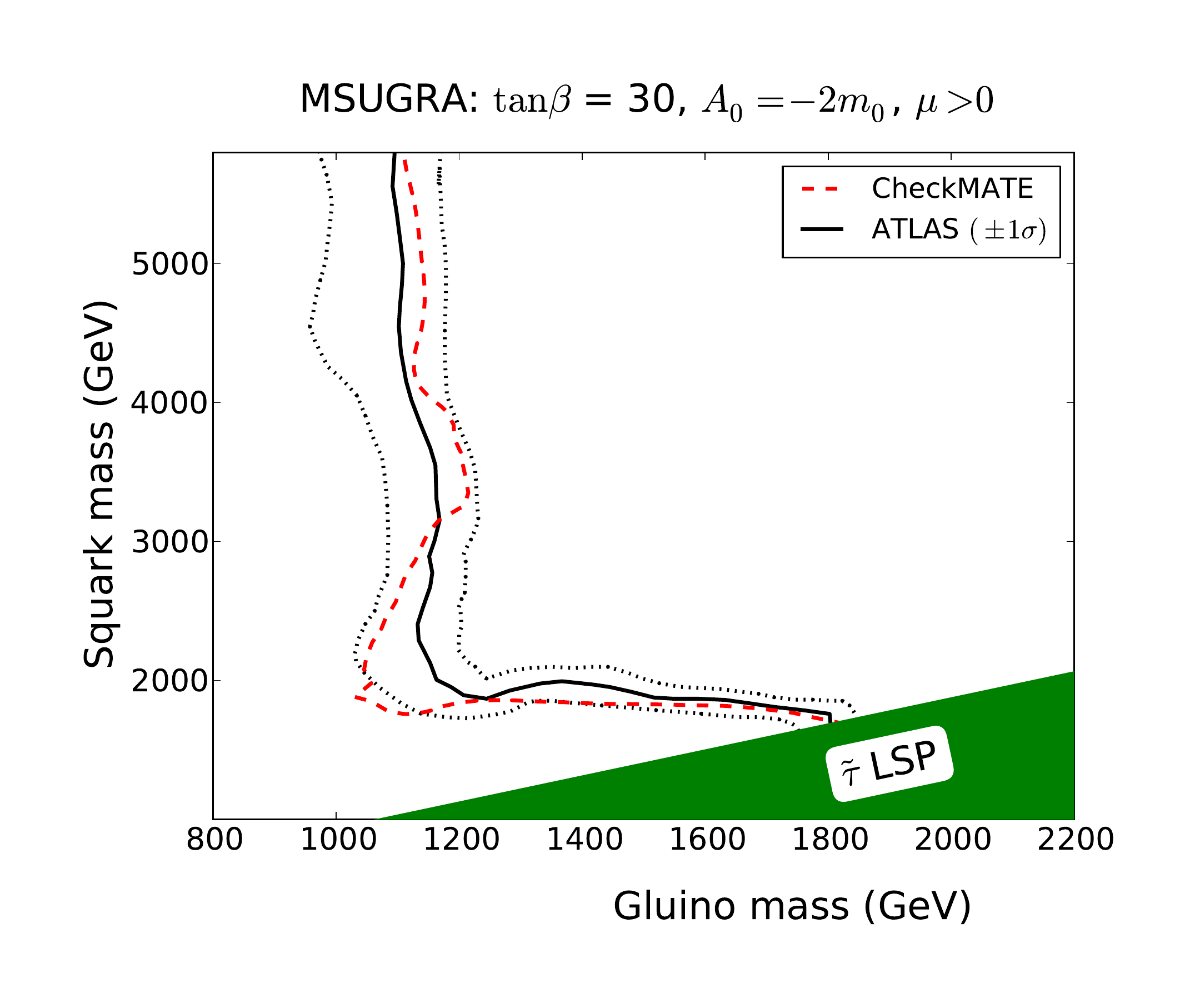}
 \caption{Exclusion curve for the CMSSM (mSUGRA) model for analysis atlas\_conf\_2013\_047, in the
 $(m_0,m_{1/2})$ (left) and $(m_{\tilde{q}},m_{\tilde{g}})$ (right) planes.}
 \label{fig:atlas_mSugra}
 \end{figure}

Many of the experimental analyses give exclusion curves for some popular (or
simplified) models. Whenever possible these have been replicated by \Checkmate{} 
(\Cref{sec:app:analysis_val}) and in general the results lie within the 1$\sigma$ theoretical uncertainty
predicted by the model. Again, in the rare occasions that the required accuracy has not been met,
possible reasons for the discrepancy are given. An example parameter scan is shown in 
\Cref{fig:atlas_mSugra} for the commonly studied SUSY model mSUGRA for the ATLAS jets and \etmiss\;
search (atlas\_conf\_2013\_047) that shows agreement within the 1$\sigma$ theoretical uncertainty
for almost the whole plane.

\subsection{Computing Performance}

As an example to show the relative computing time of \Checkmate{} we consider an estimate from the ATLAS trilepton
study (atlas\_conf\_2013\_049). The generation of 1000 $\tilde\chi_1^+\tilde \chi_1^-$ events 
in the HepMC format takes about 121 seconds of CPU time\footnote{The CPU time performance has 
been tested on an Intel Core 2, Quad CPU with 2.4 GHz and 8 GB RAM.} with Herwig++ 2.6.3. 
The Delphes 3.0.10 detector simulation with the relevant detector cards from \Checkmate{} 
requires 11 seconds of CPU time. Once the detector simulation is finished, 
running the ATLAS trilepton analysis only takes about 1 second for all five 
signal regions. Thus  the running time of the analysis is negligible 
compared to the event generation and detector simulation. This justifies our choice
to closely match the cutflows of the experiments and thus add reliability 
and not to optimise the analyses for computing time  performance.


\section{Outlook}
We would like to emphasise that while \Checkmate{} is fully functional, it remains a work in progress.
We will improve and extend the program over time.

At the moment, we have concentrated on searches for supersymmetry from ATLAS. However, the various 
supersymmetric analyses already cover multiple final 
state configurations and thus should be sensitive to many 
non--supersymmetric BSM physics scenarios containing invisible particles. Nevertheless, we want to implement as
many of the new exotic searches from ATLAS and CMS as possible. 
We hope that the user will have access to an even larger selection of analyses in the near future.

Furthermore, we have decided to implement the results of the conferences notes rather than taking 
analyses from journal publications. Our decision is justified due to the fact that the 
majority of published results have not so far taken into account the 8 \TeV data and we 
are convinced that the implementation of studies with the best kinematical reach is the 
better alternative. As soon as the published results become available, we will update our studies.

\Checkmate{} already provides a large number of validated analyses and many more only require final validation. 
Moreover, the 
structure of the program allows  for easy implementation of additional analyses 
by any user. If an analysis has been programmed we would welcome the opportunity
to include the study in the official \Checkmate{} library. After the new analysis has been validated,
it would then be made available to the whole user community.

As  discussed, we have significantly improved the detector simulation
performance of Delphes. However, we would like to implement even 
better detector tunings. In particular, when ATLAS and CMS publish updated detector performances, 
it is planned that we will revise the detector tunings to take advantage of the new 
information\footnote{We would like to take 
this opportunity to plead with the collaborations to provide more detail in regards to detector performance.
In particular, final state object maps for both identification and reconstruction should be provided in 
terms of $\pT$ and $\eta$ for all final state objects.}.

In addition to improving the analysis base of \Checkmate, we further plan to present applications 
(case studies) to investigate models that the collaborations have not yet been able to complete and which we hope
illustrate the power of \Checkmate. When these studies are available we will publish
programs that enable the user to repeat these studies and easily run new parameter scans themselves.

Further into the future we plan to significantly extend the functionality of \Checkmate{} to aid usability. 
Firstly we plan to include a Monte-Carlo generator into the program so that users no longer have to
generate event files themselves. Along with this innovation, will be a supersymmetry SLHA file reader requiring only  a supersymmetric spectrum as input. For models other than supersymmetry we plan a close
integration with FeynRules \cite{Alloul:2013bka} and SARAH \cite{Staub:2013tta} so that the user can begin 
their \Checkmate{} investigations armed only with a Lagrangian.
%
\section{Summary}
We have introduced the program \Checkmate{}, which is a tool to automatically check the 
compatibility of BSM models with LHC data. The program takes input in the form of 
Monte Carlo event files and cross-sections. These event files are given to a detector 
simulation and the result passed to implementations of LHC analyses. The outcome of
the analyses is then compared to LHC results to see if the given model can be excluded.

The program has been designed for easy installation and operation. In addition, the structure
allows the user to include new experimental analyses in a straightforward manner. However, it 
must be emphasised that the current analysis status in \Checkmate{} is in no way considered 
complete. It is foreseen that many additional LHC new physics searches will
be included in future upgrades.

The code can be obtained from the \Checkmate{} webpage
\begin{center}
http://checkmate.hepforge.org
\end{center}
as well as the installation instructions, more detailed technical informations and 
frequently asked questions. 
\section*{Acknowledgements}
\noindent We would especially like to thank Andreas Hoecker for many detailed discussions 
regarding ATLAS analyses. In addition, we would like to thank Philip Bechtle, Jamie Boyd,
Geraldine Conti, Carolina Deluca, Klaus Desch, Monica D'Onofrio, Frank Filthaut, Eva Halkiadakis, Emma Kuwertz, Zachary Marshall,
Antoine Marzin, Alaettin Serhan Mete, Marija Vranjes Milosavljevic, Maurizio Pierini, 
Tina Potter, George Redlinger, Steven Worm and Frank Wuerthwein for helpful discussions regarding 
different analyses. On the theoretical side, Tim Stefaniak provided useful advice from 
his work with `HiggsBounds/Signals'. The work has been supported
by the BMBF grant 00160200.  The work of M. Drees has been supported by the SFB TR33 "The Dark Universe" 
funded by the Deutsche Forschungsgemeinschaft (DFG). The work of J. S. Kim has been partially supported by the MICINN, Spain, under contract FPA2010- 17747; Consolider-Ingenio CPAN CSD2007-00042 and by the ARC Centre of Excellence for Particle Physics at the Terascale. J. S. Kim also thanks the Spanish MINECO Centro de excelencia Severo Ochoa Program under grant SEV-2012-0249.

\clearpage
\appendix
\begin{appendices}
  \crefalias{section}{appsec}
\section{Getting Started}
\label{sec:app:installation}
This section guides the user through the installation of \Checkmate{} and outlines how to check 
and potentially install updated versions of all the necessary prerequisites. The program has been tested and will run 
on both Linux and MacOS X\footnote{The installation on MacOS X systems require the Xcode Command Line Tools. Details for the
installation can be found on \url{http://railsapps.github.io/xcode-command-line-tools.html} and \url{https://developer.apple.com/xcode/}. Please note that 
wget is not pre-installed on MacOS X systems. However, it can be obtained from the MacPorts system \url{https://www.macports.org} or by downloading the respective files by hand.}. An alternative online step--by--step tutorial can be found under \url{http://checkmate.hepforge.org/tutorial/start.php}.

\subsubsection{Setting up Python}

\Checkmate{} requires Python 2.7.X with $X \geq 3$ on your system. Most systems already come with a Python installation, which
you can easily check by typing \verb@python -V@ and hitting `enter' in a terminal. If Python is installed, it should 
start and immediately tell you the version number. If the installation is too old, too 
new\footnote{\Checkmate{} does not work with Python 3.}  or if there is no Python at all, please install it 
either manually from \url{http://www.python.org/download/} or using the software management of your system.

\subsubsection{Setting up \Root{}}
\Checkmate{} uses \Root{} for a variety of tasks, therefore it is necessary that 
every user has a fully working \Root{} installation available on their system. 
Furthermore, \Checkmate{} uses some \Root{} packages which are not installed automatically and which may 
need to be added. Due to the large size of the \Root{} source code, we refrained from including it in 
our package and only provide a step-by-step tutorial for how to install it from scratch or add 
potentially missing obligatory packages to an existing installation. Running \verb@whereis root@ inside the terminal tells you if there already exists an installation on your system and determines how to continue with the installation:
\begin{description}
\item[Case 1: There is \textbf{no} \textsc{Root} installed yet]
    You must download the source and install \Root{} from scratch. Please do not use the \Root{} pre-compiled binaries but follow these instructions, since we encountered internal linking problems with the binary version of \Root{}.
    Start by downloading the latest version from \url{http://root.cern.ch/drupal/content/downloading-root} with:

\begin{Verbatim}        
wget ftp://root.cern.ch/root/root_xyz.source.tar.gz
gzip -dc root_xyz.source.tar.gz | tar -xf -
cd root
\end{Verbatim}

    After downloading \Root{} you can choose where you would like the installation located.
    If you don't want to do a system-wide installation in \verb@/usr/bin@ (e.g.\ because you do not 
    have administrator rights), you have to use \verb@--prefix@ and \verb@--etcdir@ to declare the installation 
    folder. Should you use a local Python installation (i.e.\ if the Python binary is not located 
    in \verb@/usr/bin@), you have to give the positions of both \verb@/include@ and \verb@/lib@ by 
    adding \verb@--with-python-incdir=[path_to_python]/include/python2.7@ \verb@--with-python-libdir=[path_to_python]/lib@ with:
 
\begin{Verbatim}
./configure --enable-python --enable-roofit --enable-minuit2  {--prefix=[desired_root_path]} \
            {--with-python-incdir=[path_to_python]/include/python2.7 --with-python-libdir=[path_to_python]/lib}            
make
\end{Verbatim}

    \Root{} is large, so go and have a (big) cup coffee in the meanwhile. After a successful completion, please hit:
\begin{Verbatim}
make install
\end{Verbatim}

\item[Case 2: There \textbf{is already} a \Root{} installation on the system.]
    If you already have \Root{} on your system, let us first check if it includes all necessary packages. In 
    the \textsc{Root} base directory, please run the following commands:
\begin{Verbatim}
        ./bin/root-config --has-minuit2
        ./bin/root-config --has-roofit
        ./bin/root-config --has-python
\end{Verbatim}
    If all three commands return `yes', your \Root{} installation includes everything \Checkmate{} needs and you can safely skip to the `Setting up \Checkmate{}' section. If not, you have to recompile the code and add the missing packages. Follow the instructions mentioned for `Case 1', but make sure that
\begin{itemize}
\item you download the same \Root{} version the system has (you can use \verb@./root-config --version@ to find out),
\item you install into the same root-directory by choosing \verb@--prefix@ accordingly. Alternatively, you can always download and install a second root version locally according to step 1. Make sure that you set up \Checkmate{} with this local \Root{} version explicitly.
\end{itemize}
        In case you are using the standalone \Root{} binaries, beware that we sometimes encountered problems during the compilation of \Checkmate{}. Should you find the same problems, i.e.\ that all the aforementioned checks were positive but still there are \Root{} linking errors occurring, please consider using a proper installation from source code as explained above.
\end{description}

\subsubsection{Setting up \Checkmate{}}

Contrary to many other tools, \Checkmate{} does not have a separate \verb@make install@ routine and 
will set itself up in the directory you put it. You should therefore begin the following procedure 
from within the folder you want \Checkmate{} to be located.

Start by downloading the tarball either by hand or from within the terminal
\begin{Verbatim}
   wget http://www.hepforge.org/archive/checkmate/CheckMATE-Current.tar.gz
\end{Verbatim}
Extract the tarball and have a look at the \Checkmate{} folder:

\begin{Verbatim}
   tar -xvf CheckMATE-Current.tar.gz
   cd CheckMATE-X.Y.Z
   ls
aclocal.m4      bin        configure.ac  depcomp     m4           missing  tools
AUTHORS         ChangeLog  COPYING       INSTALL     Makefile.am  NEWS     VERSION
autom4te.cache  configure  data          install-sh  Makefile.in  README
\end{Verbatim}
In order to run \Checkmate{} we have to compile the code, connect the Python scripts 
to the \Root{} libraries and create the binary.  The configuration step will make sure 
that you followed the aforementioned sections to properly setup \Root{} and have a valid 
Python interpreter. The easiest way to proceed is to run \verb@./configure@ and follow the 
instructions in case of an error. After successful configuration, type \verb@make@\footnote{Depending upon your
system the \texttt{make} command can generate various warning messages. These can be safely ignored unless the 
final output is an error message (e.g. \texttt{make: *** [all-recursive] Error 1}) in which case the 
installation will have failed. If you are unable to understand the cause of the problem, please check the 
installation FAQ on the \Checkmate{} website.}, which will 
create a \verb@Checkmate@ binary in \verb@bin/@. 
\begin{Verbatim}
    ./configure {--with-rootsys=[path_to_root]} {--with-python=[path_to_python]/bin/python}
    make
\end{Verbatim}
Note for experienced \Root{} users: \Checkmate{} should work even if you haven't set up \verb&$ROOTSYS& (and similar) environmental variables since it takes all information from the \verb@--with-rootsys@ parameter. However in case of any unexpected errors, run \verb@thisroot.(c)sh@ inside the \Root{} binary directory and retry. 

To test the installation, run the test parameter file in the \verb@bin@ directory.
\begin{Verbatim}
    cd bin
    ./CheckMATE testparam.dat
\end{Verbatim}
The program will ask if the automatically chosen output directory is correct. After 
answering '\verb@y@' the program should run and finish with the statement that the 
input is allowed. If it doesn't, check whether your configuration and compilation 
produced any errors or warnings. Otherwise, you finished setting up \Checkmate{} and you 
can continue with \Cref{sec:tutorial} to learn how you can run the code.
\vfill
\pagebreak

\section{Adding Analyses}
\label{sec:app:addanalyses}
\Checkmate{} is not only meant to be simple to use and transparent. The philosophy of the program 
is that it is also supposed to be easy to extend with new ideas for analyses 
or users who need a particular analysis, which has not been implemented yet. In this section we give a brief introduction into the most important steps that are necessary to implement a new analysis.

\subsection{Using the Analysis Manager}
\Checkmate{} comes with an \emph{Analysis Manager}, i.e.\ a binary which asks the user for all the necessary details and takes care of the internal setup. In order to not confuse the normal user, this binary is not created automatically  with the  ordinary compilation of the code. However, it can easily be added by typing (after having already configured the code according to \Cref{sec:app:installation})
\begin{Verbatim}
make AnalysisManager
\end{Verbatim}
This should create a second binary \verb@AnalysisManager@ within the \verb@bin/@ folder. Running it will reveal three possible options:
\begin{itemize}
\item Show a list of all currently implemented analyses,
\item add a new analysis/modify an existing analysis or
\item remove an analysis.
\end{itemize}
Choosing the second option will guide the user through a large list of 
questions regarding the analysis. These gather information with respect to the 
general properties of the analysis like name and luminosity, the setup of the detector
simulation (see \Cref{sec:delphes_sim_set}) and all the considered signal regions 
with their respective experimental numbers for observation and background.
These settings are written in human--readable form in \verb@data/ANALYSIS_var.j@ and can be 
changed afterwards, by re-running the `adding' step with the 
AnalysisManager\footnote{Adding a new analysis will not only produce the described analysis variable file, but will also save the analysis information and the signal region numbers in separate auxiliary files, which are read by the respective \Checkmate{} modules. 
Changing the analysis variable file will \emph{not} update these auxiliary files automatically.}. Furthermore, the AnalysisManager creates ready--to--compile source and header files in \verb@tools/analysis/src@ and \verb@tools/analysis/include@ and adds them to the configuration files, such that they are compiled with the \verb@make@ routine.
\subsection{Analysis Code}
\label{sec:app:analysis_code}
After the source files have been created, they have to be filled with the 
actual analysis code. For this, \Checkmate{} provides a general C++ analysis 
framework using a globally defined base class\footnote{The concept of the code 
structure is based on the Rivet analysis framework \cite{Buckley:2010ar}.}. This class 
provides particle containers for all the different types of final state objects, like 
leptons or jets, alongside with various methods that are commonly needed in many analyses, 
like phase space reduction and overlap removals.

To keep this manual at a reasonable volume, we will only show a minimal example to 
illustrate the basic structure of the code in \Cref{fig:app:example_analysis}. For 
more details on the full list of implemented procedures, see the \Checkmate{} webpage or the 
source code of the already implemented analyses.
\begin{figure}
\centering
\scriptsize
\begin{SaveVerbatim}{Example_analysis}
#include "example.h"                                                        // The header file can stay untouched.

void Example::initialize() {                                                // This function is called once before the event loop.
  setAnalysisName("example");                                               // These set the header of all output files.
  setInformation(""                                                           
    "@#Example Analysis\n"
  "");
  setLuminosity(10*units::INVFB);                                           // Events are normalised to this integrated lumi.
  ignore("towers");                                                         // These won't read tower or track information from the
  ignore("tracks");                                                         //  Delphes output branches to save computing time.
  bookSignalRegions("jets;jets_plus_e;jets_plus_m");                        // All signal- and cutflow regions to be used later have
  bookCutflowRegions("singlelep;twojets");                                  //  to be defined here using a unique name.
}

void Example::analyze() {                                                   // This function is called once for each event.
  electronsMedium = filterPhaseSpace(electronsMedium, 20, -2.5, 2.5, true); //  'medium' electrons, pT > 20 GeV, |eta| < 2.5,
                                                                            //  excluding the region with 1.37 < |eta| < 1.52.
  std::vector<Electron*> isoElecs = filterIsolation(electronsMedium);       // Whatever isolation cuts have been defined in the 
                                                                            //  analysis definition are applied here and the result
                                                                            //  stored in a new container.

  muonsCombined = filterPhaseSpace(muonsCombined, 25, -2.0, 2.0);           // 'combined' muons, pT > 25 GeV, |eta| < 2.0.
  std::vector<Muon*> isoMuons = filterIsolation(muonsCombined);                       
  missingET->addMuons(isoMuons);                                            // Muons are excluded from the missing energy calculation 
                                                                            //  in Delphes and have to be added manually.

  jets = filterPhaseSpace(jets, 50);                                        // Jets, pT > 50 GeV.
  jets = overlapRemoval(jets, electronsMedium, 0.2);                        // Jets that overlap with an isolated electron with
                                                                            //   dR < 0.2 are discarded.
  
  if (isoElecs.size() + isoMuons.size() == 1)                               // Veto event if it does not contain exactly one lepton.
    return;    
  countCutflowEvent("singlelep");                                           // Count this event for the first cutflow step.

  if (jets.size() < 2)
    return;    
  countCutflowEvent("twojets");                                             // Veto on events with less than 2 jets.

  double E_tot = 0;
  for(int j = 0; j < jets.size(); j++)                                      // This loops over all jets and sums up their scalar ET.
    E_tot += jets[j]->PT;    

  if (E_tot >= 150) {                                                       // Signal region 1: Sum(Scalar ET) > 150 GeV.
     countSignalEvent("jets");
     if ((isoElecs.size() == 1)&&
         (fabs(isoElecs[0]->P4().DeltaPhi(missingET->P4())) < 0.3))
       countSignalEvent("jets_plus_e");                                     // Signal Region 2: SR1 + an electron that lies close to  
                                                                            //  the missing momentum vector.
     else if ((isoMuons.size() == 1)&&
              (fabs(isoElecs[0]->P4().DeltaPhi(missingET->P4())) < 0.15))
       countSignalEvent("jets_plus_m");                                     // Signal Region 3: SR1 + a muon that lies even closer to  
                                                                            //  the missing momentum vector.
  }
}

void Example::finalize() {                                                  // This function is called once after the event loop. 
}                                                                           //  However, usually there is nothing to do here.
\end{SaveVerbatim}

\fbox{\BUseVerbatim{Example_analysis}} 
\caption{Example code to illustrate the most basic features of a \Checkmate{} analysis.}
\label{fig:app:example_analysis}
\end{figure}
The shown example consists of a fictitious analysis with three signal regions and a two-step cutflow. It considers isolated electrons with $\pT > \unit[20]{\GeV}, |\eta| < 2.5$, passing the `medium' identification requirement and not being in the overlap region $1.37 < |\eta| < 1.52$. Furthermore it considers `combined' muons and jets with different kinematical cuts. Jets that overlap with an isolated electron within $\Delta R_{ej} < 0.2$ are discarded. The event is also vetoed if it does not contain exactly one lepton and at least two jets. One signal region demands the total scalar sum of all isolated jet energies to be larger than \unit[150]{\GeV}. The second and third signal region requires an isolated electron or muon, respectively, that has a small angular separation from the missing momentum vector. 

We finish this section with some important information regarding analyses in general: 
\begin{itemize}
\item Available particle containers are \verb@electronsLoose, electronsMedium, electronsTight, muonsCombinedPlus@,
\verb@muonsCombined, jets, jets2, photons, tracks@ and \verb@towers@. \verb@jets2@ only contains data if a second jet type has been defined in the creation of the analysis. \verb@towers@ and \verb@tracks@ are not needed by many analyses and are hence ignored in the standard analysis initialisation, meaning that the \Root{} branch is not read.
\item These containers are of the standard C++ vector type, string pointers to the respective particle objects.
\item All vectors are sorted according to the objects' $\pT$, with e.g.\ \verb@jets[0]@ having the highest transverse momentum.
\item The final numbers for all signal--, control-- and cutflow regions are automatically saved in the corresponding output file after the analysis run, listed in alphabetical order.
\item Calling the \verb@P4()@ method returns the particle's 4--vector in the lab frame, 
using  \Root{'s} TLorentzVector format which comes with a convenient list of applicable methods.
\item The header file (\verb@example.h@ in this case) can stay untouched in most cases. However, to save computing time it might be reasonable to not define local variables in the \verb@analyze()@ method but instead define them as members of the analysis class within the header.
\end{itemize}

\pagebreak
\section{ATLAS Detector Tunings}
\label{sec:app:detector_tune}

\subsection{Jets and Photons}

For both jets and photons, the standard Delphes parametrisations for reconstruction and identification have 
been used. Please see \cite{deFavereau:2013fsa} for more details.

\begin{figure}[h!]
\centering
\begin{subfigure}{0.49\textwidth}
\includegraphics[width=1.0\textwidth]{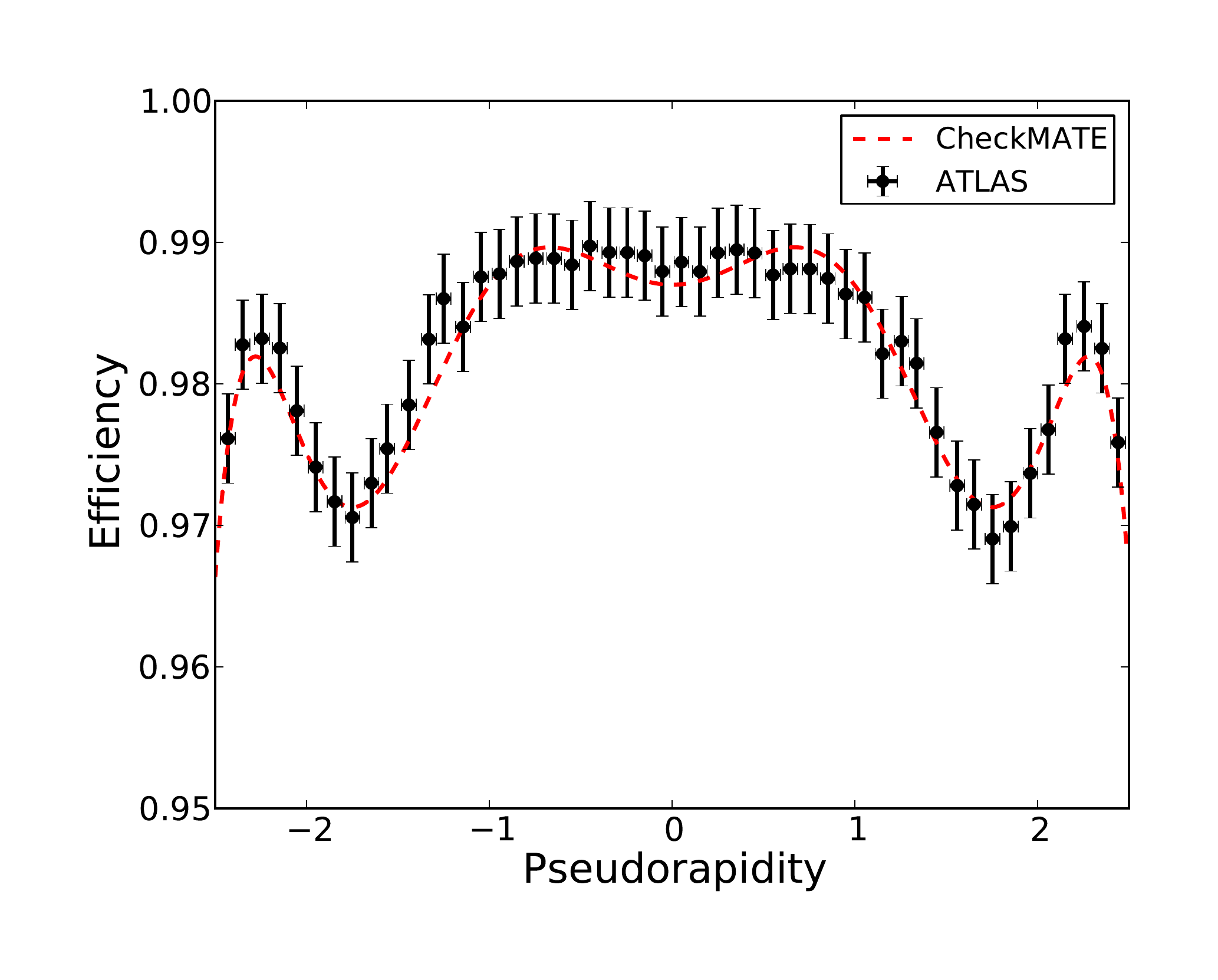}%
\caption{Efficiency for the reconstruction of an electron object inside the
calorimeter \cite{ATL-COM-PHYS-2013-1287}.}
\label{fig:tunings:el:eff:a}
\end{subfigure}%
\begin{subfigure}{0.49\textwidth}
\includegraphics[width=1.0\textwidth]{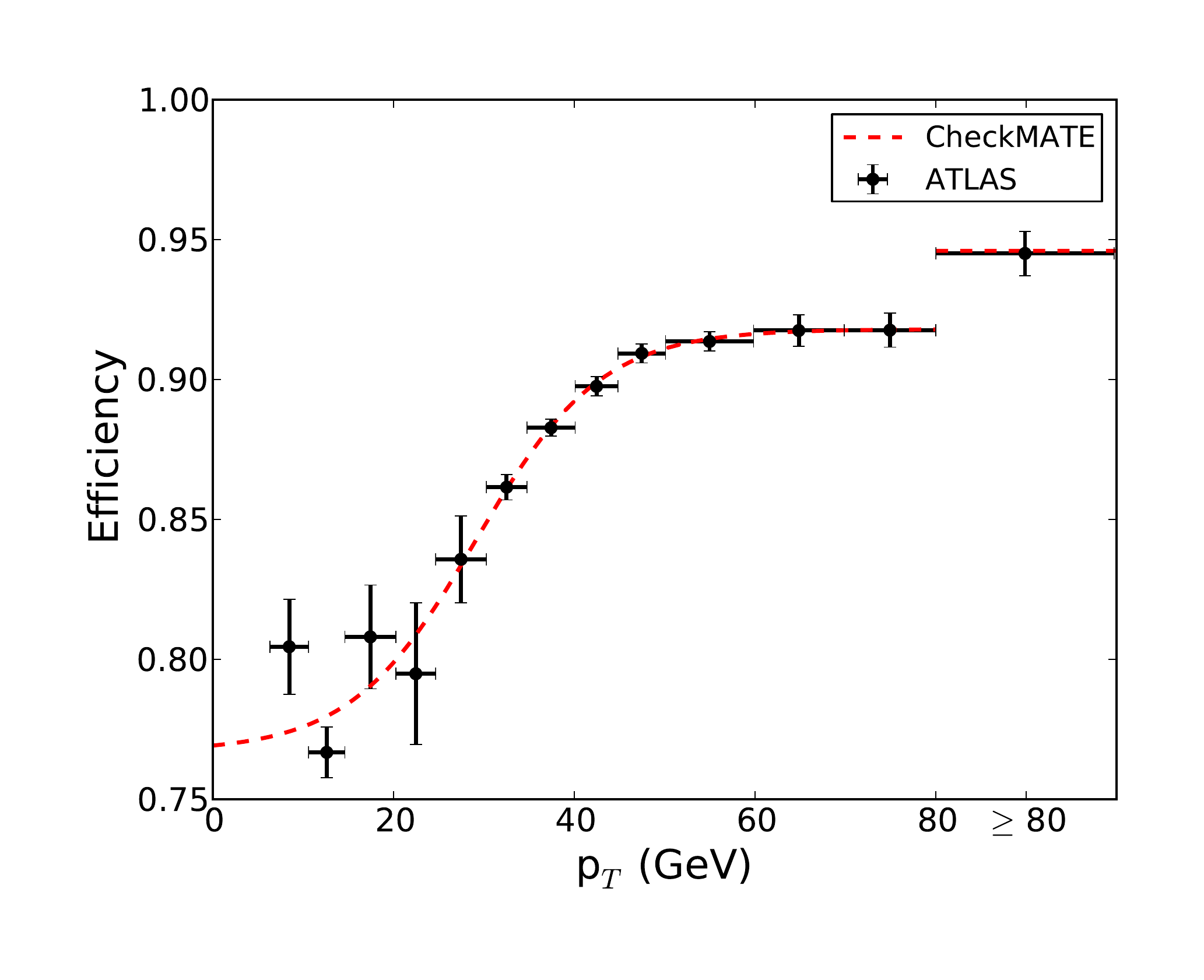}
\caption{Identification efficiency for `medium' electrons \cite{ATL-COM-PHYS-2013-1287}.}
\label{fig:tunings:el:eff:b}
\end{subfigure}

\begin{subfigure}{0.49\textwidth}
\includegraphics[width=1.0\textwidth]{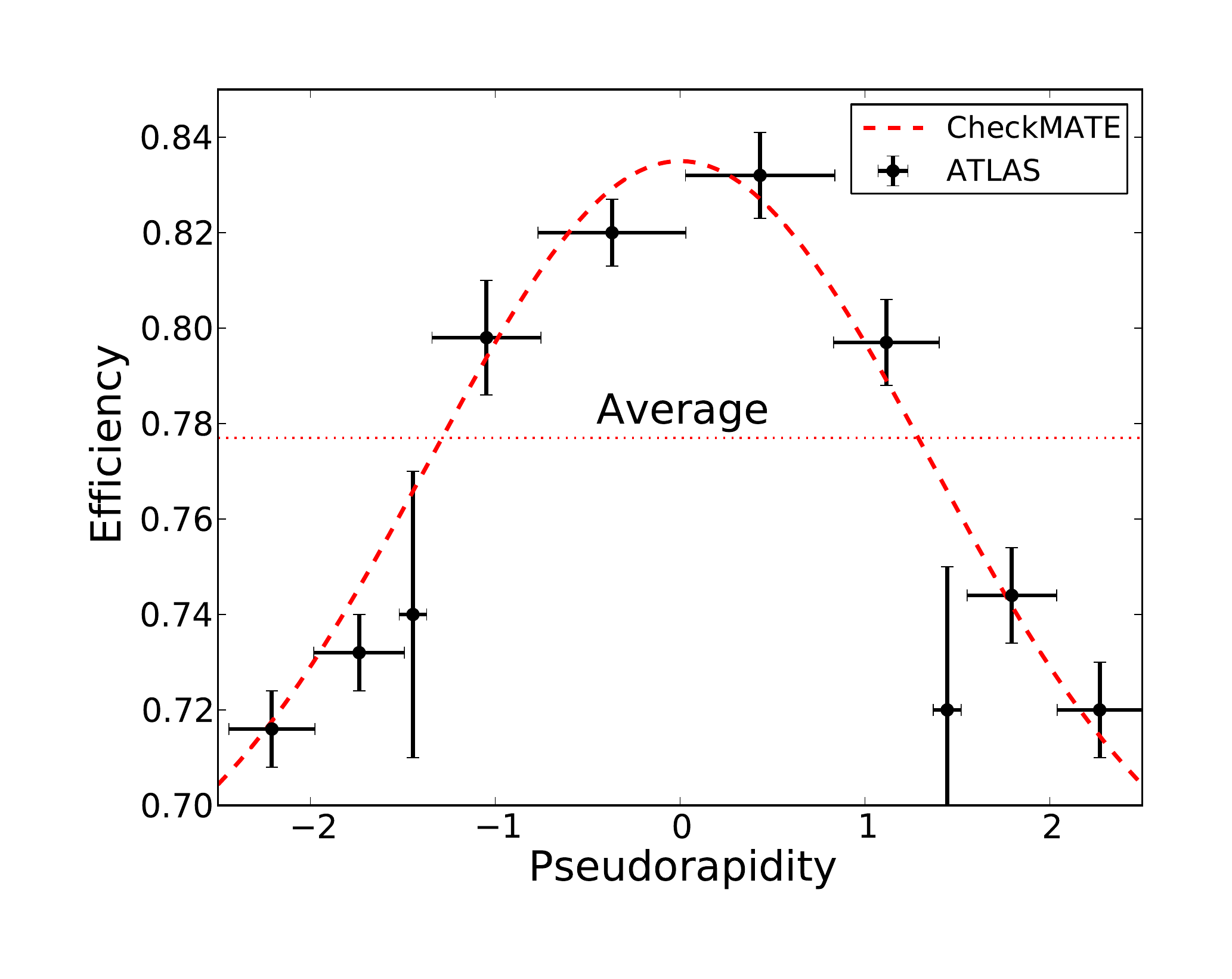}
\caption{Identification efficiency for `tight' electrons as a function of the candidate's 
pseudorapidity \cite{Aad:2011mk}.}
\label{fig:tunings:el:eff:c}
\end{subfigure}
\begin{subfigure}{0.49\textwidth}
\includegraphics[width=1.0\textwidth]{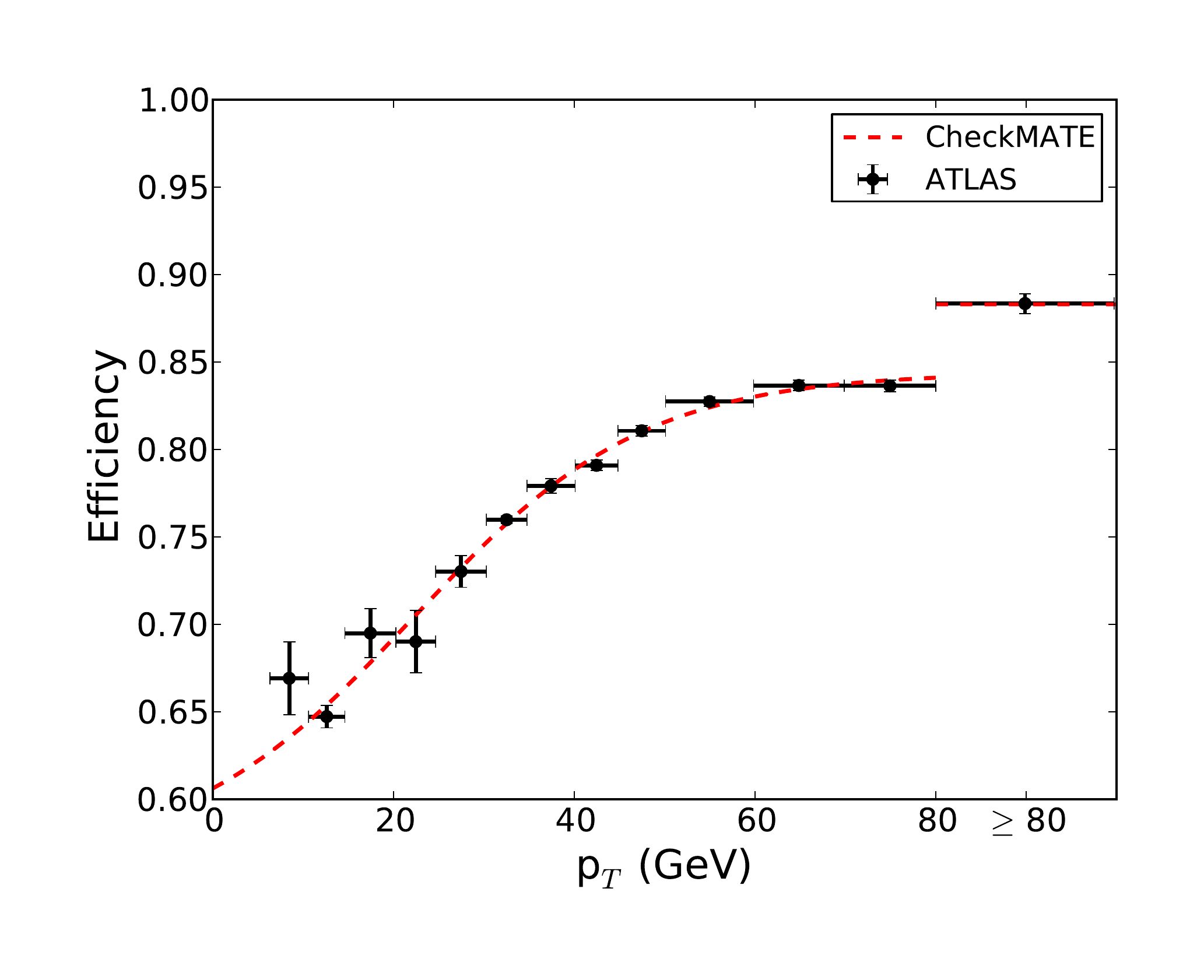} 
\caption{Identification efficiency for `tight' electrons as a function of the 
candidate's $p_T$ \cite{ATL-COM-PHYS-2013-1287}.}
\label{fig:tunings:el:eff:d}
\end{subfigure}
\caption{Electron efficiency distributions used in \Checkmate{}. The functions that have 
been used to fit the data are given in \Cref{eqn:app:el:eff1,eqn:app:el:eff2,eqn:app:el:eff3,eqn:app:el:eff4}. For 
the tight identification efficiency, \Checkmate{} uses the absolute value from the $p_T$ dependent efficiency shown in d), and multiplies it with the $\eta$ dependent efficiency  in c) normalised to an average value of $1$. }
\label{fig:tunings:el:eff}
\end{figure}

\subsection{Electrons}
\begin{figure}[b]
\centering
\begin{subfigure}{0.49\textwidth}
\includegraphics[width=1.0\textwidth]{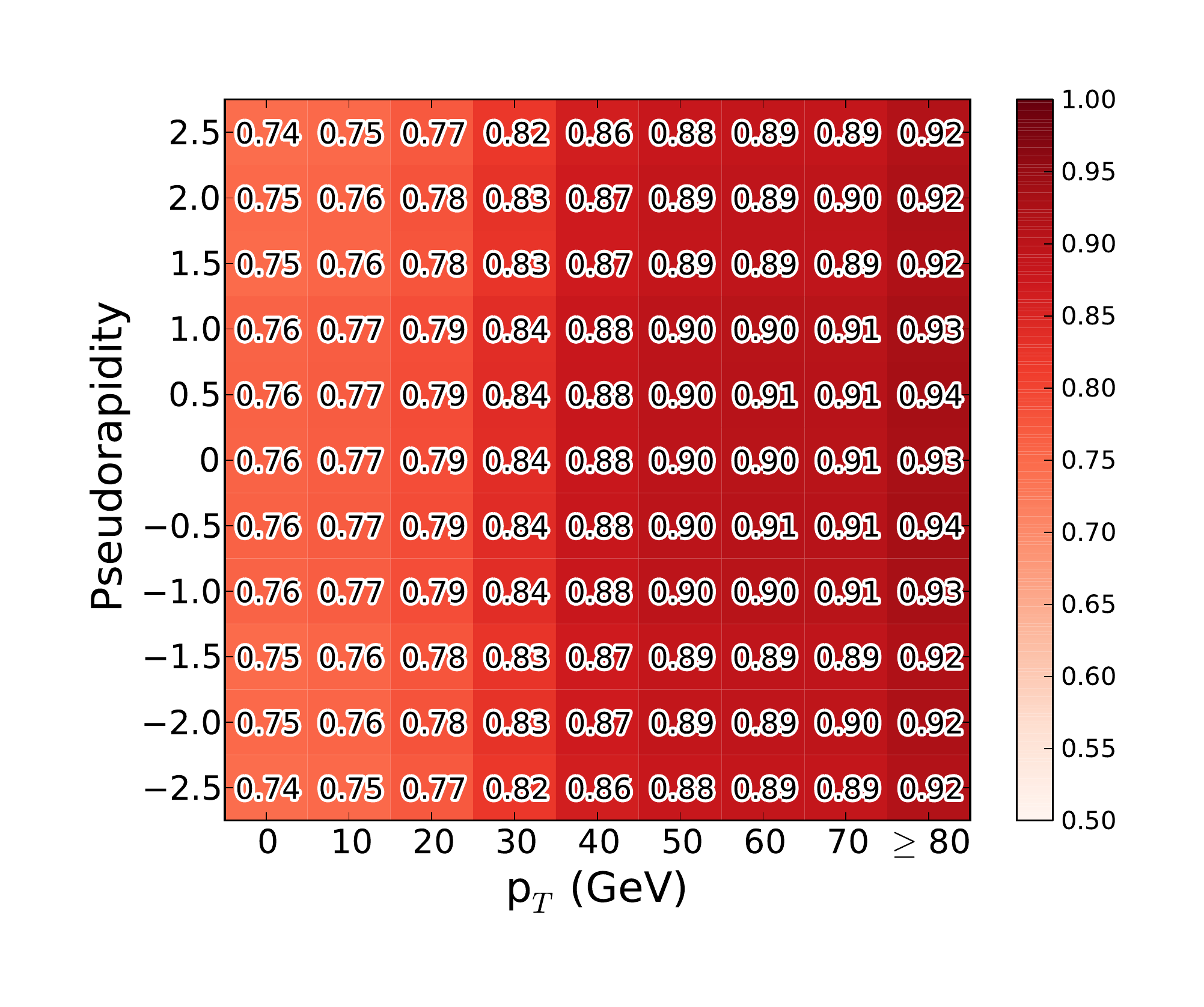}
\caption{Medium Electrons}
\label{fig:tunings:el:tot:a}
\end{subfigure}
\begin{subfigure}{0.49\textwidth}
\includegraphics[width=1.0\textwidth]{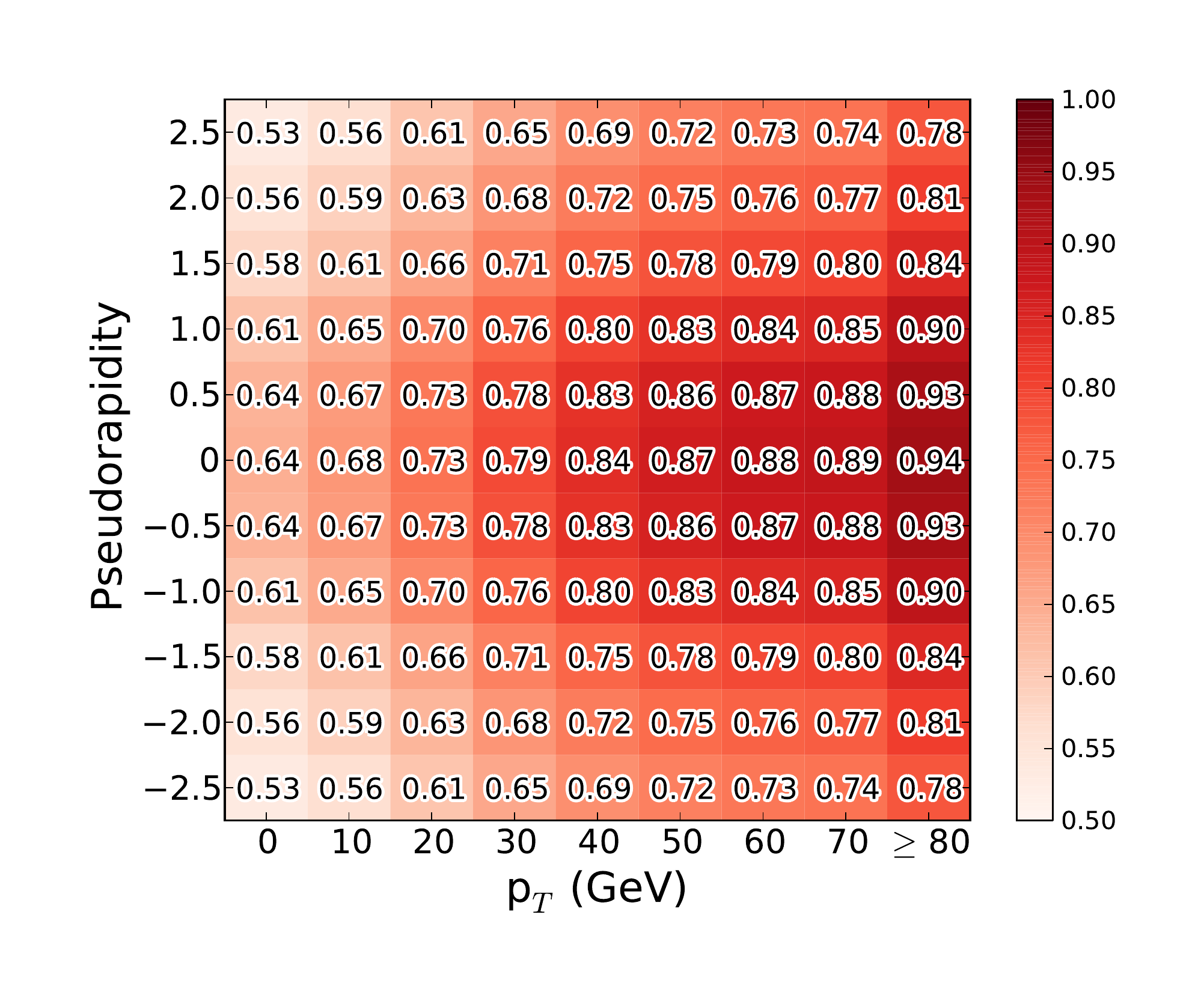}
\caption{Tight Electrons}
\label{fig:tunings:el:tot:b}
\end{subfigure}
\caption{Total efficiency map with respect to pseudorapidity and energy. This 
discretised version is only meant for illustrative purposes;  \Checkmate{} uses the full functional behaviour as shown in  \Cref{eqn:app:el:eff1,eqn:app:el:eff2,eqn:app:el:eff3,eqn:app:el:eff4}.}
\label{fig:tunings:el:tot}
\end{figure}

The efficiencies for electron objects to be reconstructed as such are combined from the 
individually measured reconstruction and 
identification efficiency, for which the latter distinguishes between two kinds of 
electrons: `medium' and `tight'\footnote{The 'loose' electron efficiency for identifying a truth-level electron 
is estimated by 
implementing a weak calorimeter isolation that mimics the ATLAS reconstruction. We require that 
within a cone size $dR<0.2$ around the electron, at least 80\% of energy deposition is due to the
electron.}. The reconstruction 
efficiency does not show any significant dependence on the candidate's momentum but mostly on the 
pseudorapidity, as can be seen in \Cref{fig:tunings:el:eff:a}. Conversely, the identification efficiency for medium electrons 
mostly depends on the transverse energy of the electron, as shown in \Cref{fig:tunings:el:eff:b}, and insignificantly on its position. The 
total efficiency is given as the product of the two contributions and can be seen in a representative two-dimensional grid in \Cref{fig:tunings:el:tot:a}.

For `tight' electrons, there exist measurements for both the pseudorapidity and the 
$p_T$ dependence of the identification efficiency individually, for which 
the corresponding other variable has been integrated 
over (see \Cref{fig:tunings:el:eff:c} and \Cref{fig:tunings:el:eff:d}). We 
normalise the pseudorapidity dependent function to an average 
value of unity and use as the overall identification efficiency the absolute 
efficiency with respect to $p_T$, multiplied by the renormalised efficiency with respect to pseudorapidity. 
Combining this with the reconstruction efficiency described before leads to the total efficiency distribution shown in \Cref{fig:tunings:el:tot:b}.

The following functional behaviour for the efficiencies is used ($p_T$ in \GeV):
\begin{align}
      \epsilon_\text{rec}(\eta) &= 0.987 + 1.28 \times 10^{-2} \cdot \eta^2  - 1.76 \times 10^{-2} \cdot \eta^4 + 5.21 \times 10^{-3} \cdot \eta^6 - 4.49 \times 10^{-4} \cdot \eta^8, \label{eqn:app:el:eff1}\\ 
\epsilon_{\text{id, medium}}(\pT) &= \left\{
  \begin{array}{lr}
   \displaystyle 0.767+\frac{0.151}{1+\mathrm{e}^{-0.145 \cdot (\pT-29.1)}} &  \quad  \text{if } \pT < 80, \\
\\
    0.946 &  \quad \text{if } \pT \geq 80,
  \end{array}
\right. \label{eqn:app:el:eff2}\\
\epsilon_{\text{id, tight}}(\pT) &= \left\{
  \begin{array}{lr}
   \displaystyle 0.565+\frac{0.279}{1+\mathrm{e}^{-0.0786 \cdot (\pT-22.3)}} &  \quad \text{if } \pT < 80, \\
\\
    0.883 &  \quad \text{if } \pT \geq 80,
  \end{array}
\right. \label{eqn:app:el:eff3}\\
\epsilon_{\text{id, tight}}^\text{normalised}(\eta) &= \frac{0.675+0.160 \cdot \exp\left(-\left(\frac{\eta}{1.92}\right)^2\right)}{0.777}. \label{eqn:app:el:eff4}
\end{align}
\subsection{Muons}
\begin{figure}
\centering
\begin{subfigure}{0.49\textwidth}
\includegraphics[width=1.0\textwidth]{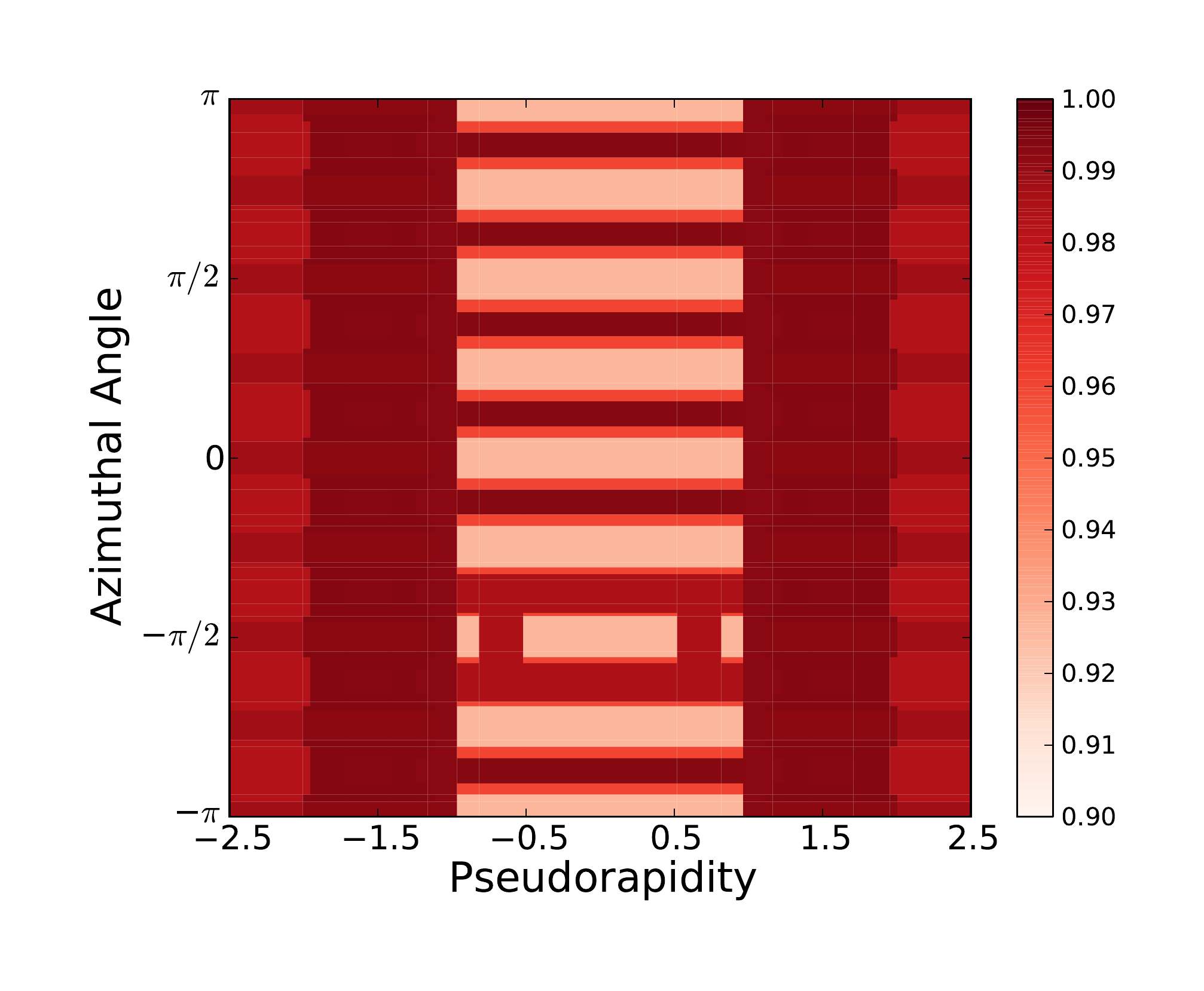}
\caption{Combined + Standalone Muons}
\label{app:fig:mu:a}
\end{subfigure}
\begin{subfigure}{0.49\textwidth}
\includegraphics[width=1.0\textwidth]{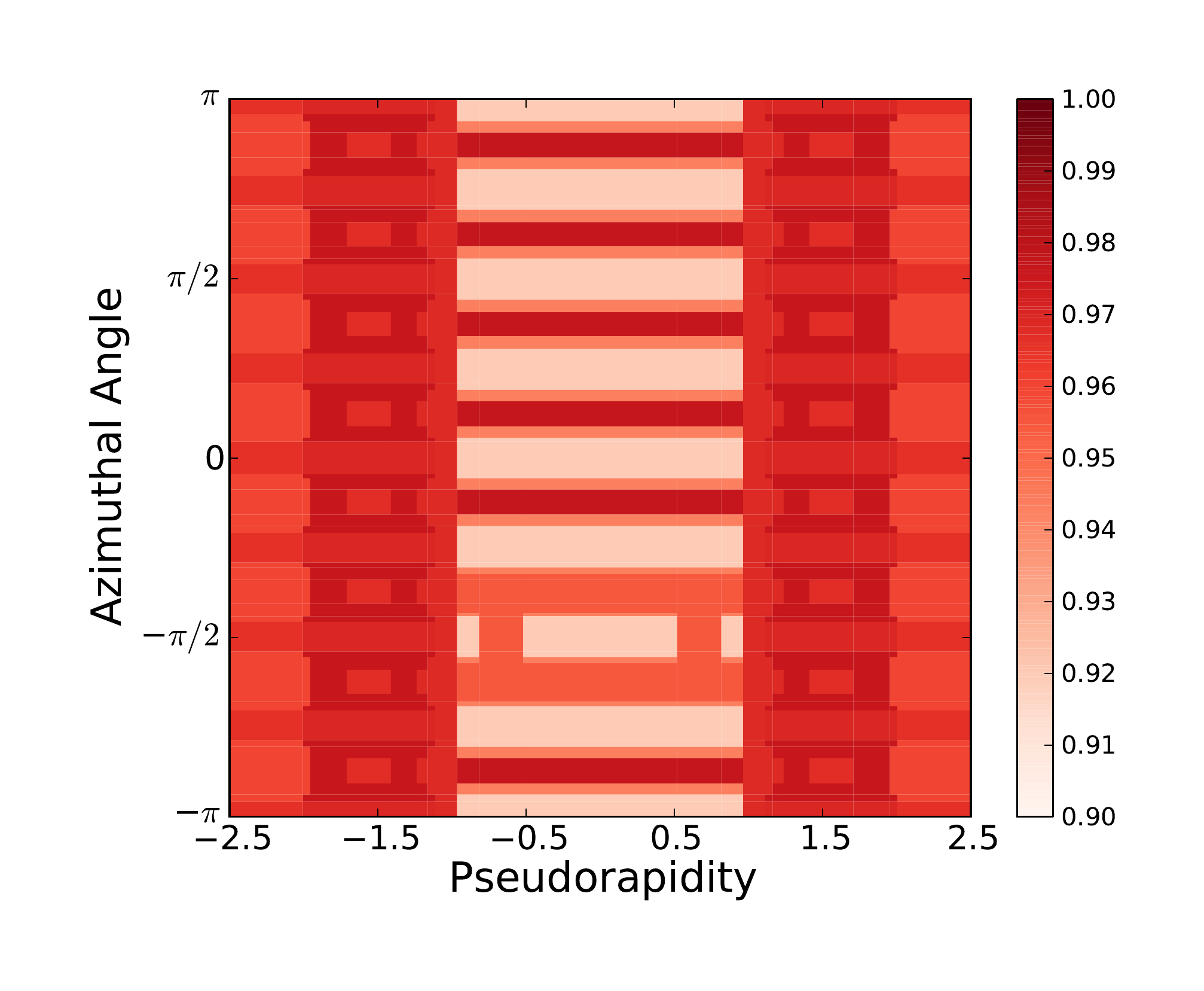}
\caption{Combined Muons}
\label{app:fig:mu:b}
\end{subfigure}
\caption{Efficiencies for muon objects, depending on the position in the detector. Detector map and component--dependent efficiencies are taken from \cite{ATLAS-CONF-2011-063}.}
\label{app:fig:mu}
\end{figure}
For muons, the main effect that has to be taken into account is the efficiency 
to reconstruct an object in the muon chambers and associate it to a track in the 
inner detector region. There exist two main quality criteria for this 
reconstruction, namely `Combined' (requires both a track in the inner detector and 
muon chambers) and `Standalone' (only requires a track in the muon chamber). The latter 
is usually used in combination with the first to maximise the muon efficiency and 
is hence called `Combined+Standalone' (a standalone track is used when a combined track is not reconstructed).

The muon chambers in ATLAS consist of different components, which each have a different reconstruction probability, mainly caused by different types and quantities of material and the geometry inside the full detector. In \Checkmate{}, we parametrised a detector-component map in the $\eta$--$\phi$ plane and associate a particular efficiency to each detector type. The resulting two-dimensional grid\footnote{We refrain from showing the functional description of the full map inside the main text, as the fine and irregular segmentation leads to a hard-to-read $21 \times 53$ matrix. If further information is desired, the user should feel free to contact the authors.}  is 
shown in \Cref{app:fig:mu}.
\subsection{B--Tagger}

\begin{figure}[t]
\begin{subfigure}{0.49\textwidth}
\includegraphics[width=1.0\textwidth]{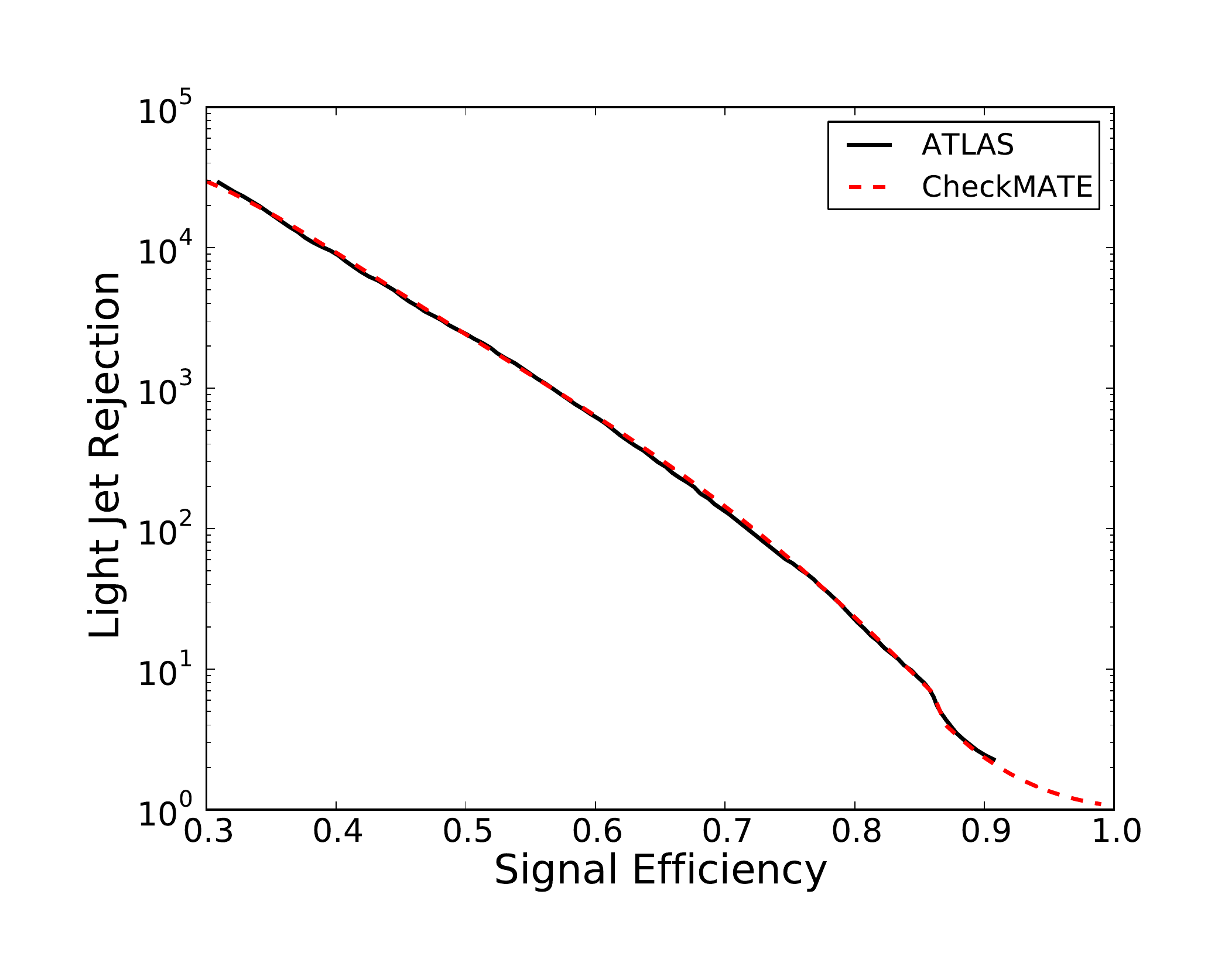}
\caption{Rejection curve for jets containing light quarks only.}
\label{fig:app:tune:btag:roc:a}
\end{subfigure}
\begin{subfigure}{0.49\textwidth}
\includegraphics[width=1.0\textwidth]{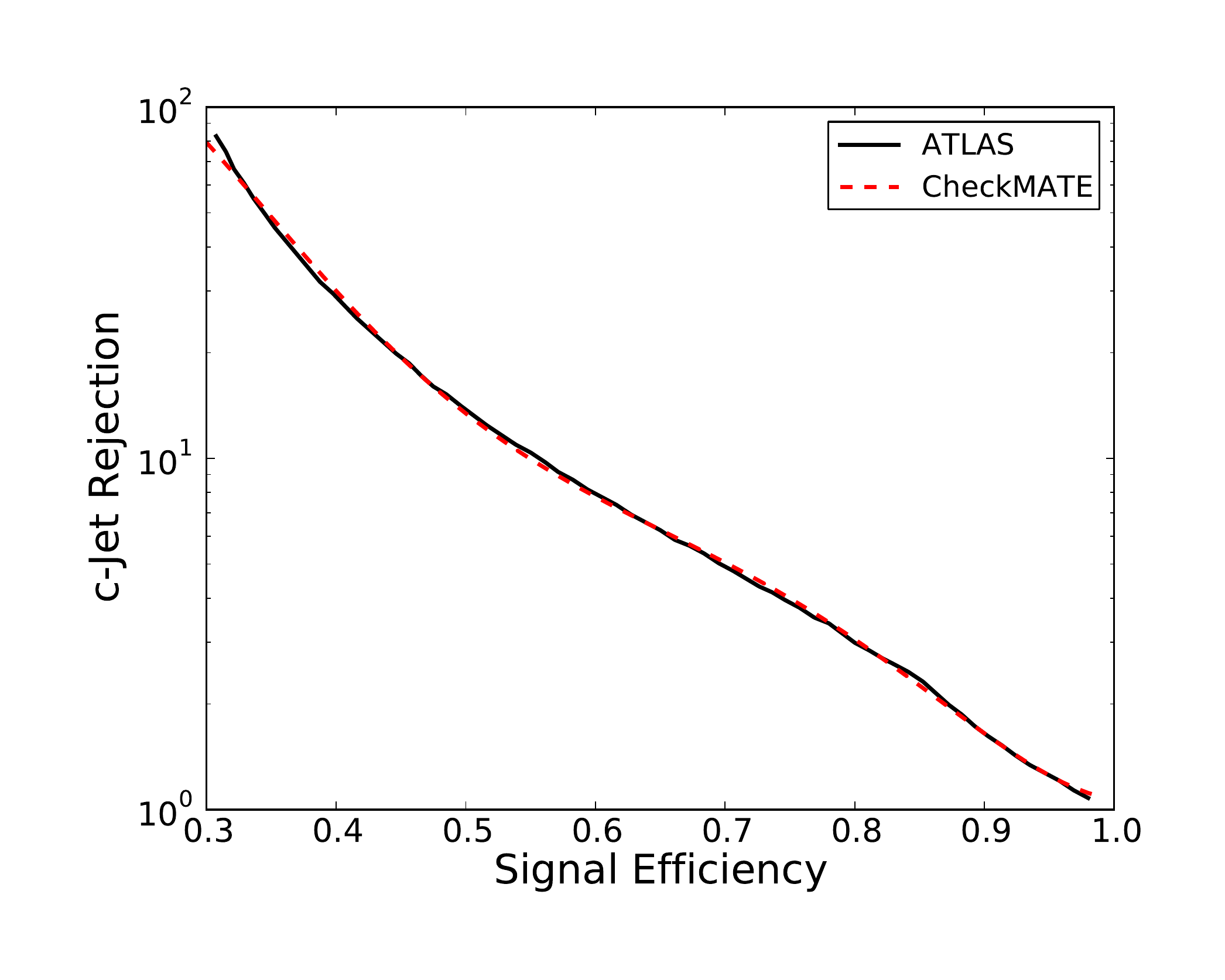}
\caption{Rejection curve for jets with charm content.}
\label{fig:app:tune:btag:roc:b}
\end{subfigure}

\caption{Receiver Operation Characteristic curves for the dependence of the background rejection for jets with different quark contents on the chosen signal efficiency working point of the b-tagger. \cite{ATLAS-CONF-2012-043}}
\label{fig:app:tune:btag:roc}
\end{figure}

The quality of algorithms that try to filter jets containing b--quarks from others 
is determined by two main quantities: The signal efficiency describes the probability 
to assign a tag to a jet that actually contains a b--quark, whereas the background
efficiency is a measure for the relative amount of jets that are tagged even though 
they don't have any bottom quark content. Since the background efficiency is 
usually small, it is common to use the inverse value, called rejection, for 
illustrative purposes. Also, one usually distinguishes between rejections against 
jets with charm--content and other jets that only contain light quarks, as the first 
are harder to distinguish from the signal. 

\begin{figure}[h]
\begin{subfigure}{0.49\textwidth}
\includegraphics[width=1.0\textwidth]{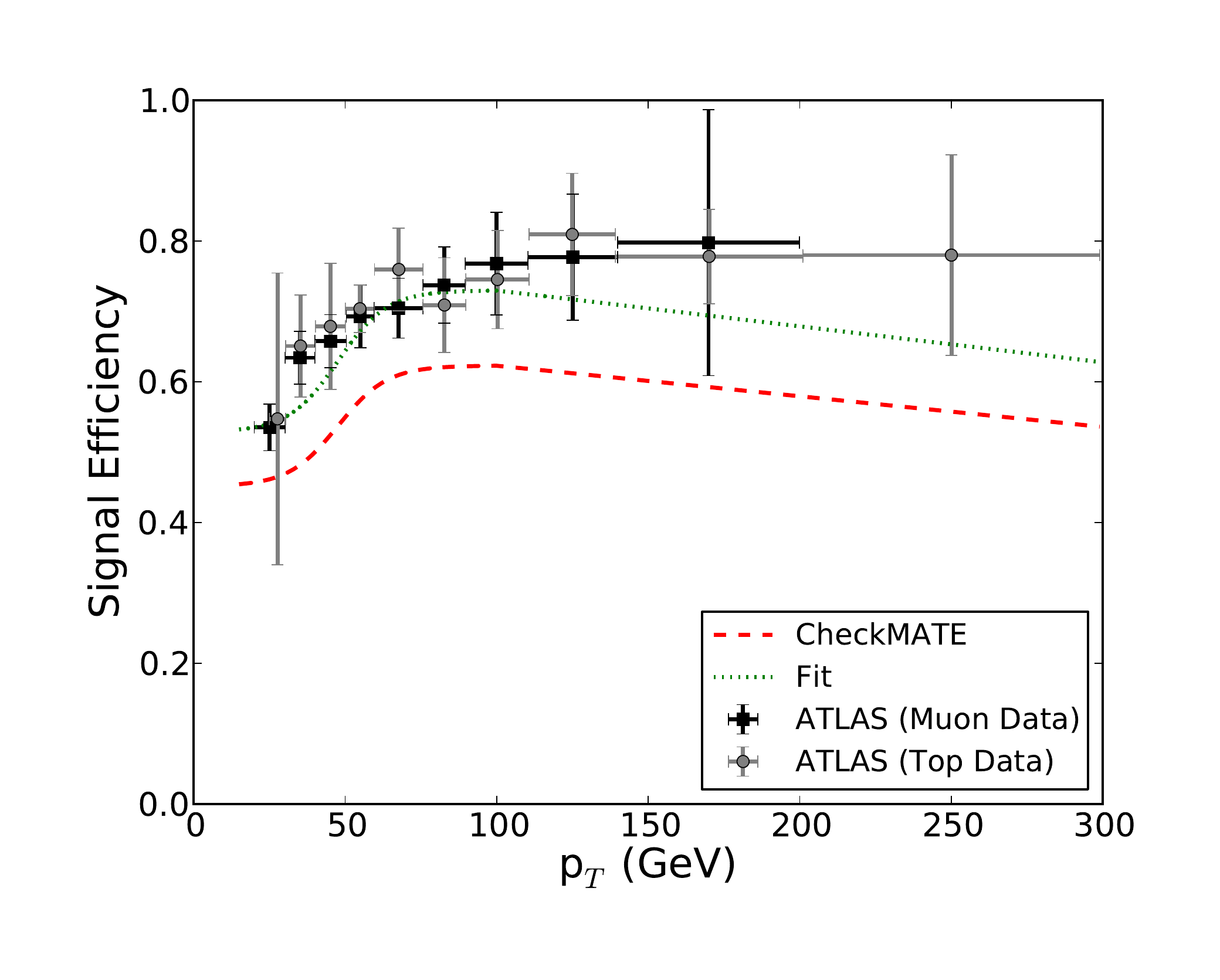}
\caption{Signal efficiencies for b--tagging , determined by combining the information of two different search 
channels for $\bar{\epsilon}_\text{S} = 0.7$ \cite{ATLAS-CONF-2012-043,ATLAS-CONF-2012-097}.}
\label{fig:app:tun:btag:pt:a}
\end{subfigure} \begin{subfigure}{0.49\textwidth}
\includegraphics[width=1.0\textwidth]{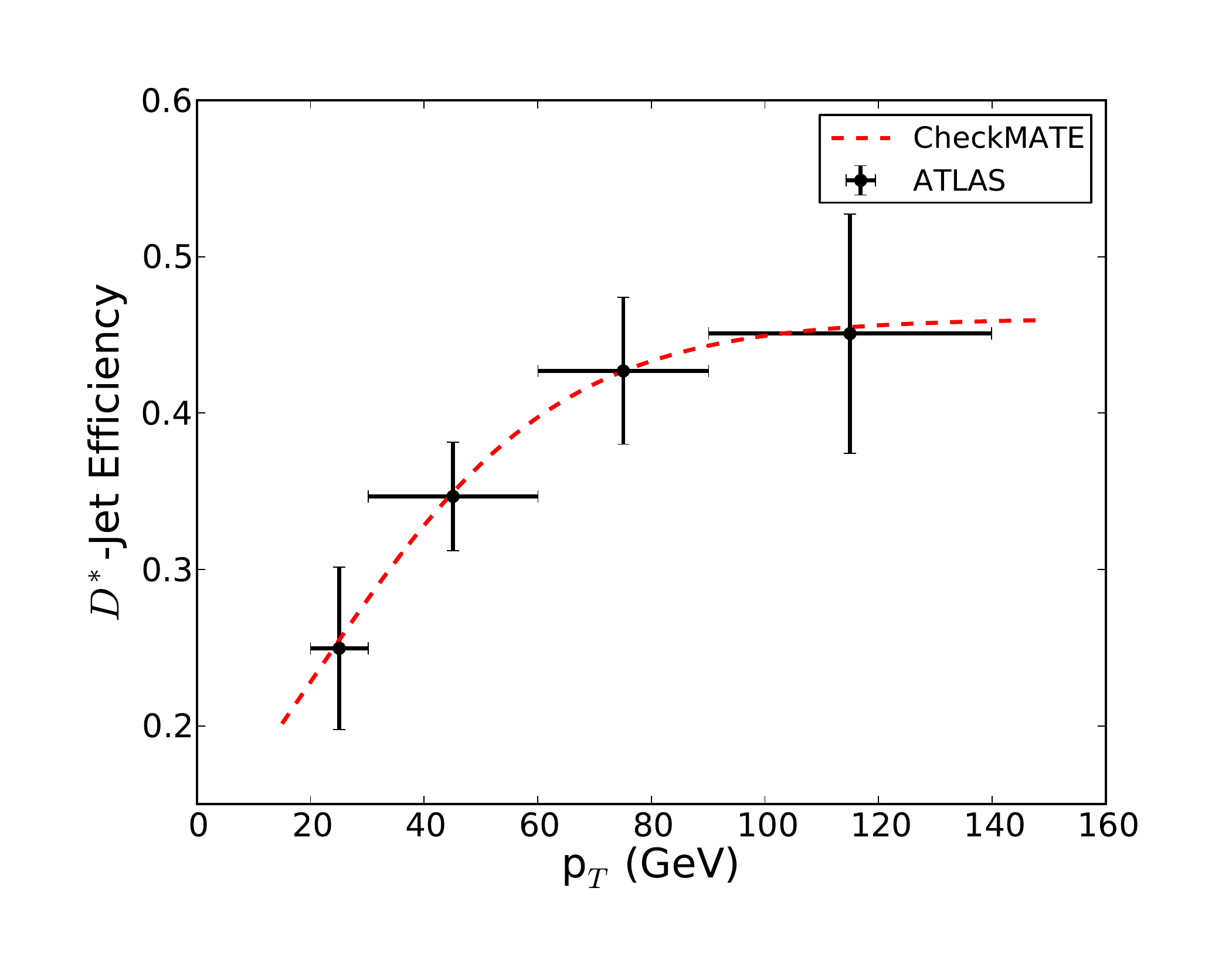} 
\caption{B--tagging efficiency of jets containing $D^*$ mesons for $\bar{\epsilon}_\text{S} = 0.7$ \cite{ATLAS-CONF-2012-097}. The inclusive efficiency on c--jets is assumed to be \unit[40]{\%} of this $D^*$ efficiency.}
\label{fig:app:tun:btag:pt:b}
\end{subfigure}

\begin{subfigure}{0.49\textwidth}
\includegraphics[width=1.0\textwidth]{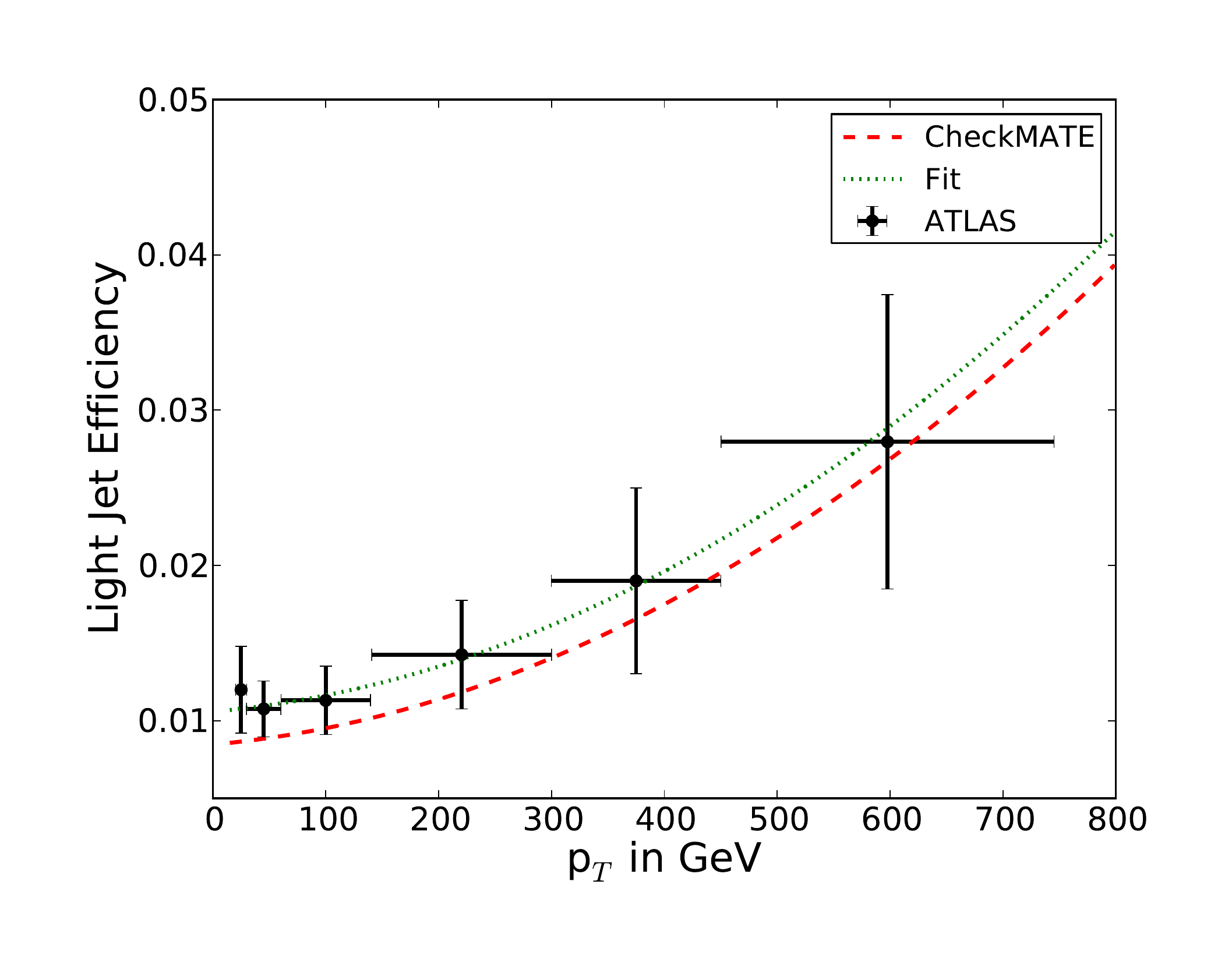}
\caption{B--tagging efficiency of jets containing light quarks for $\bar{\epsilon} _\text{S}= 0.7$ and $|\eta| < 1.3$ \cite{ATLAS-CONF-2012-040}.}
\label{fig:app:tun:btag:pt:c}
\end{subfigure}
\begin{subfigure}{0.49\textwidth}
\includegraphics[width=1.0\textwidth]{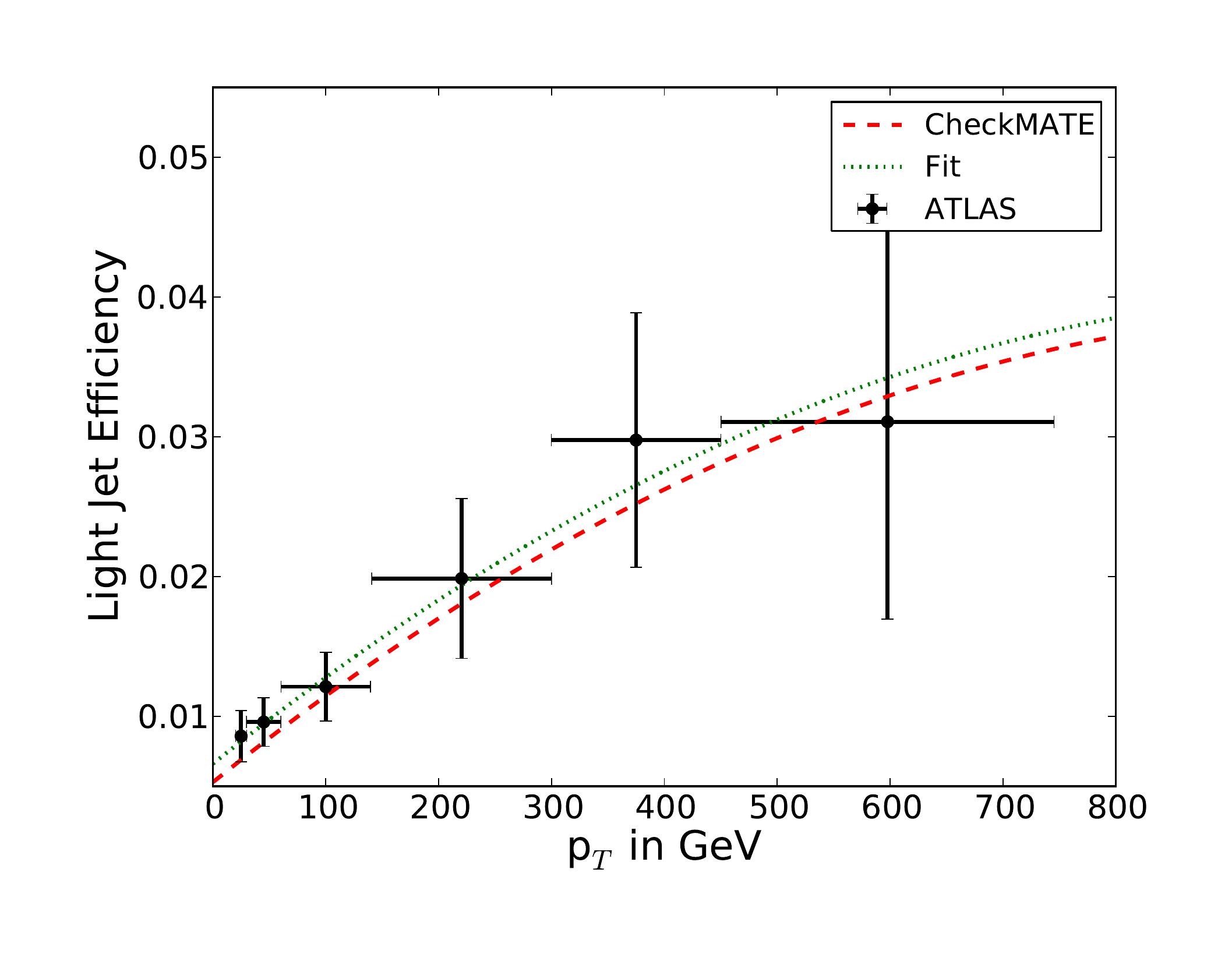}
\caption{Same as (c) for  $1.3 \leq |\eta| < 2.5$ \\ } 
\label{fig:app:tun:btag:pt:d}
\end{subfigure}
\caption{Dependence on the signal and light--jet / $c-$jet background efficiencies for b--tagging  on the transverse momentum of the jet candidate.}
\label{fig:app:tun:btag:pt}
\end{figure}

Since the rejection gets weaker with increasing signal efficiency, one has to find a 
balance between signal quantity and signal purity, which depends crucially on the 
details of the respective analysis. For this purpose, one uses the 
ROC (Receiver Operation Characteristic) curve that describes the relation between these 
two quantities. We show the ROC curves for light-jet rejection and c-jet 
rejection separately in \Cref{fig:app:tune:btag:roc} and internally parametrise 
these as follows:
\begin{align}
\log_{10}\left[\bar{r}_\text{light}(\bar{\epsilon}_\text{S})\right] &= \left\{
  \begin{array}{lr}
   \displaystyle 52.8\cdot(\bar{\epsilon}_\text{S}-4.045\cdot\bar{\epsilon}_\text{S}^2+7.17\cdot\bar{\epsilon}_\text{S}^3-6.14\cdot\bar{\epsilon}_\text{S}^4+2.01\cdot\bar{\epsilon}_\text{S}^5) &  \quad \text{if } \bar{\epsilon}_\text{S} < 0.87, \\
    -75.4\cdot(\bar{\epsilon}_\text{S}-1.07)^3 &  \quad \text{if }  \bar{\epsilon}_\text{S} \geq 0.87,
  \end{array}
\right. \\
\log_{10}\left[\bar{r}_\text{c}(\bar{\epsilon}_\text{S})\right] &= 29.3\cdot(\bar{\epsilon}_\text{S}-4.572\cdot\bar{\epsilon}_\text{S}^2+8.496\cdot\bar{\epsilon}_\text{S}^3-7.253\cdot\bar{\epsilon}_\text{S}^4+2.33\cdot\bar{\epsilon}_\text{S}^5).
\end{align}

Given a particular working point on the ROC curve, i.e.\ a specific chosen signal 
efficiency $\bar{\epsilon}_\text{S}$ and the corresponding background 
rejections $\bar{r}_\text{light/c}$, the actual tagging probabilities depend on the transverse momentum of the considered object. These have been measured 
individually for signal--, light-quark-- and $D^*$ meson\footnote{The tagging probability for
jets containing $D^*$ mesons is roughly 2 times better than for `normal' c-quarks. Using the cutflows 
from various analyses we have tuned this parameter to 0.4 to be in agreement with the
ATLAS results.} jets and we show the results in \Cref{fig:app:tun:btag:pt}. 

For the signal efficiency, we use two different data sets as they have different 
sensitivities at low and high energies (see \Cref{fig:app:tun:btag:pt:a}). In order
to agree with the cutflows of various analyses that require $b$-tagging, a reduction in the 
overall normalisation by 15\% has been applied. In addition, the 
significant decrease of the signal efficiency at large energies has been manually 
added in order to get better agreement with experimental results. 

Furthermore, the light quark 
jet rejection has been measured for two different $\eta$ regions, which we adapt in 
our parametrisations. We also perform a reduction in the light-quark tagging rates (20\%) in 
order to better agree with experimental cutflows. 

Since the $p_T$ dependent distributions are given for a 
particular working point $\bar{\epsilon}_\text{S} = 0.7$, we linearly rescale the functions to the given 
chosen signal efficiency $\bar{\epsilon}_\text{S}$, or the corresponding background efficiency given 
by the ROC curves ($p_T$ in \GeV):
\begin{align}
\epsilon_\text{S}(\pT) &= \frac{\bar{\epsilon}_\text{S} \cdot 0.85}{0.7}  \left(0.552+ \frac{0.210}{1+\mathrm{e}^{-0.123\cdot(\pT-47.6)}}\right)\frac{0.7+0.05\cdot \mathrm{e}^{-\frac{\pT}{308}}}{0.75}\left\{
  \begin{array}{ll}
  1 &  \quad \text{if }  \pT \leq 100, \\
\\
   1-7\times 10^{-4} \cdot (\pT-100)  &  \quad \text{if }  \pT > 100,
  \end{array}
\right. \\
\epsilon_{\text{light}}(\pT, \eta) &= \frac{\bar{r}_\text{light}(0.7)\times0.8}{\bar{r}_{\text{light}}(\bar{\epsilon}_\text{S})} \left\{
 \begin{array}{ll}
   1.06 \times 10^{-2} + 6.47 \times 10^{-6} \cdot \pT^2 + 4.03 \times 10^{-8} \cdot \pT^4   & \quad \text{if } |\eta| < 1.3, \\
\\
6.61 \times 10^{-3} + 6.49 \times 10^{-5} \cdot \pT^2 - 3.12 \times 10^{-8} \cdot \pT^4  & \quad \text{if } 1.3 \leq |\eta| < 2.5),
  \end{array}
\right. \\
\epsilon_{\text{c}}(\pT) &= \frac{\bar{r}_\text{c}(0.7)\times0.4}{\bar{r}_{\text{c}}(\bar{\epsilon}_\text{S})} \frac{0.461}{1+\mathrm{e}^{-0.0464\cdot(\pT-20.4)}}.
\end{align}
%
\subsection{Tau--Tagger}
\begin{figure}[b]
\begin{subfigure}{0.49\textwidth}
\includegraphics[width=1.0\textwidth]{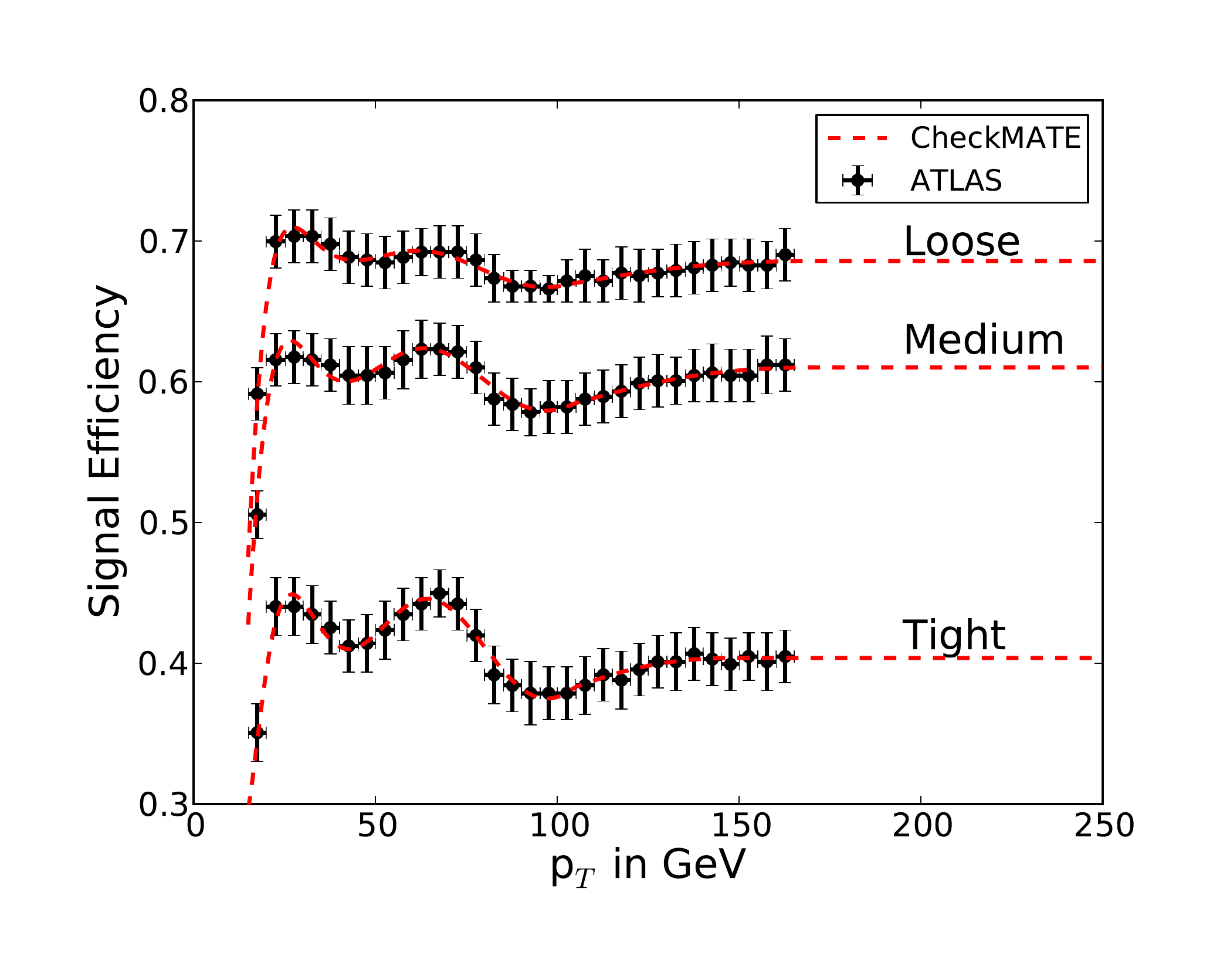}
\caption{Signal efficiencies for 1--prong candidates.}
\end{subfigure}
\begin{subfigure}{0.49\textwidth}
\includegraphics[width=1.0\textwidth]{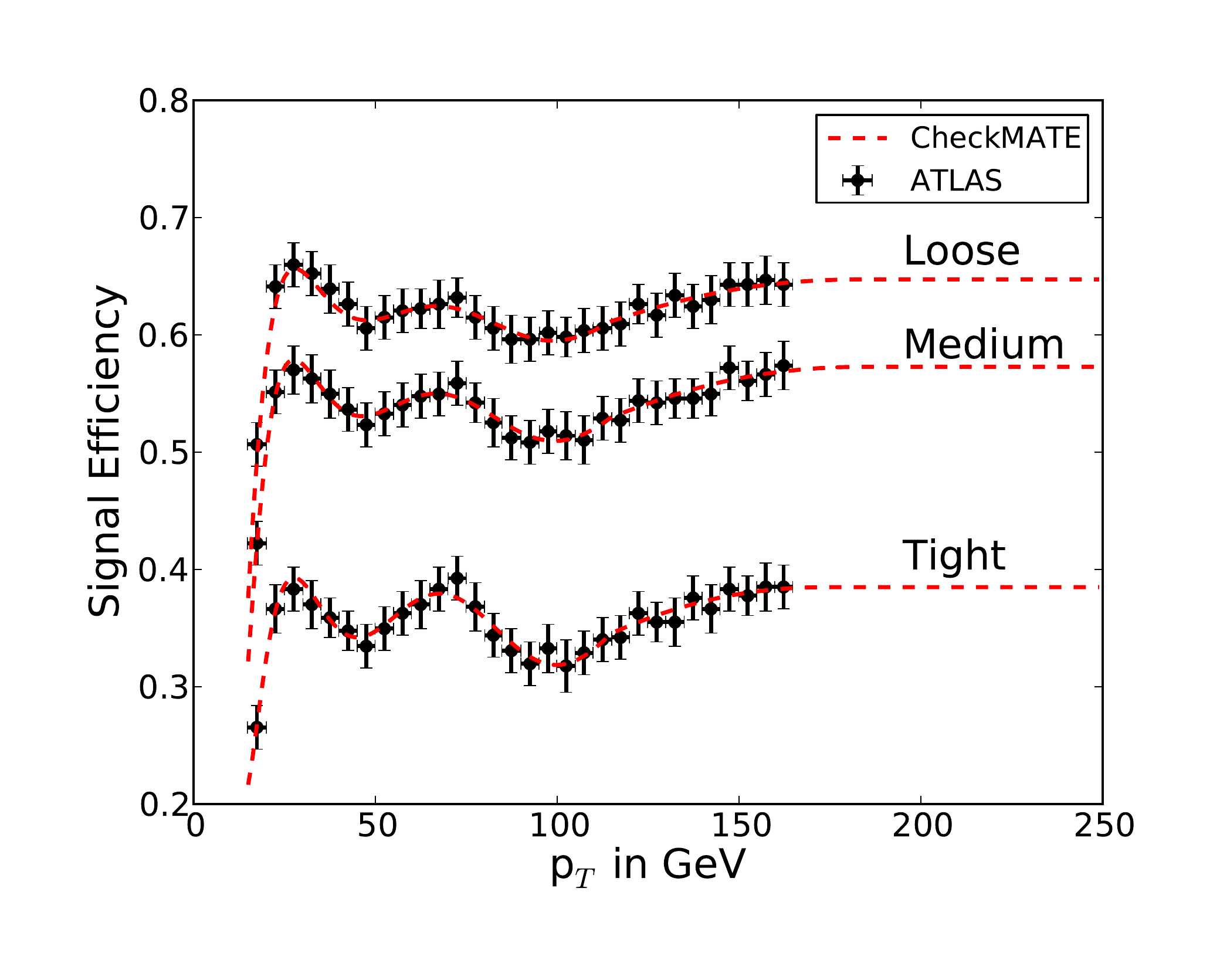}
\caption{Signal efficiencies for 3--prong candidates.}
\end{subfigure}

\begin{subfigure}{0.49\textwidth}
\includegraphics[width=1.0\textwidth]{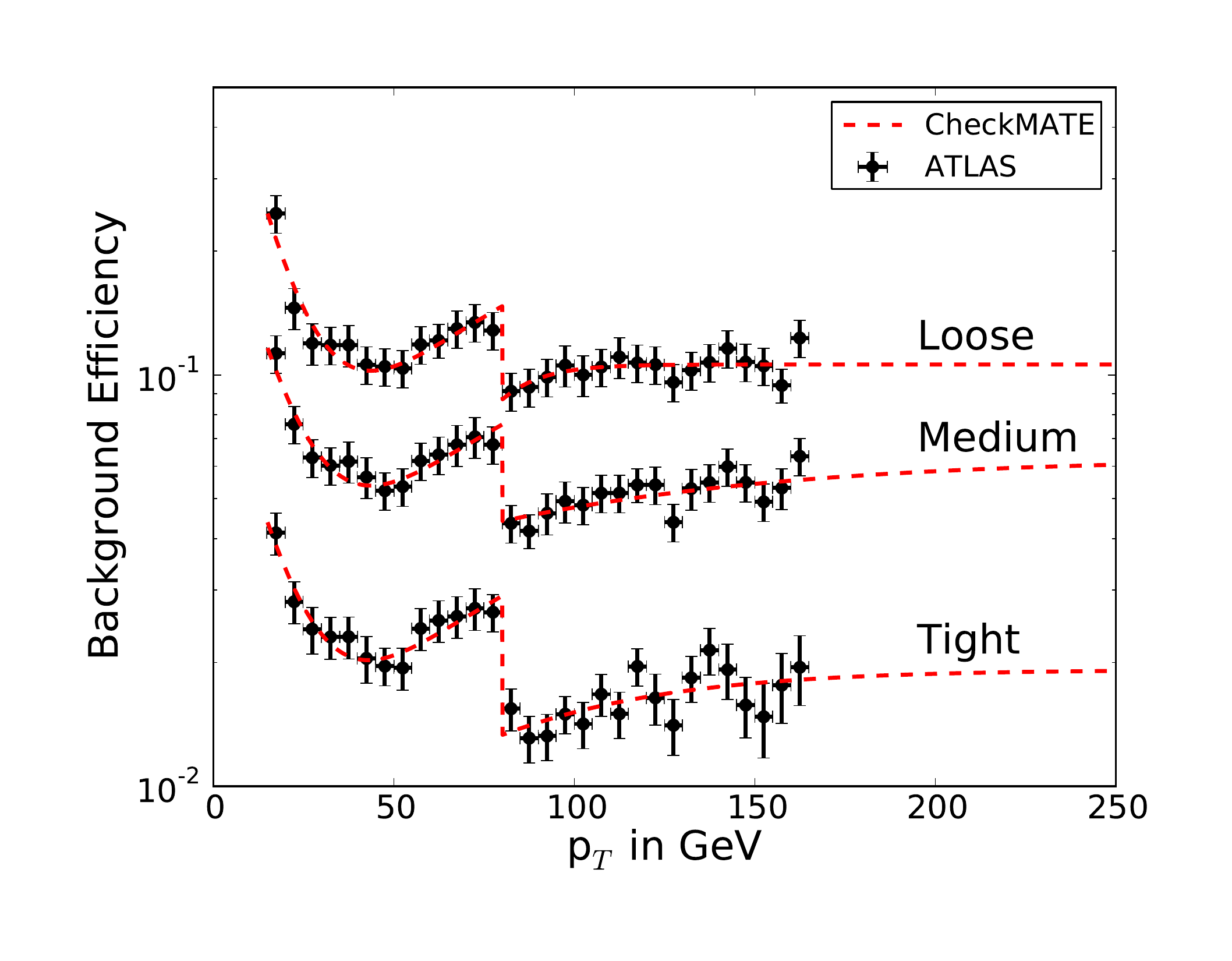}
\caption{Background efficiencies for 1--prong candidates.}
\end{subfigure}
\begin{subfigure}{0.49\textwidth}
\includegraphics[width=1.0\textwidth]{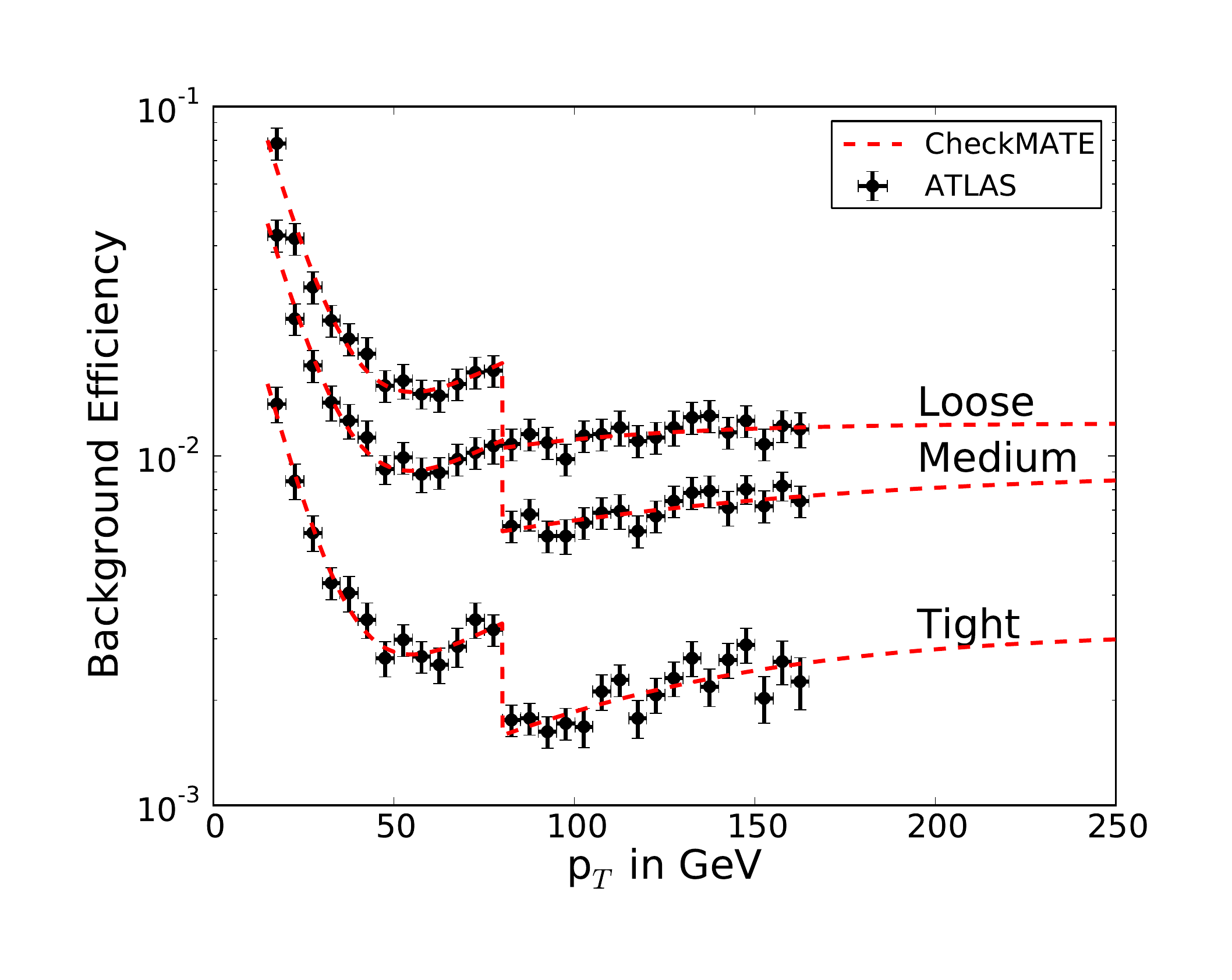}
\caption{Background efficiencies for 3--prong candidates.}
\end{subfigure}
\caption{Signal and background efficiencies used for the tau--tagging discrimination against QCD jets. The results differentiate between jet candidates with one and with three reconstructed tracks (prongs) and the three most commonly used working points `loose', `medium' and `tight \cite{ATLAS-CONF-2013-064}. Explicit functions for the \Checkmate{} parametrisations are shown in \Cref{eqn:app:tun:tau:S,eqn:app:tun:tau:B}.}
\label{sec:app:tun:tau}
\end{figure}

\begin{figure}
\begin{subfigure}{0.49\textwidth}
\includegraphics[width=1.0\textwidth]{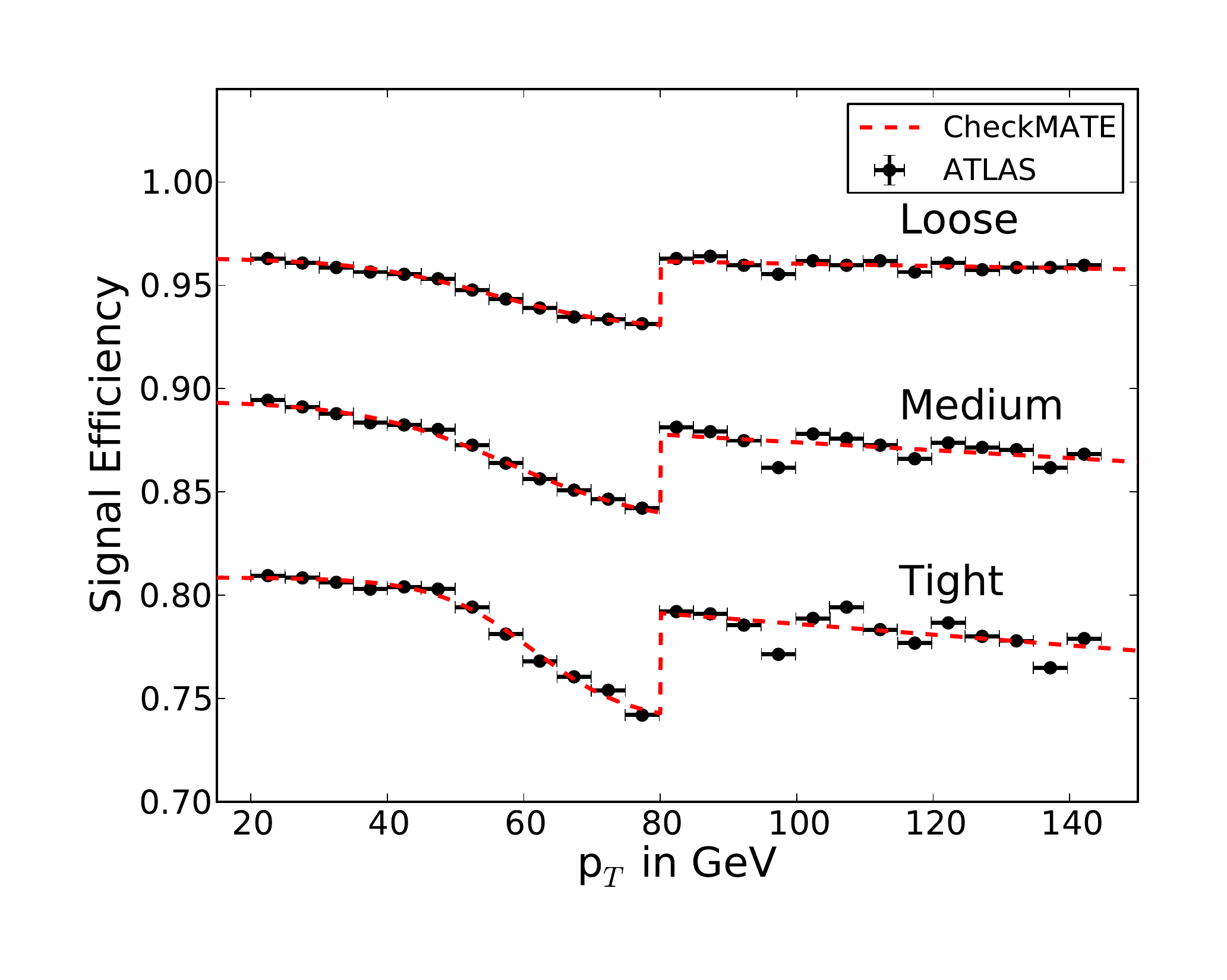}
\caption{Signal efficiency with respect to the candidate's transverse momentum.}
\label{fig:app:tun:tau:sig:el1}
\end{subfigure}
\begin{subfigure}{0.49\textwidth}
\includegraphics[width=1.0\textwidth]{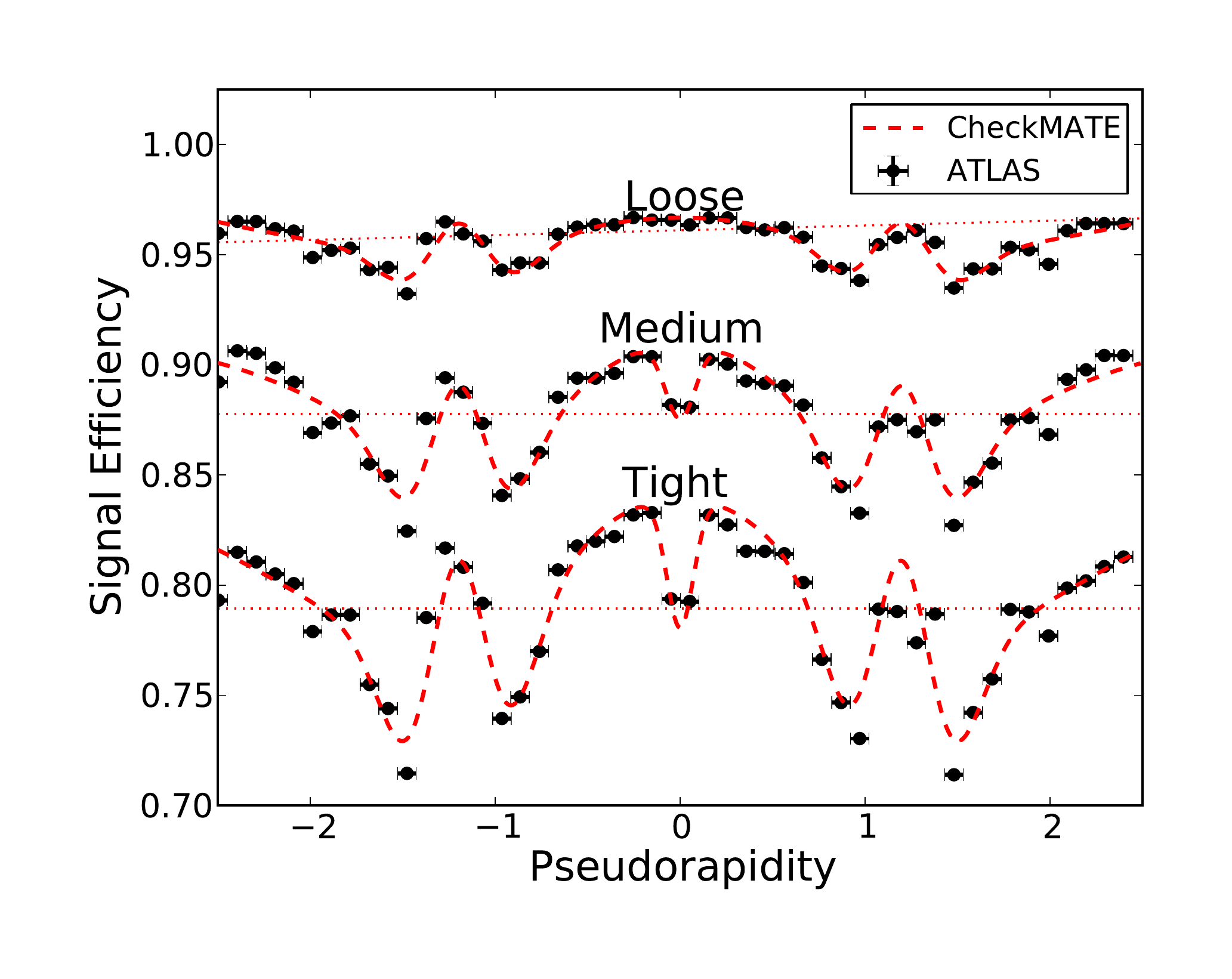}
\caption{Signal efficiency with respect to the candidate's pseudorapidity.}
\label{fig:app:tun:tau:sig:el2}
\end{subfigure}
\caption{Signal efficiencies used for the tau--tagging discrimination against electrons  \cite{ATLAS-CONF-2013-064}. These only affect 1--prong candidates, but differ between the different considered working points. Explicit functions for the \Checkmate{} parametrisations are shown in \Cref{eqn:app:tun:tau:sig:el1,eqn:app:tun:tau:sig:el2}. Dashed lines denote the respective average efficiency value.}
\end{figure}

\begin{figure}
\begin{subfigure}{0.49\textwidth}
\includegraphics[width=1.0\textwidth]{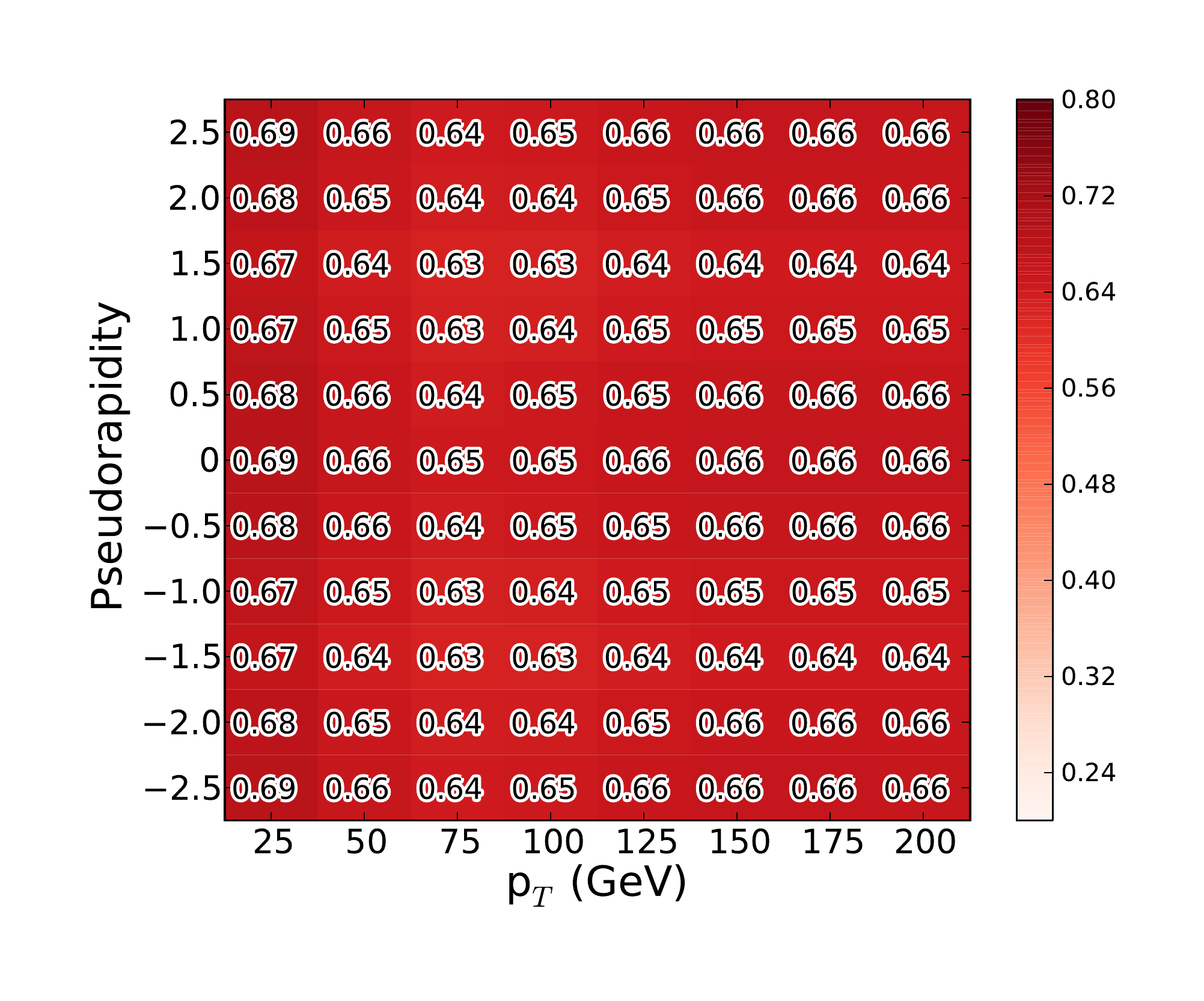}
\caption{Loose}
\label{fig:app:tun:tau:sig:tot:loose}
\end{subfigure}
\begin{subfigure}{0.49\textwidth}
\includegraphics[width=1.0\textwidth]{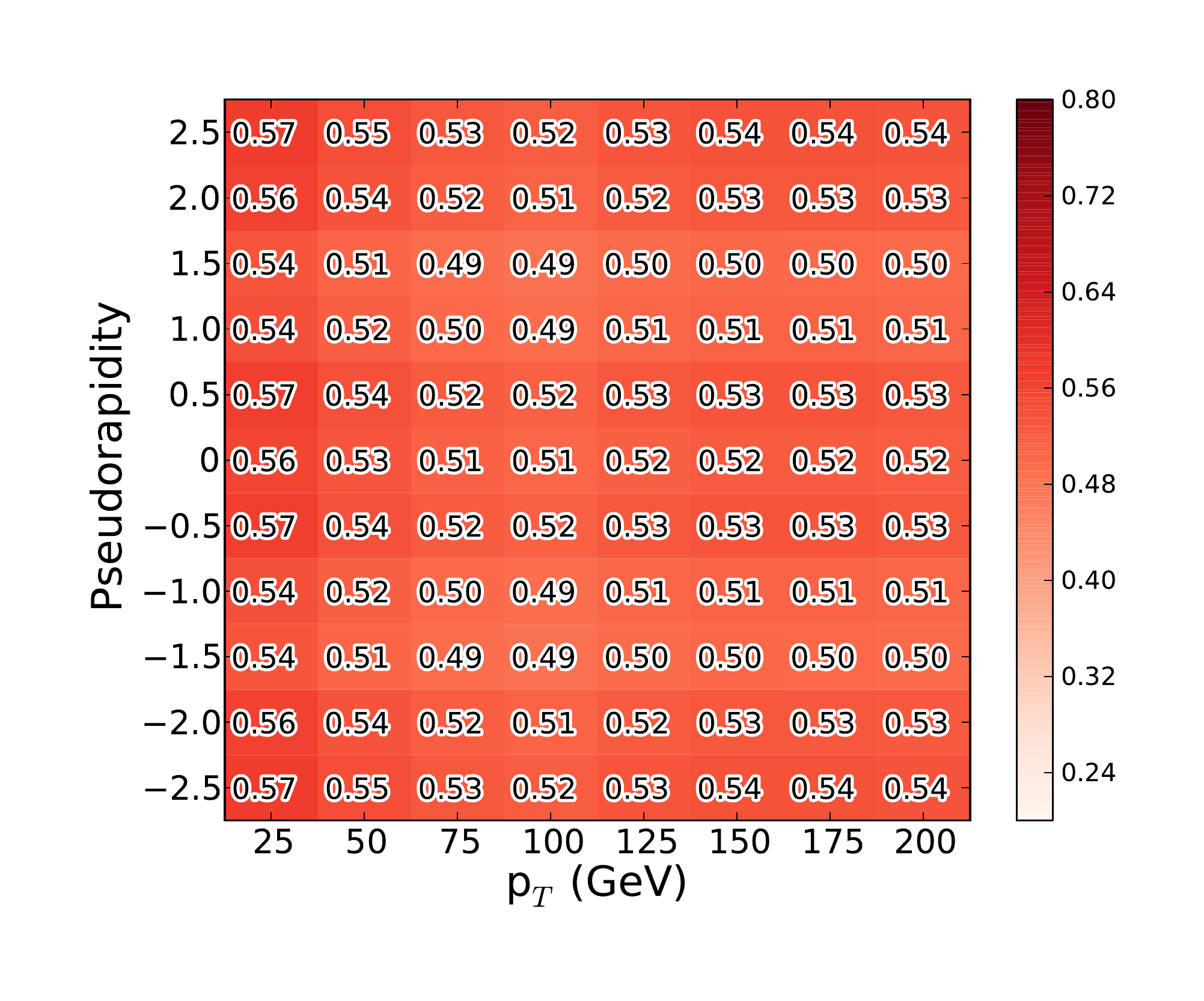}
\caption{Medium}
\label{fig:app:tun:tau:sig:tot:medium}
\end{subfigure}
\begin{subfigure}{0.49\textwidth}
\includegraphics[width=1.0\textwidth]{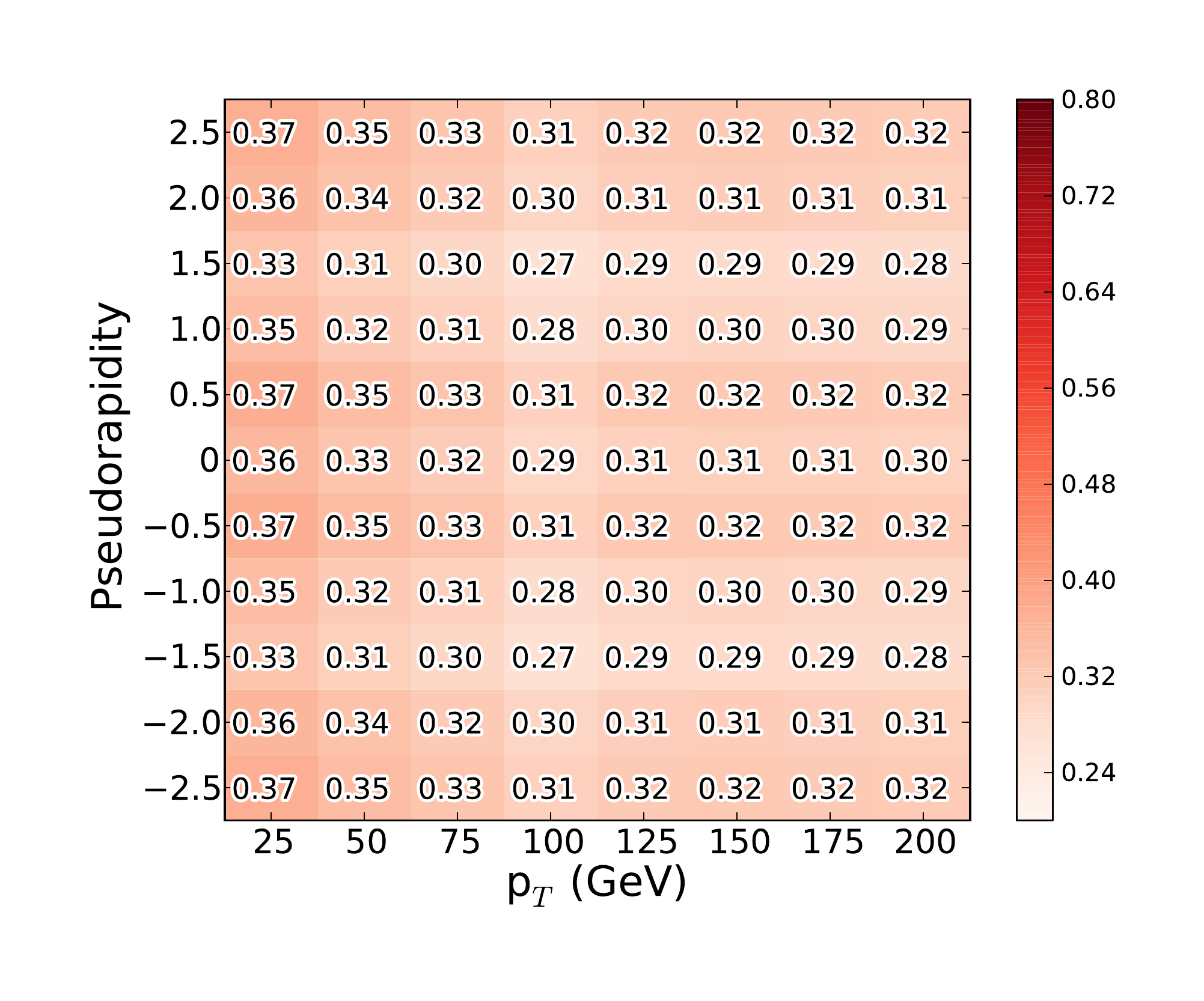}
\caption{Tight}
\label{fig:app:tun:tau:sig:tot:tight}
\end{subfigure}
\caption{Total signal 1--prong efficiencies for 1--prong signal candidates, combining the contributions from rejections against QCD jets and electrons.}
\end{figure}
Analogously to the tagging of jets containing $b-$quarks, there exist algorithms to distinguish 
jets that originated from a hadronically decaying $\tau$ 
lepton from those that originated from quarks or gluons. Due to charge conservation, 
the $\tau$ lepton can only decay into an odd number of charged objects --- mostly into 1 
or 3 --- called prongs. Since the structures of the resulting jets look rather different, 
identification algorithms usually differentiate between these two cases and hence there are 
individual efficiencies for each. 

We show the efficiencies for signal and (light parton--)background with respect to the momentum 
of the jet candidate in \Cref{sec:app:tun:tau}. Again, three common efficiency 
working points --- `loose, `medium' and `tight' --- are used in most analyses, and 
we show results for each of the three. The functional descriptions of the data points 
are shown in \Cref{eqn:app:tun:tau:S,eqn:app:tun:tau:B}, with the 
corresponding numerical values listed in \Cref{tab:app:tun:tau:S,tab:app:tun:tau:B}. For background and 3--prong jets, these are the final efficiency functions \Checkmate{} uses. Note that these efficiencies do not depend on $\eta$, but $|\eta| < 2.5$ is always implied to
fall within the angular coverage of the tracking detectors.

In addition to QCD jets, there is a second type of background which has to be taken into account: Electrons, which have been reconstructed as jets, can resemble 1--prong tau decays and need a separate tagging algorithm, which has been specifically tuned to reject these electron jets\footnote{We only consider the impact on the signal efficiencies of this algorithm. In particular, we did not implement a specific mistagging efficiency for electrons. With the given Delphes code structure, any electron that fails the identification efficiency cut will be counted as a jet. Then, it will be tagged according to the corresponding background efficiency in \Cref{eqn:app:tun:tau:B}}. Efficiencies with respect to this algorithm depend on both transverse momentum and pseudorapidity as shown in \Cref{fig:app:tun:tau:sig:el1,fig:app:tun:tau:sig:el2}. We follow the same approach as for the electron efficiencies, i.e.\ we take the absolute value from the momentum dependent efficiency distribution and 
multiply with the pseudorapidity distribution normalised to an average efficiency value of 1. 
The corresponding functions are shown in \Cref{eqn:app:tun:tau:sig:el1,eqn:app:tun:tau:sig:el2} and \Cref{tab:app:tun:tau:sig:el1,tab:app:tun:tau:sig:el2}. We show a final combination of these with the 1--prong 
signal efficiency from before in 
\Cref{fig:app:tun:tau:sig:tot:loose,fig:app:tun:tau:sig:tot:medium,fig:app:tun:tau:sig:tot:tight} for each 
of the three working points ($p_T$ in \GeV):

\allowdisplaybreaks[1]
\begin{align}
\epsilon_\text{S}(\pT) &= \epsilon_0 + A_1 \pT^{\alpha}e^{-\lambda_1 \pT^{\beta}} + 
A_2 \left\{
\begin{array}{ll}
 (\pT-80)^2e^{-\lambda_2(\pT-80)}  &  \quad \text{if } \pT < 80 + \frac{2}{\lambda_2}, \\
                                   \\
 \displaystyle \frac{4}{\lambda_2^2}e^{-2}  &  \quad \text{if } \pT \geq 80 + \frac{2}{\lambda_2},
  \end{array}
\right\}
+ \nonumber \\
& \hspace{0.25\textwidth}
A_3\left\{
\begin{array}{ll}
 \displaystyle\frac{\sin(\omega(\pT-15)+\phi)}{\pT^{\gamma}}  &  \quad \text{if }  \pT < 105, \\
                                   \\
 \displaystyle\frac{\sin(90\omega+\phi)}{105^{\gamma}}  &  \quad \text{if }  \pT \geq 105,
  \end{array}
\right\}   \label{eqn:app:tun:tau:S} \\
\epsilon_\text{B}(\pT) &= \left\{
\begin{array}{ll}
A_1\mathrm{e}^{-\lambda_1\pT}+m\cdot \pT  &  \quad \text{if }  \pT < 80, \\
                                   \\
A_2 \left(1-\frac{\displaystyle 1}{\displaystyle 1+e^{\lambda_2(\pT-\pT^0)}}\right)  &  \quad \text{if }  \pT \geq 80,
  \end{array}
\right\} \label{eqn:app:tun:tau:B} \\
\epsilon^\text{el. veto}_{\text{S}}(\pT) &= \left\{
  \begin{array}{ll}
 \epsilon_0 + \frac{\displaystyle k}{\displaystyle 1+e^{\lambda \cdot(\pT-\pT^0)}}  &  \quad \text{if }  \pT \leq 80, \\
\\
- m \cdot \pT + n   &  \quad \text{if }   \pT > 80,
  \end{array}
\right. \label{eqn:app:tun:tau:sig:el1}\\
\epsilon^\text{1p el. rej. (norm)}_{\text{S}}(\eta) &= \frac{1}{\langle \epsilon \rangle} \left(\epsilon_0 -A_1 \cdot \mathrm{e}^{-\frac{(|\eta|-1.2)^2}{\sigma_1^2}} + A_2 \cdot \mathrm{e}^{-\frac{(|\eta|-1.2)^2}{\sigma_2^2}} - A_3 \cdot \frac{\sin(\eta)}{\eta}\mathrm{e}^{-\frac{\eta^2}{\sigma_3^2}} + A_4 \cdot \eta^2 \mathrm{e}^{-\frac{\eta^2}{\sigma_4^2}}
\right). \label{eqn:app:tun:tau:sig:el2}
\end{align}

\begin{table}
\begin{subtable}{1.0\textwidth}
\centering
\begin{tabular*}{1.0\textwidth}{@{\extracolsep{\fill} } l ccccccccccc}
\hline
\hline
 & $\epsilon_0$ &  $A_1$ & $\alpha$ & $\lambda_1$ & $\beta$ & $A_2$ & $\lambda_2$ &$A_3$ & $\omega$ & $\phi$ & $\gamma$ \\
\hline
1 prong, loose &0.0223 & $-2.55$&1.16&0.427&0.846&$2.14\times10^{-5}$    &0.670&100&0.0974&2.34&  2.04    \\
1 prong, medium   & 0.0223&   $-2.69$& 1.23& 0.483& 0.791& $3.46\times10^{-5}$   & 0.586 &100& 0.0997& 2.23&  1.91 \\
1 prong, tight & 0.0292&  $-2.78$& 1.17& 0.377& 0.799& $6.17\times10^{-5}$    & 0.388& 100& 0.101& 2.16& 1.79   \\  
3 prong, loose& 0.0192  & $-6.44$& 0.143& 0.106& 1.08& $4.22\times10^{-5}$   & 0.594& 99.6& 0.0958& 2.09& 1.92 \\
3 prong, medium   & 0.0191 &  $-2.32$& 0.924& 0.233& 0.940& $5.00\times10^{-5}$   &  0.510& 100& 0.0950& 2.18& 1.86 \\
3 prong, tight  & 0.0212 & $-2.29$& 0.671& 0.104& 1.096& $6.30\times10^{-5}$      &  0.324& 100& 0.0971& 2.09& 1.78   \\
\hline
\hline
\end{tabular*}
\caption{Parameters corresponding to \Cref{eqn:app:tun:tau:S} for different signal working points and both 1--prong and 3--prong candidates.}
\label{tab:app:tun:tau:S}
\end{subtable}

\begin{subtable}{1.0\textwidth}
\centering
\begin{tabular*}{1.0\textwidth}{@{\extracolsep{\fill} } l cccccc}
\hline
\hline
 & $A_1$ & $\lambda_1$ & $m$ & $A_2$ & $\lambda_2$ & $\pT^0$  \\
\hline
1 prong, loose& 0.717& $7.89 \times 10^{-2}$& $1.82\times 10^{-3}$& 0.106& $9.73\times 10^{-2}$& 64.3 \\
1 prong, medium & 0.301& $7.20\times 10^{-2}$& $9.35\times 10^{-4}$& $6.27\times 10^{-2}$& $1.44\times 10^{-2}$& 20.0\\
1 prong, tight& 0.117& $7.42\times 10^{-2}$& $3.59\times 10^{-4}$& $1.92\times 10^{-2}$& $2.47\times 10^{-2}$& 46.8 \\
3 prong, loose & 0.265& $8.27\times 10^{-2}$& $2.26\times 10^{-4}$& $1.24\times 10^{-2}$& $2.28\times 10^{-2}$& 4.08 \\  
3 prong, medium & 0.154& $8.32\times 10^{-2}$& $1.36\times 10^{-4}$& $9.06\times 10^{-3}$& $1.19\times 10^{-2}$& 20.0 \\ 
3 prong, tight& $5.79 \times 10^{-2}$ & $8.80\times 10^{-2}$& $4.08\times 10^{-5}$     & $3.14\times 10^{-3}$& $1.73\times 10^{-2}$& 78.9 \\  
\hline
\hline
\end{tabular*}
\caption{Parameters corresponding to \Cref{eqn:app:tun:tau:B} for different working points and both 1--prong and 3--prong candidates.}
\label{tab:app:tun:tau:B}
\end{subtable}

\begin{subtable}{1.0\textwidth}
\begin{tabular*}{1.0\textwidth}{@{\extracolsep{\fill} } lccccccccc}
\hline
\hline
 & $\epsilon_0$ & $A_1$ & $\sigma_1$ & $A_2$ & $\sigma_2$ & $A_3$ & $\sigma_3$ & $A_4$ & $\sigma_4$ \\
\hline
loose & 0.94 &  0.875 &  0.294 &  0.886 &  0.286 &  $-2.68 \times 10^{-2}$ &  1.16 &  $4.19 \times 10^{-3}$  &  10.9 \\
medium & 0.91 &  0.283 &  0.303 &  0.309 &  0.256 &  $3.46 \times 10^{-2}$ &  0.113 &  $-7.96 \times 10^{-2}$ &  1.25 \\
tight & 0.84 &  0.472 &  0.304 &  0.503 &  0.258 &  $5.94 \times 10^{-2}$ &  0.102 &  $-8.51 \times 10^{-2}$ &  1.42 \\
\hline
\hline
\end{tabular*}
\caption{Parameters corresponding to \Cref{eqn:app:tun:tau:sig:el1} for different working points (1--prong only).}
\label{tab:app:tun:tau:sig:el1}
\end{subtable}

\begin{subtable}{1.0\textwidth}
\begin{tabular*}{1.0\textwidth}{@{\extracolsep{\fill} } lccccccc}
\hline
\hline
     & $\langle \epsilon \rangle$& $\epsilon_0$ & k & $\lambda$ & $\pT^0$& $m$ & $n$  \\
\hline
loose & 0.956 & 0.928  &  $3.54 \times 10^{-2}$ &  $0.0994$ &  $55.1$ &  $5.52 \times 10^{-5}$ & 0.966 \\
medium & 0.878 & 0.833  & $ 6.13 \times 10^{-2}$ & 0.0922 & 57.8 & $1.90 \times 10^{-4}$ & 0.893 \\
tight & 0.789 & 0.738 & $7.06 \times 10^{-2}$ & 0.140 & 61.4 & $2.59 \times 10^{-4}$ & 0.812   \\
\hline
\hline
\end{tabular*}
\caption{Parameters corresponding to \Cref{eqn:app:tun:tau:sig:el2} for different working points (1--prong only).}
\label{tab:app:tun:tau:sig:el2}
\end{subtable}
\caption{Numerical parameter values for the various efficiency distributions 
defined in \Cref{eqn:app:tun:tau:S,eqn:app:tun:tau:B,eqn:app:tun:tau:sig:el1,eqn:app:tun:tau:sig:el2}.}
\end{table}

\newpage

\clearpage
\section{Analysis Validation}
\label{sec:app:analysis_val}

\subsection{atlas\_conf\_2012\_104}
\noindent 1 lepton and \etmiss. \cite{ATLAS-CONF-2012-104} \\
Energy: 8 \TeV \\
Luminosity: 5.8 fb$^{-1}$ \\
Validation notes:
\begin{itemize}
\item Validation has been performed versus the published CMSSM (mSUGRA) parameter scan.
\item The exclusion in \Checkmate{} is slightly weaker than the published ATLAS result, since
ATLAS uses a different limit setting procedure. \Checkmate{} only uses the 
signal region with the best expected sensitivity to set the limit. However, in this search ATLAS used a combined likelihood including all
signal and control regions.

\end{itemize}

\begin{figure}[h]
\centering \vspace{-0.4cm}
\includegraphics[width=0.49\textwidth]{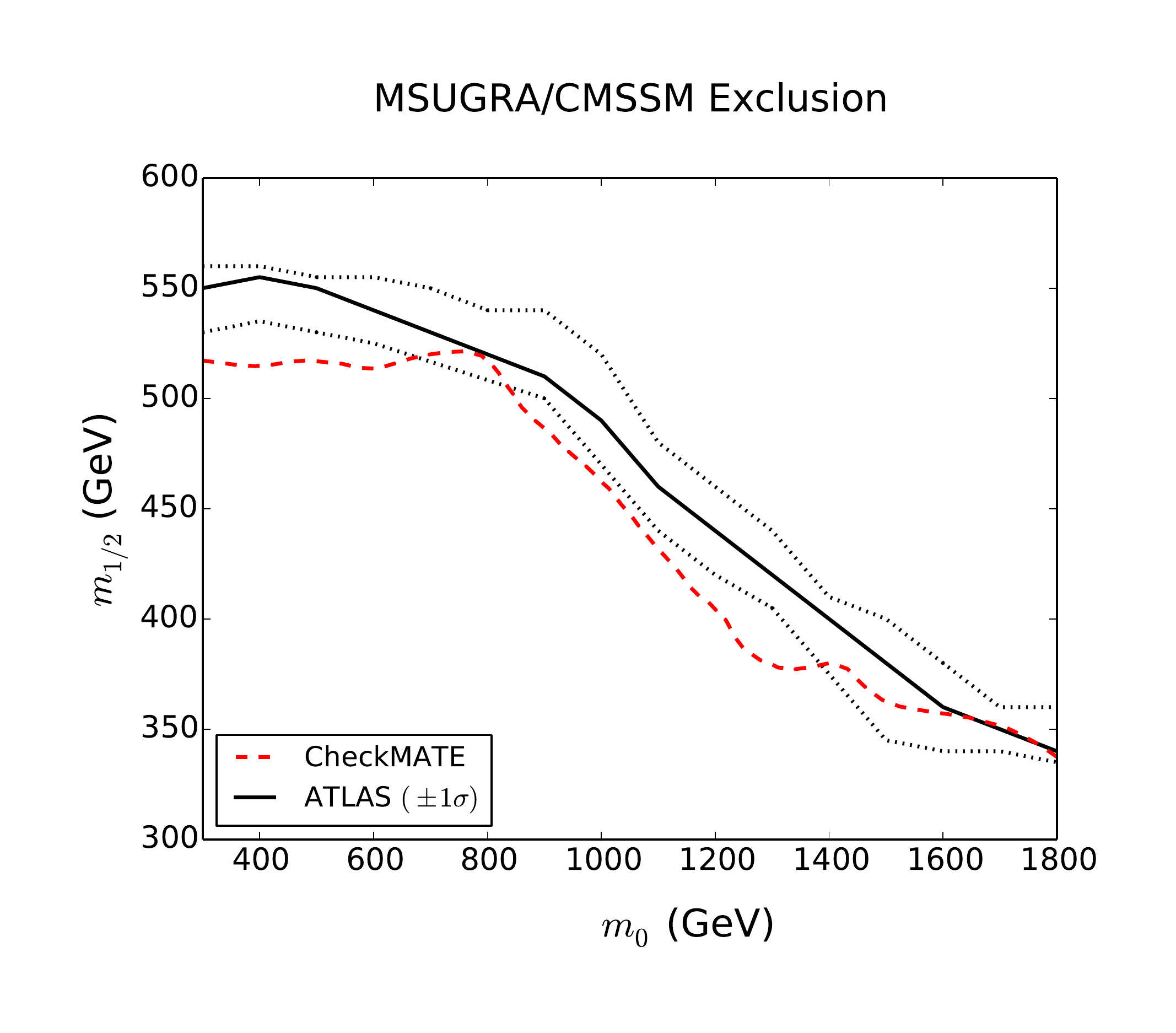}
\caption{Exclusion curve for general supersymmetric particle production in the CMSSM 
(mSUGRA) for atlas\_conf\_2012\_104.}
\label{fig:atlas_2013_104_1}
\end{figure}

 \clearpage
\subsection{atlas\_conf\_2012\_147}

\noindent Monojet search, \cite{ATLAS-CONF-2012-147} \\
Energy: 8 \TeV \\
Luminosity: 10.5 fb$^{-1}$ \\
Validation notes:
\begin{itemize}
  \item Validation was performed against standard model $W$ and $Z$ samples in signal 
  region distributions of leading jet $\pT$ and \etmiss.
  \item The validation finishes with the hardest jet $\pT<600$~\GeV due to finite Monte-Carlo statistics.
  \item No validation has yet been performed with parameter scans.
\end{itemize}

\begin{figure}[h]
\centering \vspace{-0.4cm}
\includegraphics[width=0.49\textwidth]{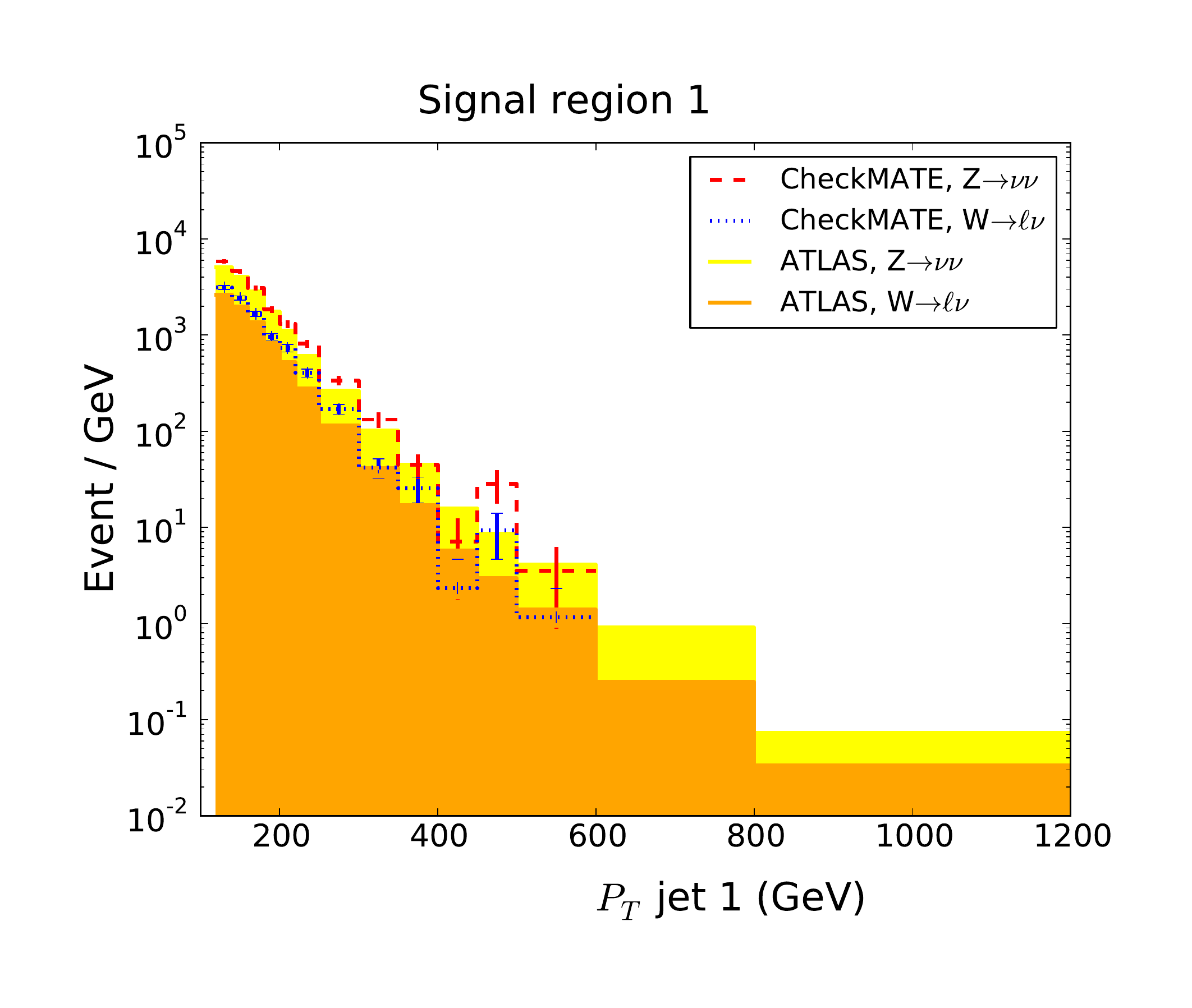}
\includegraphics[width=0.49\textwidth]{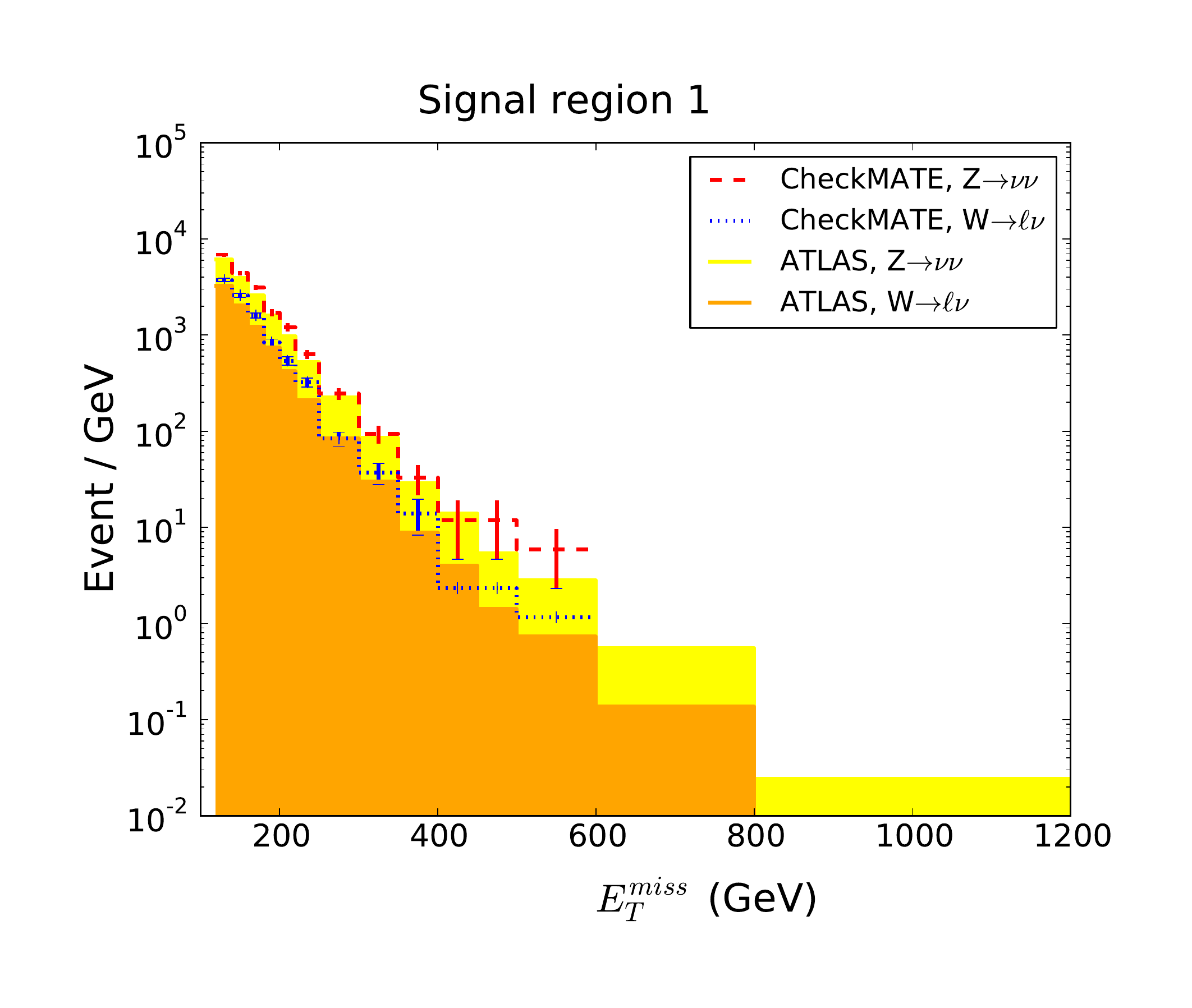}
\caption{Distributions for leading jet $\pT$ (left) and \etmiss (right) for standard model production of 
$W$ and $Z$ plus jets in signal region 1 of atlas\_conf\_2012\_147. The ATLAS $W/Z$ plus jets backgrounds are 
estimated using Monte-Carlo event samples normalised using data in control regions.}
\label{fig:atlas_conf_147_1}
\end{figure}

\begin{figure}[h]
\centering \vspace{-0.8cm}
\includegraphics[width=0.49\textwidth]{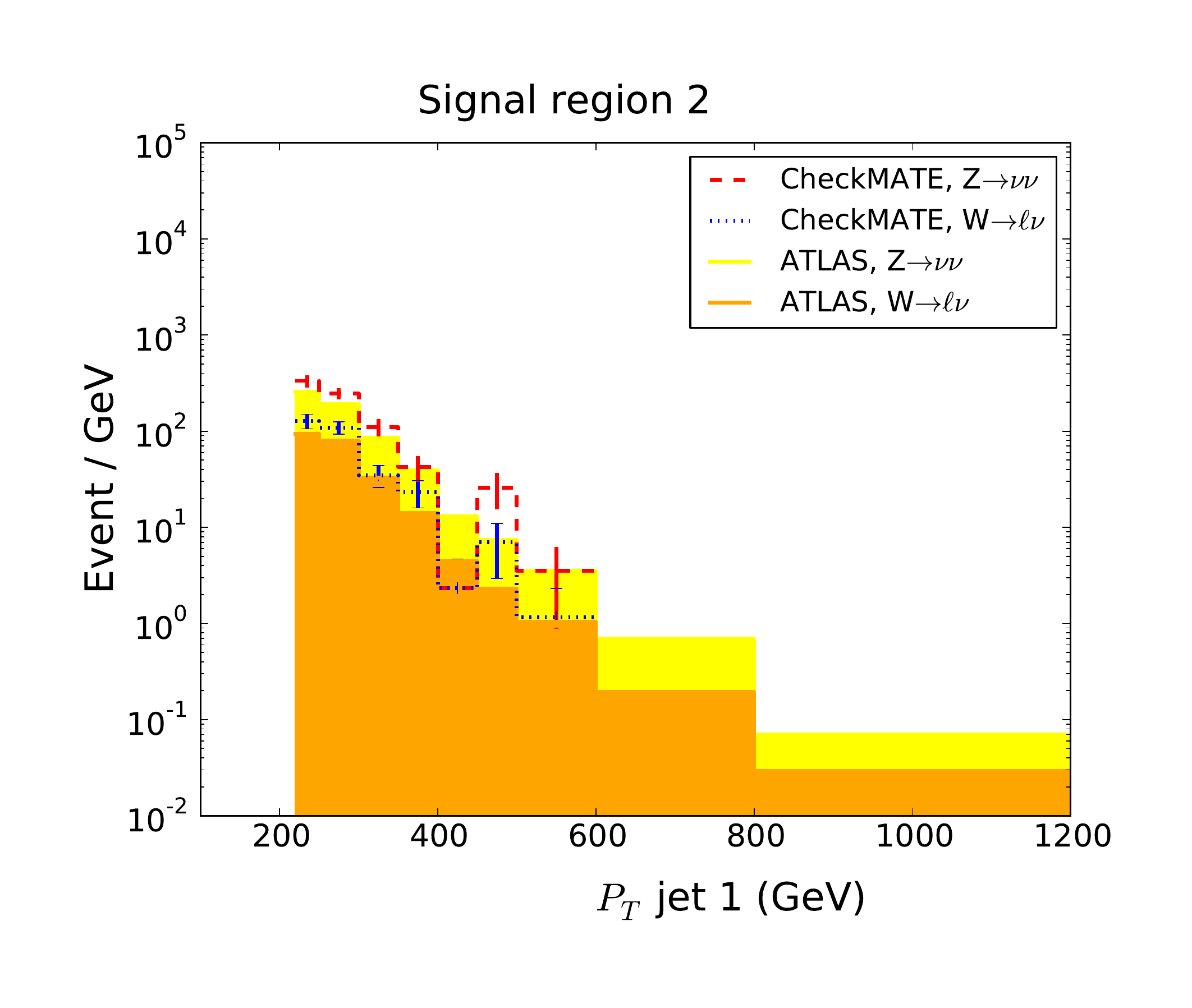}
\includegraphics[width=0.49\textwidth]{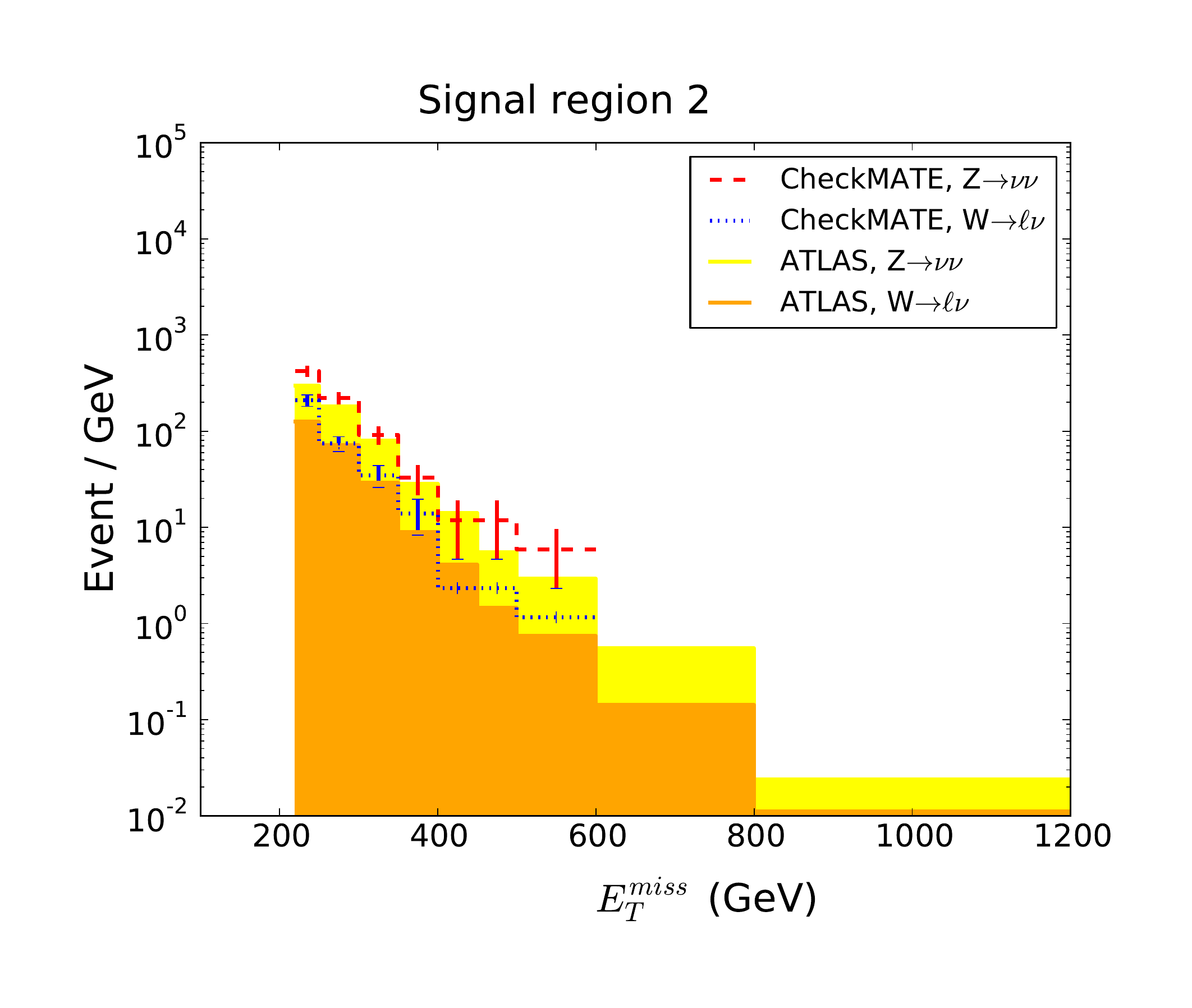}
\caption{Distributions for leading jet $\pT$ (left) and \etmiss (right) for standard model production of 
$W$ and $Z$ plus jets in signal region 2 of atlas\_conf\_2012\_147. The ATLAS $W/Z$ plus jets backgrounds 
are estimated using Monte-Carlo event samples normalised using data in control regions.}
\label{fig:atlas_conf_147_2}
\end{figure}

\newpage
\subsection{atlas\_conf\_2013\_024}

\noindent All-hadronic stop search, $t\overline{t} + E_{T}^{miss}$, \cite{ATLAS-CONF-2013-024} \\
Energy: 8 \TeV \\
Luminosity: 20.5 fb$^{-1}$ \\
Validation notes:
\begin{itemize}
  \item Validation has been performed versus all published cutflows and a simplified model 
  consisting of pure stop production followed by the decay $\tilde{t}\to t \tilde{\chi}^0_1$.
\end{itemize}

\begin{table*}[h] 
  \begin{tabularx}{1.0\textwidth}{l|>{\centering\arraybackslash}X>{\centering\arraybackslash}X|>{\centering\arraybackslash}X>{\centering\arraybackslash}X} \hline \hline
 Process						&\multicolumn{4}{c}{$\tilde{t}\tilde{t}^*$ direct}	 	   	 \\
 Point 							&\multicolumn{4}{c}{$m(\tilde{t})=600$~\GeV}	        \\ 
							&\multicolumn{4}{c}{$m(\tilde{\chi}^0_1)=0$~\GeV}	\\ 
Top Polarization						&    \multicolumn{2}{c|}{Right-handed}			&      \multicolumn{2}{c} {Left-handed}				 \\ 
  Source						&     ATLAS		&  \Checkmate{}			&     ATLAS		&  \Checkmate{}				 \\ \hline
  No selection						&	507.3		&	507.3			&	507.3		& 507.3			 \\ 
  Trigger						&	468.0		&	469.7			&    	467.8		& 468.8							  \\ 
  Primary vertex *					&	467.8		&	-			&       467.4 		& -	\\
 Event Cleaning *					&	459.0		&  	- 			&  	459.6		& -				\\ 
  Muon veto						&	381.2		&	380.1			&  	382.5		& 380.4							\\ 
  Electron veto 					&	284.4		&	297.6			& 	292.3		& 302.2					\\ 
  \etmiss$>$ 130~\GeV 				&	263.1		&	275.4			& 	270.1		& 277.7	  					\\  
   Jet multiplicity and $\pT$				&	97.7		&	95.0			&  	92.2		& 87.3	  					\\  
    $\slashed{E}_T^{\text{track}}>$ 30~\GeV				&	96.3		&	93.5			&  	90.5		& 85.9	  		\\
    $\Delta\phi($\etmiss$,\slashed{E}_T^{\text{track}})< \pi/3$	&	90.3		&	89.4			&  	84.3		& 82.2	  		\\
 $\Delta\phi(\text{jet},\slashed{E}_T^{\text{track}})< \pi/3$ 		&	77.1		&	76.0			&  	72.0		& 69.2	  				\\	  
  Tau veto 						&	67.4		&	66.6			&  	61.9		& 59.3	  				\\  
  $2 \geq b$-tagged jets 				&	29.5		&	28.6			&       31.5		& 27.9	 				\\  
  $m_T(b-jet,$\etmiss$) >$ 175~\GeV 			&	20.2		&	20.2			& 	23.6		& 21.3	  				 \\ 
 80~\GeV $< m^0_{jjj} <$ 175~\GeV 			&	17.8		&	18.4			& 	20.4		& 19.3	  		\\
 80~\GeV $< m^1_{jjj} <$ 175~\GeV  			&	10.9		&	10.8			& 	11.9		& 10.9	  		\\
 \etmiss$>$ 150~\GeV 					&	10.8		&	10.7			& 	11.8		& 11.8	  		\\
 \etmiss$>$ 200~\GeV 	(SR1)				&	10.3 $\pm 0.1$	&	10.2	$\pm 0.2$	& 	11.2 $\pm 0.1$	& 10.5 $\pm 0.2$	  		\\
 \etmiss$>$ 250~\GeV 					&	9.2 		&	9.3 			& 	10.0 		& 9.3  	  		\\
 \etmiss$>$ 300~\GeV 	(SR2)				&	7.8 $\pm 0.1$	&	8.1 $\pm 0.1$		& 	8.3 $\pm 0.1$	& 8.1	  $\pm 0.1$		\\
 \etmiss$>$ 350~\GeV 	(SR3)				&	6.1 $\pm 0.1$	&	6.2 $\pm 0.1$		& 	6.6 $\pm 0.1$	& 6.5	 $\pm 0.1$ 		\\  
\hline \hline
  \end{tabularx}
\caption{In the left (right) column, the tops from the stop decay are right-handed (left-handed) in 95$\%$ (100$\%$) of the decays. Shown are the number of events after each selection cut, normalised to 20.5 fb$^{-1}$. Final error is from Monte Carlo statistics for both ATLAS and \Checkmate{}. *\textit{No vertex finding or event cleaning is performed by 
\Checkmate{}. Instead, a flat efficiency factor is included to account 
for these effects.} \label{tab:atlas_2013_024_CF1} }
\end{table*}
\begin{figure}[h]
\centering \vspace{-0.4cm}
\includegraphics[width=0.49\textwidth]{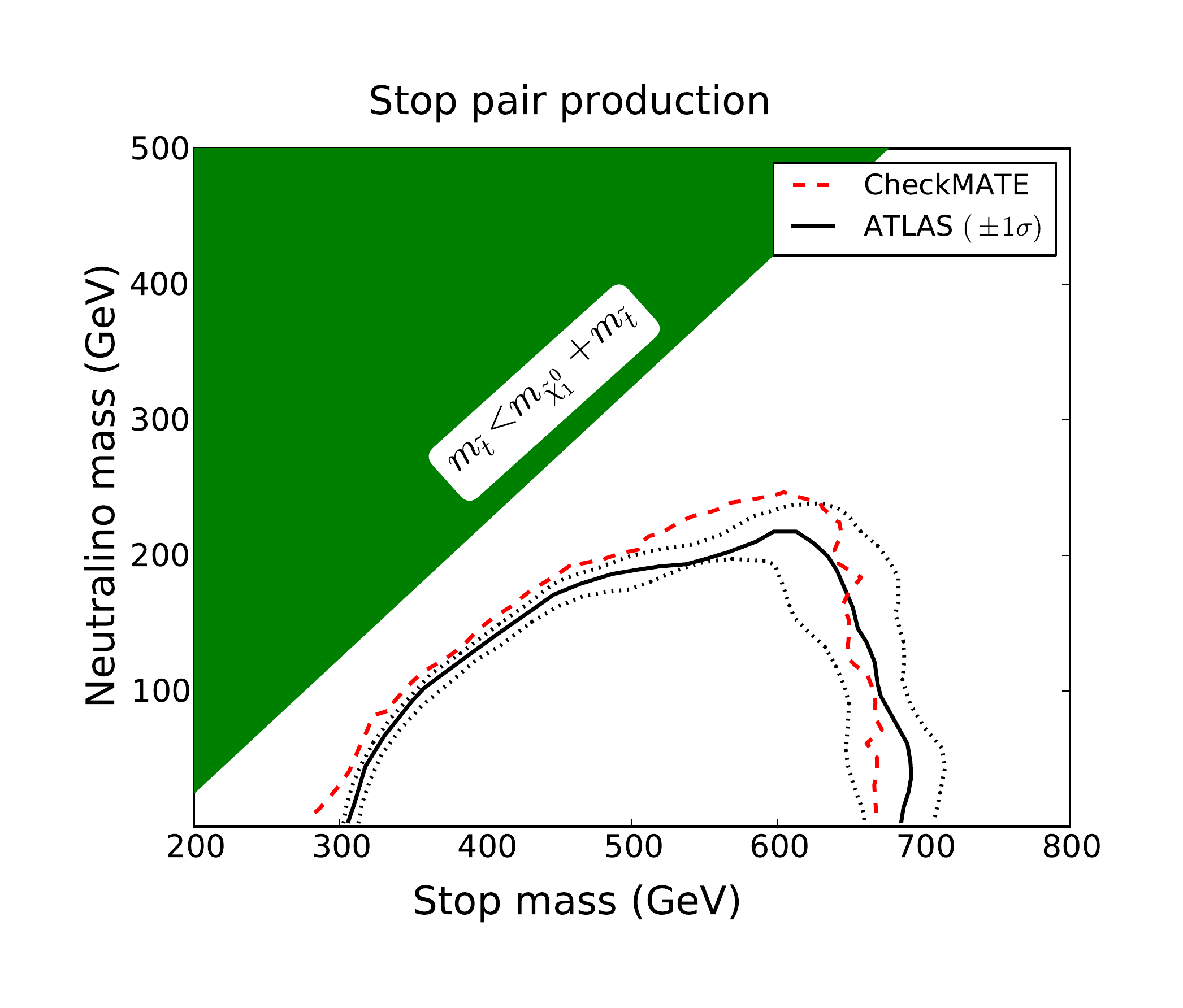}
\caption{Exclusion curve for the stop pair production with simplified 
decay, $\tilde{t}\to t \tilde{\chi}^0_1$ for analysis atlas\_conf\_2013\_024. The top produced in the decay has right--handed polarization in 95$\%$ of the decays.}
\label{fig:atlas_2013_024_1}
\end{figure}

\clearpage
\subsection{atlas\_conf\_2013\_035}

\noindent Trilepton, 0 jets + $E_{T}^{miss}$, \cite{ATLAS-CONF-2013-035} \\
Energy: 8 \TeV \\
Luminosity: 20.7 fb$^{-1}$ \\
Validation notes:
\begin{itemize}
  \item Validation has been performed versus all published cutflows.
\end{itemize}

\begin{table*}[h] 
\begin{tabularx}{1.0\textwidth}{ l|>{\centering\arraybackslash}X>{\centering\arraybackslash}X|>{\centering\arraybackslash}X>{\centering\arraybackslash}X|>{\centering\arraybackslash}X>{\centering\arraybackslash}X} \hline \hline
Process 		&\multicolumn{6}{c}{$\tilde{\chi}^0_2+\tilde{\chi}^\pm_1$ production; decay via $W$/$Z$ or slepton} \\ 

Point 		&\multicolumn{2}{c|}{Simplified $\tilde\ell_L$ }	&\multicolumn{2}{c|}{Simplified WZ}  	&\multicolumn{2}{c}{Simplified $\tilde\ell_L$}\\
&\multicolumn{2}{c|}{$m_{\tilde\chi_1^0}=157.5$ \GeV} & \multicolumn{2}{c|}{$m_{\tilde\chi_1^0}=75$ \GeV} & \multicolumn{2}{c}{$m_{\tilde\chi_1^0}=0$ \GeV} \\
&\multicolumn{2}{c|}{$m_{\tilde\chi_2^0}=m_{\tilde\chi_1^\pm}=192.5$ \GeV} & \multicolumn{2}{c|}{$m_{\tilde\chi_2^0}=m_{\tilde\chi_1^\pm}=150$ \GeV} & \multicolumn{2}{c}{$m_{\tilde\chi_2^0}=m_{\tilde\chi_1^\pm}=500$ \GeV} \\
&\multicolumn{2}{c|}{$m_{\tilde\nu}=m_{\tilde l_L}=(m_{\tilde \chi_1^0}+m_{\tilde\chi_1^\pm})/2$} & \multicolumn{2}{c|}{} & \multicolumn{2}{c}{$m_{\tilde\nu}=m_{\tilde l_L}=(m_{\tilde \chi_1^0}+m_{\tilde\chi_1^\pm})/2$} \\

	 		& \multicolumn{2}{c|}{SRnoZa }	&\multicolumn{2}{c|}{SRnoZb}  	&\multicolumn{2}{c}{SRnoZc}	\\
\hline
Source			&     ATLAS	&  \Checkmate{}	&     ATLAS	&  \Checkmate{}		&     ATLAS	&  \Checkmate{}	\\ 
Generated events	&	25000	&	40000	&	20000	&	40000		&	40000	&	40000	 \\ \hline
Lepton multiplicity 	&	537.1	&	688.1	& 	227.3	& 	278.8		&	28.5		&	28.1		\\ 
SFOS requirement 	&	536.3	&	615		& 	226.5	& 	272.5		&	28.1		&	27.9		\\  
b veto			&	491.0	&	557.4	&  	211.0	&  	251			&	24.9		&	24.8			\\  
Z veto/request 		&	476.3	&	537		&  	196.6	&   	235			&	24.1		&	24	  \\ 
$E_T^{\rm miss}$ 	&	161.2	&	181.3	&  	53.8		&   	56.2			&	22.1		&	22	  \\
$m_{\rm SFOS}$ 	&	141.2	&	154.2	&  	27.1		&   	26.8			&	-		&	-	  \\
$m_T$ 			&	-		&	-		&  	-		&   	-			&	19.2		&   19 	\\	
$\pT^{\rm third\, lepton}$ &  -		&	-		&  	-		&   	-			&	$18.4\pm0.3$		&   $18\pm0.3$ 	\\
SRnoZc veto 		&	$141.2\pm9.9$	&$154.2\pm8.1$&  	$26.3\pm1.4$		&   	$26.8\pm1.0$			&	-		&	-	  \\ \hline \hline
\end{tabularx}
\caption{Shown are the number of events after each selection cut, normalised to 20.7 fb$^{-1}$. Final error is from Monte Carlo statistics for both ATLAS and \Checkmate{}.  \label{tab:atlas_2013_035_SRnoZ} } 
\end{table*}

 \begin{table*}[h] 
\begin{tabularx}{1.0\textwidth}{l|>{\centering\arraybackslash}X>{\centering\arraybackslash}X|>{\centering\arraybackslash}X>{\centering\arraybackslash}X|>{\centering\arraybackslash}X>{\centering\arraybackslash}X} \hline \hline
Process			&\multicolumn{6}{c}{$\tilde{\chi}^0_2+\tilde{\chi}^\pm_1$ production; decay via $W$/$Z$ or slepton}	 	   	 \\
Point 			&\multicolumn{2}{c|}{Simplified WZ }	&\multicolumn{2}{c|}{Simplified WZ}  	&\multicolumn{2}{c}{Simplified WZ}		\\
&\multicolumn{2}{c|}{$m_{\tilde\chi_1^0}=0$ \GeV} & \multicolumn{2}{c|}{$m_{\tilde\chi_1^0}=0$ \GeV} & \multicolumn{2}{c}{$m_{\tilde\chi_1^0}=0$ \GeV} \\
&\multicolumn{2}{c|}{$m_{\tilde\chi_2^0}=m_{\tilde\chi_1^\pm}=100$ \GeV} & \multicolumn{2}{c|}{$m_{\tilde\chi_2^0}=m_{\tilde\chi_1^\pm}=150$ \GeV} & \multicolumn{2}{c}{$m_{\tilde\chi_2^0}=m_{\tilde\chi_1^\pm}=250$ \GeV} \\
&\multicolumn{2}{c|}{} & \multicolumn{2}{c|}{} & \multicolumn{2}{c}{$m_{\tilde\nu}=m_{\tilde l_L}=(m_{\tilde \chi_1^0}+m_{\tilde\chi_1^\pm})/2$} \\
				& \multicolumn{2}{c|}{SRZa }	&\multicolumn{2}{c|}{SRZb}  	&\multicolumn{2}{c}{SRZc}	\\\hline

Source				&     ATLAS	&  \Checkmate{}	&     ATLAS	&  \Checkmate{}		&     ATLAS	&  \Checkmate{}	\\ 
Generated events		&	15000	&	40000	&	20000	&	40000		&	20000	&	40000	 \\\hline 
Lepton multiplicity 		&	1071.4	&	1118.9      & 	259.8 	& 	276.9		& 	40.0		&	44.2		\\ 
SFOS requirement 		&	1067.5	&	1109.5	& 	258.0	&	273.9		& 	39.7		&	43.7		\\  
b veto				&	989.4	&	1039		&  	240.0	&	254.6		&	36.4		&	39.8			\\  
Z veto/request 			&	912.7	&	866.5	&  	227.2	&	227.1 		&	34.4		&	34.9	  \\ 
$E_T^{\rm miss}$ 		&	170.7	&	135.9	&  	67.7		&	66.5			&	17.7		&	17.3	  \\
$m_{\rm SFOS}$ 		&	-		&	-		&  	-		&	-			& 	-		&	-	  \\
$m_T$ 				&	$159.3\pm8.6$	&$133\pm5.2$	&  	$27.8\pm1.4$		&	$26\pm1.0$		&	$12.0\pm0.34$		&	$11.2\pm0.23$	  \\	
$\pT^{\rm third\, lepton}$ &	-		&	-		&  	-	  	&	- 			&	-		&	-	  \\
SRnoZc veto 			&	-		&	-		&  	-		&	-			&	-		&	-	  \\\hline \hline
\end{tabularx}
\caption{Shown are the number of events after each selection cut, normalised to 20.7 fb$^{-1}$.  Final error is from Monte Carlo statistics for both ATLAS and \Checkmate{}.  \label{tab:atlas_2013_035_SRZ} } 
\end{table*}

\clearpage
\subsection{atlas\_conf\_2013\_047}
\noindent 0 lepton + $\geq$ 2-6 jets + $E_{T}^{miss}$, \cite{ATLAS-CONF-2013-047} \\
Energy: 8 \TeV \\
Luminosity: 20.3 fb$^{-1}$ \\
Validation notes:
\begin{itemize}
\item Validation has been performed versus all published cutflows.
\item Additional validation has been performed with the parameter scan of the CMSSM (mSUGRA) model shown in \Cref{fig:atlas_mSugra}.
\end{itemize}
\begin{table}[h] 
\begin{tabularx}{1.0\textwidth}{l|>{\centering\arraybackslash}X>{\centering\arraybackslash}X|>{\centering\arraybackslash}X>{\centering\arraybackslash}X|>{\centering\arraybackslash}X>{\centering\arraybackslash}X} \hline \hline
 Process						&\multicolumn{6}{c}{$\tilde{q}\tilde{q}$ direct}	 	   	 \\
 Point 							&\multicolumn{2}{c|}{$m(\tilde{q})=450$~\GeV}	     	& \multicolumn{2}{c|}{$m(\tilde{q})=850$~\GeV}    	& \multicolumn{2}{c}{$m(\tilde{q})=662$~\GeV}   \\ 
							&\multicolumn{2}{c|}{$m(\tilde{\chi}^0_1)=400$~\GeV}	&\multicolumn{2}{c|}{$m(\tilde{\chi}^0_1)=100$~\GeV}	& \multicolumn{2}{c}{$m(\tilde{\chi}^0_1)=287$~\GeV}\\ 
 Signal Region						&\multicolumn{2}{c|}{A-medium} 			&	\multicolumn{2}{c|}{A-medium}			& \multicolumn{2}{c}{C-medium} \\  \hline
  Source						&     ATLAS		&  \Checkmate{}			&     ATLAS		&  \Checkmate{}			&     ATLAS		&  \Checkmate{}	 \\
  Generated events					&	20000		&	50000			&	5000		&	50000			&	5000		&	50000	 \\ \hline
  Jet Cleaning *					&	99.7		&  	- 			&  	99.6		& -				&	99.6		&	-		\\ 
  0-lepton	*					&	89.9		&	-			&  	98.5		& -				&	98.2		&	-				\\ 
  \etmiss$ >$ 160~\GeV *				&	15		&	-			& 	89.9		& -				&	80.7		&	-			\\ 
  $\pT(j_1) >$ 130~\GeV 					&	12.9		&	12.9			& 	89.7		& 89.5	  			&	80.0		&	79.3			\\  
   $\pT(j_2) >$ 130~\GeV					&	9.0		&	8.4			&  	87.4		& 87.1	  			&	75.6		&	75.3			\\  
    $\pT(j_3) >$ 0-60~\GeV				&	9.0		&	8.4			&  	87.4		& 87.1	  			&	35.3		&	35.6	\\
     $\pT(j_4) >$ 0-60~\GeV				&	9.0		&	8.4			&  	87.4		& 87.1	  			&	11.5		&	11.3	\\
 $\Delta\phi(j_i>40,$\etmiss$)>0.4$ 			&	7.0		&	6.8			&  	79.2		& 79.0	  			&	10.1		&	9.9			\\	  
  $\Delta\phi(j_i>40~\mathrm{GeV},$\etmiss$)>0-0.2$ 	&	7.0		&	6.8			&  	79.2		& 79.0	  			&	9.3		&	9.2			\\  
  \etmiss$/\sqrt{H_T}>0-15$ 			&	2.6		&	1.8			&  	49.9		& 48.0	 			&	9.3		&	9.2			\\  
  \etmiss$/m_{\text{eff}}(N_j)>0.15-0.4$ 			&	2.6		&	1.8			& 	49.9		& 48.0	  			&	7.2		&	6.8			 \\ 
  $m_{\text{eff}}(\mathrm{incl.})>1-2.2$~\TeV 			&	$0.1\pm0.02$	&	$0.08\pm0.01$		& 	$16.5\pm0.6$	& $18.3\pm0.2$			&	$3.0\pm0.2$	&	$3.1\pm0.1$				   \\  \hline
  \end{tabularx}
\caption{The cutflow is given as an absolute efficiency in $\%$ for each step of event selection. Final error is from Monte Carlo statistics for both ATLAS and \Checkmate{}. *\textit{Variable trigger efficiencies mean that the results are only comparable after 
both an \etmiss and jet $\pT$ cut have been applied.} \label{tab:atlas_2013_047_CF1} }
\end{table}
\begin{table}[h] 
\begin{tabularx}{1.0\textwidth}{l|>{\centering\arraybackslash}X>{\centering\arraybackslash}X|>{\centering\arraybackslash}X>{\centering\arraybackslash}X|>{\centering\arraybackslash}X>{\centering\arraybackslash}X} \hline \hline
 Process						&\multicolumn{4}{c|}{$\tilde{q}\tilde{g}$ direct}	 	   						&\multicolumn{2}{c}{$\tilde{g}\tilde{g}$ direct} \\
 Point 							&\multicolumn{2}{c|}{$m(\tilde{g})=1425$~\GeV}	     	& \multicolumn{2}{c|}{$m(\tilde{g})=1612$~\GeV}    	& \multicolumn{2}{c}{$m(\tilde{g})=1162$~\GeV}   \\ 
							&\multicolumn{2}{c|}{$m(\tilde{\chi}^0_1)=525$~\GeV}	&\multicolumn{2}{c|}{$m(\tilde{\chi}^0_1)=37$~\GeV}	& \multicolumn{2}{c}{$m(\tilde{\chi}^0_1)=337$~\GeV}\\
 Signal Region						&\multicolumn{2}{c|}{B-medium} 			&	\multicolumn{2}{c|}{B-tight}			& \multicolumn{2}{c}{D} \\ \hline
  Source						&     ATLAS		&  \Checkmate{}			&     ATLAS		&  \Checkmate{}			&     ATLAS		&  \Checkmate{}	 \\
  Generated events					&	5000		&	50000			&	5000		&	50000			&	5000		&	50000	 \\ \hline
  Jet Cleaning *					&	99.7		&  	- 			&  	99.6		& -				&	99.8		&	-		\\ 
  0-lepton	*					&	98.0		&	-			&  	98.8		& -				&	98.5		&	-				\\ 
  \etmiss$ >$ 160~\GeV *				&	93.3		&	-			& 	95.9		& -				&	88.9		&	-			\\ 
  $\pT(j_1) >$ 130~\GeV 					&	93.3		&	93.9			& 	95.8		& 96.0	  			&	88.8		&	88.1			\\  
   $\pT(j_2) >$ 130~\GeV					&	92.4		&	92.7			&  	95.2		& 95.1	  			&	88.8		&	88.1			\\  
    $\pT(j_3) >$ 0-60~\GeV				&	68.5		&	67.0			&  	75.7		& 73.5	  			&	87.1		&	86.8	\\
     $\pT(j_4) >$ 0-60~\GeV				&	68.5		&	67.0			&       75.7		& 73.5	  			&	74.1		&	74.4	\\
      $\pT(j_5) >$ 0-60~\GeV				&	68.5		&	67.0			&       75.7		& 73.5	  			&	40.9		&	36.0	\\
 $\Delta\phi(j_i>40,$\etmiss$)>0.4$ 			&	60.4		&	58.7			&  	66.2		& 64.2	  			&	34.2		&	30.1			\\	  
  $\Delta\phi(j_i>40~\mathrm{GeV},$\etmiss$)>0-0.2$ 	&	60.4		&	58.7			&  	66.2		& 64.2	  			&	28.6		&	25.9			\\   
  \etmiss$/m_{\text{eff}}(N_j)>0.15-0.4$ 			&	44.8		&	41.8			& 	31.8		& 28.1	  			&	22.1		&	18.9			 \\ 
  $m_{\text{eff}}(\mathrm{incl.})>1-2.2$~\TeV 			&	$27.5\pm 0.7$	&	$25.8\pm0.2$		& 	$22.8\pm 0.7$	& $20.5\pm0.2$			&	$13.4\pm 0.5$	&	$13.0\pm0.2$				   \\  \hline \hline
  \end{tabularx}
\caption{The cutflow is given as an absolute efficiency in $\%$ for each step of event selection. Final error is from Monte Carlo statistics for both ATLAS and \Checkmate{}.  *\textit{Variable trigger efficiencies mean that the results are only comparable after 
both an \etmiss and jet $\pT$ cut have been applied.} \label{tab:atlas_2013_047_CF2} }
\end{table}
\begin{table}[h] 
\begin{tabularx}{1.0\textwidth}{l|>{\centering\arraybackslash}X>{\centering\arraybackslash}X|>{\centering\arraybackslash}X>{\centering\arraybackslash}X} \hline \hline
 Process						&\multicolumn{4}{c}{$\tilde{g}\tilde{g}$ one-step ($\tilde{g}$ decay via $\tilde{\chi}^{\pm}$)}	 	   						 \\
 Point 							&\multicolumn{2}{c|}{$m(\tilde{g})=1065$~\GeV}	     	& \multicolumn{2}{c}{$m(\tilde{g})=1265$~\GeV}    	  \\ 
							&\multicolumn{2}{c|}{$m(\tilde{\chi}^{\pm}_1)=785$~\GeV}	&\multicolumn{2}{c}{$m(\tilde{\chi}^\pm_1)=865$~\GeV}	\\
							&\multicolumn{2}{c|}{$m(\tilde{\chi}^0_1)=525$~\GeV}	&\multicolumn{2}{c}{$m(\tilde{\chi}^0_1)=465$~\GeV}	\\
 Signal Region						&\multicolumn{2}{c|}{D} 			&	\multicolumn{2}{c}{E-tight}			\\ 
\hline
  Source						&     ATLAS		&  \Checkmate{}			&     ATLAS		&  \Checkmate{}				 \\ 
  Generated events					&	20000		&	50000			&	20000		&	50000				 \\ \hline
  Jet Cleaning *					&	99.8		&  	- 			&  	99.8		& -						\\ 
  0-lepton	*					&	63.7		&	-			&  	63.5		& -							\\ 
  \etmiss$ >$ 160~\GeV *				&	50.0		&	-			& 	55.6		& -							\\ 
  $\pT(j_1) >$ 130~\GeV 					&	49.3		&	47.7			& 	55.6		& 54.4	  				\\  
   $\pT(j_2) >$ 130~\GeV					&	49.2		&	47.6			&  	55.6		& 54.4	  					\\  
    $\pT(j_3) >$ 0-60~\GeV				&	48.6		&	47.1			&  	55.4		& 54.2	  			\\
     $\pT(j_4) >$ 0-60~\GeV				&	44.5		&	43.8			&       53.4		& 52.8	  				\\
      $\pT(j_5) >$ 0-60~\GeV				&	34.4		&	34.8			&       46.3		& 46.6	  				\\
        $\pT(j_5) >$ 0-60~\GeV				&	34.4		&	34.8			&       31.7		& 33.0	  	    \\
 $\Delta\phi(j_i>40,$\etmiss$)>0.4$ 			&	29.2		&	29.5			&  	26.5		& 27.5	  						\\	  
  $\Delta\phi(j_i>40~\mathrm{GeV},$\etmiss$)>0-0.2$ 	&	24.6		&	24.7			&  	21.3		& 22.4	  					\\   
  \etmiss$/m_{\text{eff}}(N_j)>0.15-0.4$ 			&	21.6		&	21.2			& 	12.0		& 11.2	  					 \\ 
  $m_{\text{eff}}(\mathrm{incl.})>1-2.2$~\TeV 			&	$2.0\pm 0.1$	&	$1.9\pm0.06$		& 	$7.9\pm 0.2$	& $8.2\pm0.1$						   \\  \hline \hline
  \end{tabularx}
\caption{The cutflow is given as an absolute efficiency in $\%$ for each step of event selection. Final error is from Monte Carlo statistics for both ATLAS and \Checkmate{}. *\textit{Variable trigger efficiencies mean that the results are only comparable after 
both an \etmiss and jet $\pT$ cut have been applied.} \label{tab:atlas_2013_047_CF3} }
\end{table}

 \begin{figure}[h]
 \centering \vspace{-0.4cm}
 \includegraphics[width=0.49\textwidth]{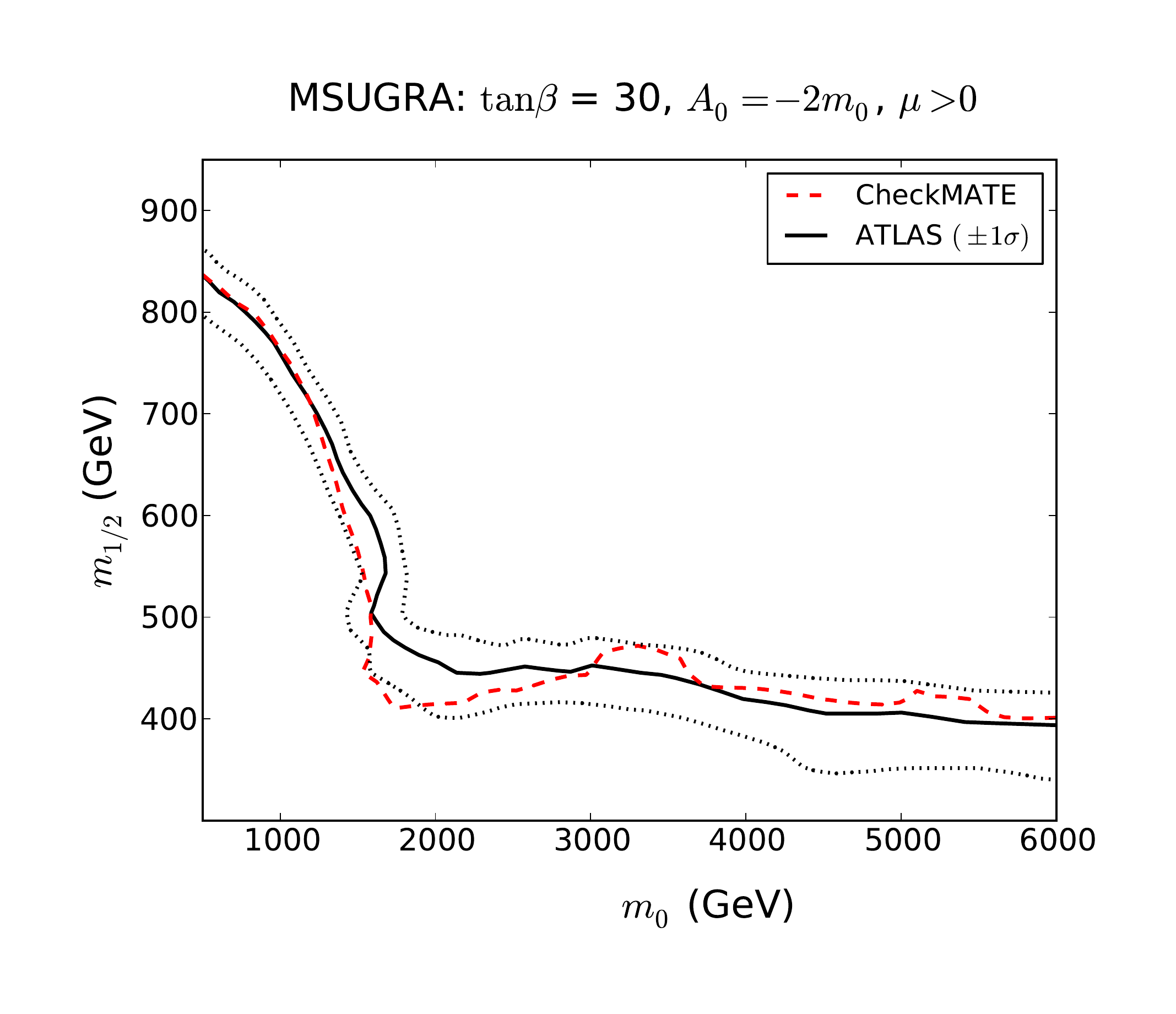}
 \includegraphics[width=0.49\textwidth]{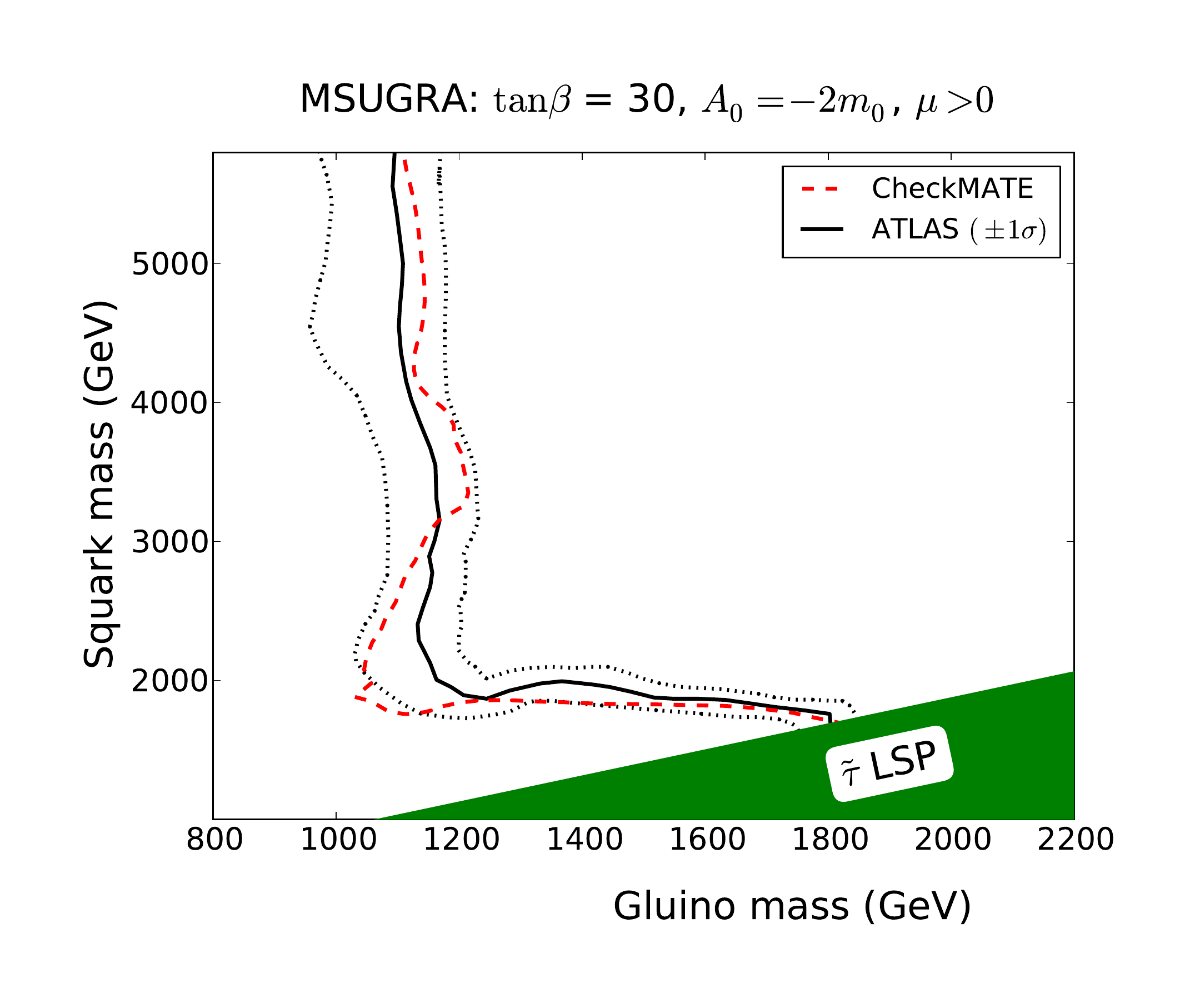}
 \caption{Exclusion curve for the CMSSM (mSUGRA) model for analysis atlas\_conf\_2013\_047. The 
 exclusion in the parameter space $M_0$, $M_{1/2}$ (left) and $m_{\tilde{q}}$, 
 $m_{\tilde{g}}$ are shown (right).}
 \label{fig:atlas_2013_047_1}
 \end{figure}

 \begin{figure}[h]
 \centering \vspace{-0.0cm}
 \includegraphics[width=0.49\textwidth]{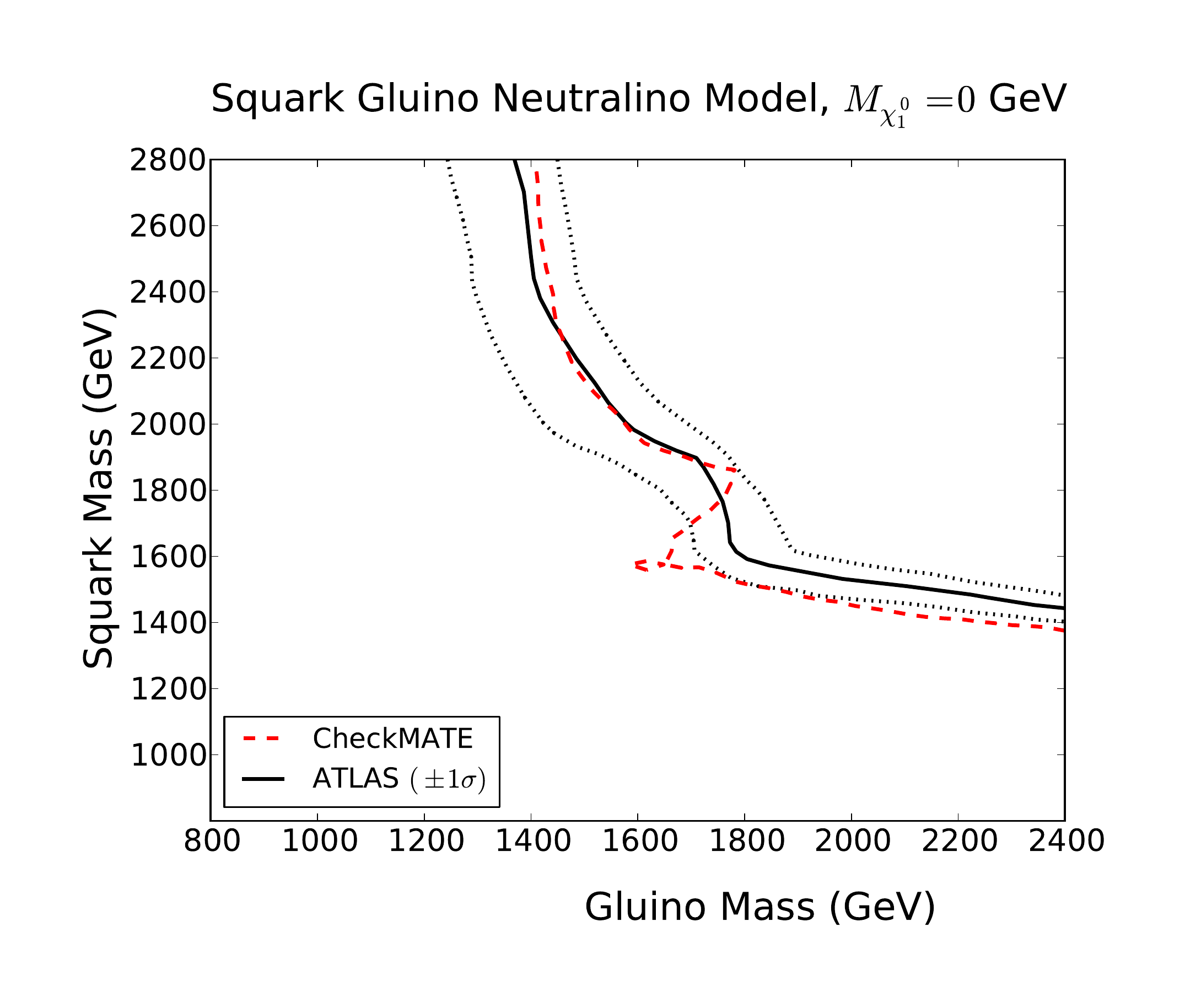}
 \includegraphics[width=0.49\textwidth]{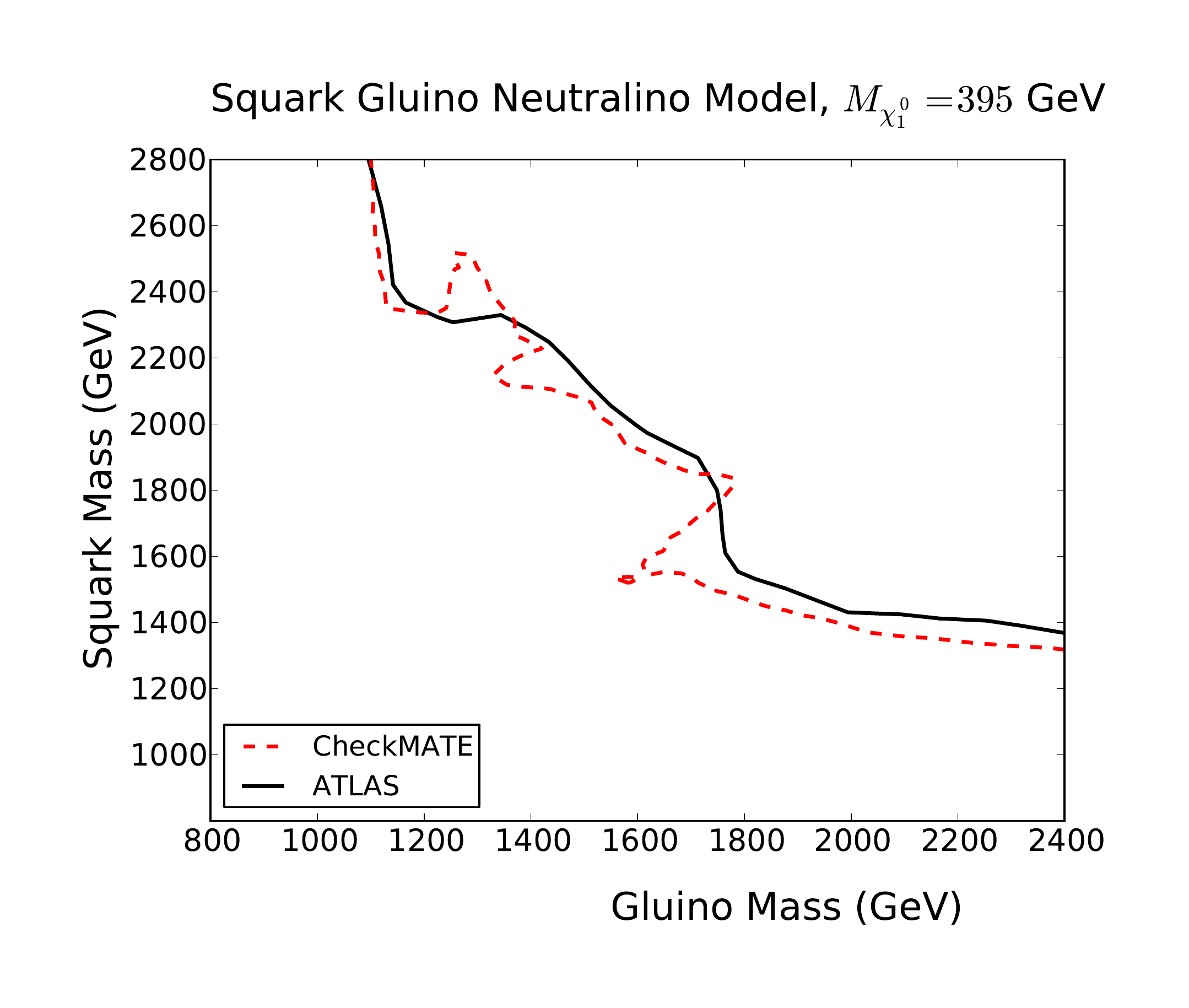}
 \caption{Exclusion curve for the simplified model with only strong production of 
 gluinos and first- and second-generation squarks. The lightest supersymmetric particle 
 has mass, $(m_{\tilde{\chi}^0_1})=0$~\GeV (left) or $(m_{\tilde{\chi}^0_1})=395$~\GeV (right). Jumps in exclusion limit of \Checkmate{} are due to the change of signal region.}
 \label{fig:atlas_2013_047_2}
 \end{figure}

\begin{figure}[h]
 \centering \vspace{-1.4cm}
 \includegraphics[width=0.49\textwidth]{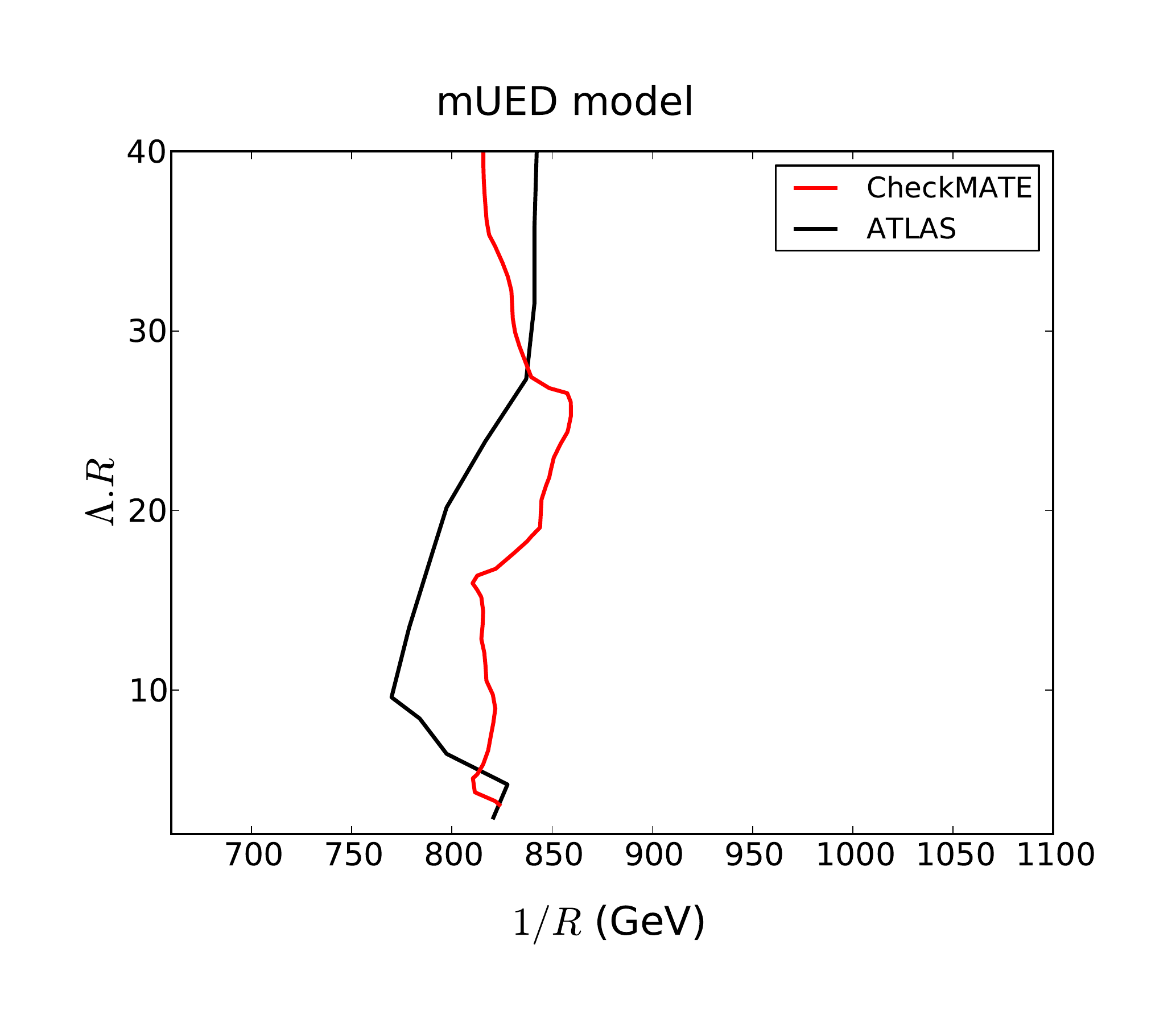} 
 \caption{Exclusion curve for the minimal Universal Extra Dimensions (mUED) model. Jumps in exclusion limit of \Checkmate{} are due to the change of signal region.}
 \label{fig:atlas_2013_047_3}
 \end{figure}

\clearpage
\subsection{atlas\_conf\_2013\_049}
\noindent 2 opposite-sign leptons, 0 jets + $E_{T}^{miss}$, \cite{ATLAS-CONF-2013-049} \\
Energy: 8 \TeV \\
Luminosity: 20.3 fb$^{-1}$ \\
Validation notes:
\begin{itemize}
  \item Validation has been performed versus all published cutflows.
  \item Jet veto conditions have been tightened by 25\% compared to the corresponding stated ATLAS values. This was done
  to match cutflow values and account for pile-up effects that \Checkmate{} cannot simulate.
  \item WARNING: For simplified models with chargino production and weak boson mediated 
  decay, \Checkmate{} systematically underestimates signal region cutflows. Despite much effort, no reason for this discrepancy has been found. The 
  corresponding setting is conservative since it will not lead to spurious model exclusion. 
  \item No validation has yet been performed with parameter scans.
\end{itemize}

\begin{table*}[h] 
\begin{tabularx}{1.0\textwidth}{l|>{\centering\arraybackslash}X>{\centering\arraybackslash}X|>{\centering\arraybackslash}X>{\centering\arraybackslash}X|>{\centering\arraybackslash}X>{\centering\arraybackslash}X|>{\centering\arraybackslash}X>{\centering\arraybackslash}X} \hline \hline
 Process					&\multicolumn{8}{c}{$\tilde{\ell}^+\tilde{\ell}^-$ production}	 	   	 \\ 
 Point 						&\multicolumn{4}{c|}{$m(\tilde{\ell_L})=m(\tilde{\ell_R})=191$~\GeV}				&\multicolumn{4}{c}{$m(\tilde{\ell_L})=m(\tilde{\ell_R})=250$~\GeV}    \\  
						&\multicolumn{4}{c|}{$m(\tilde{\chi}^0_1)=90$~\GeV}			&\multicolumn{4}{c}{$m(\tilde{\chi}^0_1)=10$~\GeV}	\\ 
						&\multicolumn{2}{c|}{$e^+e^-$}	&\multicolumn{2}{c|}{$\mu^+\mu^-$}  	&\multicolumn{2}{c|}{$e^+e^-$}	&\multicolumn{2}{c}{$\mu^+\mu^-$}	\\ \hline
  Source					&     ATLAS	&  \Checkmate{}	&     ATLAS	&  \Checkmate{}		&     ATLAS	&  \Checkmate{}	&     ATLAS	&  \Checkmate{}\\ 
  Generated events				&	5000	&	50000	&	5000	&	50000		&	5000	&	50000	&	5000	&	50000 \\ \hline
  Trigger 					&	150	&	153	& 	159	& 	153		&	55	&	55	&	50	&	50	\\ 
  $e^{\pm}\mu^{\pm}$ 				&	139	&	143	& 	148	& 	142		&	54	&	52	&	49	&	48		\\  
  Jet veto					&	58	&	62	&  	62	&  	61		&	20	&	22	&	20	&	21		\\  
  $(\pT^{\ell1},\pT^{\ell2})>(35,20)$~\GeV	&	45	&	47	&  	50	&   	47		&	17	&	19	&	17	&	18   \\ 
  SR-$m_{T2,90}$				&$21.6\pm1.1$	&$20.9\pm0.4$	&  $21.6\pm1.1$	&$20.6\pm0.3$		&$12.2\pm0.5$	&$12.3\pm0.2$	&$12.5\pm0.5$	&	$12.8\pm0.2$    \\ 
  SR-$m_{T2,110}$				& $12.3\pm0.9$	&$11.6\pm0.3$	&  $12.0\pm0.8$	& $11.8\pm0.3$		&$10.5\pm0.4$	&$10.0\pm0.1$	&$11.2\pm0.5$	&	$11.0\pm0.2$		\\	  \hline \hline
  \end{tabularx}
\caption{Shown are the number of events after each selection cut, normalised to 20.7 fb$^{-1}$. Final error is from Monte Carlo statistics for both ATLAS and \Checkmate{}. \label{tab:atlas_2013_049_CF1} } 
\end{table*}

\begin{table*}[h] 
\begin{tabularx}{1.0\textwidth}{l|>{\centering\arraybackslash}X>{\centering\arraybackslash}X|>{\centering\arraybackslash}X>{\centering\arraybackslash}X|>{\centering\arraybackslash}X>{\centering\arraybackslash}X} \hline \hline
 Process					&\multicolumn{6}{c}{$\tilde{\chi}^+\tilde{\chi}^-$ production, slepton decay}	 	   	 \\
 Point 						&\multicolumn{6}{c}{$m(\tilde{\chi}^{\pm})=350$~\GeV}				   \\
  						&\multicolumn{6}{c}{$m(\tilde{\ell_L})=m(\tilde{\nu})=175$~\GeV}				   \\  
						&\multicolumn{6}{c}{$m(\tilde{\chi}^0_1)=0$~\GeV}				\\
						&\multicolumn{2}{c|}{$e^+e^-$}	&\multicolumn{2}{c|}{$\mu^+\mu^-$}  	&\multicolumn{2}{c}{$e^{\pm}\mu^{\mp}$}		\\ \hline
  Source					&     ATLAS	&  \Checkmate{}	&     ATLAS	&  \Checkmate{}		&     ATLAS	&  \Checkmate{}	\\
  Generated events				&	40000	&	50000	&	40000	&	50000		&	40000	&	50000	 \\ \hline
  Trigger 					&	52	&	50      & 	48	& 	49		&	79	&	74		\\ 
  $e^{\pm}\mu^{\pm}$ 				&	48	&	47	& 	45	& 	46		&	74	&	69		\\  
  Jet veto					&	20	&	20	&  	19	&  	20		&	30	&	29			\\  
  $(\pT^{\ell1},\pT^{\ell2})>(35,20)$~\GeV	&	17	&	17	&  	17	&   	17		&	25	&	25	  \\
  SR-$m_{T2,90}$				&$11.7\pm0.4$	&$11.4\pm0.4$	&  $10.5\pm0.4$	&$10.6\pm0.4$		&$16.6\pm0.5$	&$16.4\pm0.5$	   \\ 
  SR-$m_{T2,110}$				& $9.5\pm0.4$	&$9.4\pm0.3$	&  $8.7\pm0.4$	& $8.7\pm0.3$		&$14.0\pm0.4$	&$13.5\pm0.4$		\\	  \hline \hline
  \end{tabularx}
\caption{Shown are the number of events after each selection cut, normalised to 20.7 fb$^{-1}$. Final error is from Monte Carlo statistics for both ATLAS and \Checkmate{}. \label{tab:atlas_2013_049_CF2} } 
\end{table*}
 
 \begin{table*}[h] 
\begin{tabularx}{1.0\textwidth}{l|>{\centering\arraybackslash}X>{\centering\arraybackslash}X|>{\centering\arraybackslash}X>{\centering\arraybackslash}X|>{\centering\arraybackslash}X>{\centering\arraybackslash}X} \hline  \hline
 Process					&\multicolumn{6}{c}{$\tilde{\chi}^+\tilde{\chi}^-$ production, slepton decay}	 	   	 \\
 Point 						&\multicolumn{6}{c}{$m(\tilde{\chi}^{\pm})=425$~\GeV}				   \\
  						&\multicolumn{6}{c}{$m(\tilde{\ell_L})=m(\tilde{\nu})=212.5$~\GeV}				   \\  
						&\multicolumn{6}{c}{$m(\tilde{\chi}^0_1)=75$~\GeV}				\\
						&\multicolumn{2}{c|}{$e^+e^-$}	&\multicolumn{2}{c|}{$\mu^+\mu^-$}  	&\multicolumn{2}{c}{$e^{\pm}\mu^{\mp}$}		\\ \hline
  Source					&     ATLAS	&  \Checkmate{}	&     ATLAS	&  \Checkmate{}		&     ATLAS	&  \Checkmate{}	\\ 
  Generated events				&	40000	&	50000	&	40000	&	50000		&	40000	&	50000	 \\\hline 
  Trigger 					&	20	&	21      & 	20      & 	20		&	31	&	30		\\ 
  $e^{\pm}\mu^{\pm}$ 				&	19	&	20	& 	19	& 	19		&	29	&	28		\\  
  Jet veto					&	7	&	8	&  	7	&  	8		&	11	&	11			\\  
  $(\pT^{\ell1},\pT^{\ell2})>(35,20)$~\GeV	&	6	&	7	&  	6	&   	7		&	9	&	10	  \\
  SR-$m_{T2,90}$				&$4.3\pm0.2$	&$4.9\pm0.2$	&  $4.4\pm0.2$	&$4.6\pm0.1$		&$6.7\pm0.2$	&$7.0\pm0.2$	   \\ 
  SR-$m_{T2,110}$				& $3.7\pm0.1$	&$4.3\pm0.1$	&  $3.8\pm0.1$	& $3.9\pm0.1$		&$5.7\pm0.2$	&$5.9\pm0.2$		\\	  \hline \hline
  \end{tabularx}
\caption{Shown are the number of events after each selection cut, normalised to 20.7 fb$^{-1}$. Final error is from Monte Carlo statistics for both ATLAS and \Checkmate{}. \label{tab:atlas_2013_049_CF3} } 
\end{table*}

\begin{table*}[h] 
\begin{tabularx}{1.0\textwidth}{l|>{\centering\arraybackslash}X>{\centering\arraybackslash}X|>{\centering\arraybackslash}X>{\centering\arraybackslash}X|>{\centering\arraybackslash}X>{\centering\arraybackslash}X} \hline  \hline
 Process					&\multicolumn{6}{c}{$\tilde{\chi}^+\tilde{\chi}^-$ production, $WW$ decay}	 	   	 \\ 
 Point 						&\multicolumn{2}{c|}{$m(\tilde{\chi}_1^{\pm})=100$~\GeV}	&\multicolumn{2}{c|}{$m(\tilde{\chi}_1^{\pm})=140$~\GeV} & \multicolumn{2}{c}{$m(\tilde{\chi}_1^{\pm})=200$~\GeV}   \\ 
						&\multicolumn{2}{c|}{$m(\tilde{\chi}^0_1)=0$~\GeV}	&\multicolumn{2}{c|}{$m(\tilde{\chi}^0_1)=20$~\GeV}	& \multicolumn{2}{c}{$m(\tilde{\chi}^0_1)=0$~\GeV}\\ \hline
  Source					&     ATLAS	&  \Checkmate{}				&     ATLAS		&  \Checkmate{}			&     ATLAS		&  \Checkmate{}	 \\
  Generated events				&	20000	&	50000				&	20000		&	50000			&	20000		&	50000	 \\ \hline
  No Cuts					&	11003	&	-				&   	3393 		& 	-			&	749		&	-					  \\ 
  All Cleaning *				&	10691	&  	 10673				&  	3299		&    	3289			&	732		&	727		\\ 
  Two signal leptons 				&	3178	&	2610				&  	1060		& 	898			&	261		&	218				\\ 
  Trigger 					&	2559	&	1977				& 	872		& 	684			&	214		&	167			\\ 
  $e^{\pm}\mu^{\pm}$ 				&	861	&	803				& 	296		& 	288			&	71		&	69			\\  
  Jet veto					&	443	&	437				&  	139		&  	152			&	31		&	34			\\  
  $(\pT^{\ell1},\pT^{\ell2})>(35,20)$~\GeV	&	310	&	302				&  	103		&   	111			&	25		&	27	\\
  SR-WWa					&$31.5\pm4.1$	&	$21.6\pm2.4$			&  	-		&	  -			&	-		&	-	\\ 
  SR-WWb 					&	-	&	-				&  	$8.2\pm1.2$	&   	$4.5\pm0.6$		&	-		&	-			\\	  
  SR-WWc		 			&	-	&	-				&  	-		& 	 -  			&	$3.3\pm0.4$	&	$2.6\pm0.2$			\\  \hline \hline
  \end{tabularx}
\caption{Shown are the number of events after each selection cut, normalised to 20.7 fb$^{-1}$. Final error is from Monte Carlo statistics for both ATLAS and \Checkmate{}. *\textit{No cleaning cuts are performed by \Checkmate{}, instead a flat efficiency factor is applied to 
simulate this effect.} \label{tab:atlas_2013_049_CF4} } 
\end{table*}

 \clearpage
\subsection{atlas\_conf\_2013\_061}
\noindent Search for strong production with at least three $b$-jets + $E_{T}^{miss}$, \cite{ATLAS-CONF-2013-061} \\
Energy: 8 \TeV \\
Luminosity: 20.1 fb$^{-1}$ \\
Validation notes:
\begin{itemize}
  \item Validation has been performed versus all published cutflows.
  \item B-tagging efficiency was reduced by 3\% compared with the nominal value in the experimental paper to better agree with cutflow data. 
  \item No validation has yet been performed with parameter scans.
\end{itemize}

\begin{table*}[h] 
\begin{tabularx}{1.0\textwidth}{l|>{\centering\arraybackslash}X>{\centering\arraybackslash}X|>{\centering\arraybackslash}X>{\centering\arraybackslash}X|>{\centering\arraybackslash}X>{\centering\arraybackslash}X} \hline \hline
 Process						&\multicolumn{6}{c}{$pp\to\tilde{g}\tilde{g}, \tilde{g} \to b \overline{b} \tilde{\chi}^0_1$}	 	   	 \\ 
 Point 							&\multicolumn{6}{c}{$m(\tilde{g})=1300$~\GeV, $m(\tilde{\chi}^0_1)=100$~\GeV}	        \\ 
  Channel						&\multicolumn{6}{c}{0 Lepton, 4 Jets}	\\  \hline
  Source						& \multicolumn{3}{c}{ATLAS}		&  \multicolumn{3}{c}{\Checkmate{}}						 \\ \hline
  No selection						& \multicolumn{3}{c}{100~\%}		&  \multicolumn{3}{c}{-~\%}		 \\ 
  Jet and Event cleaning *				& \multicolumn{3}{c}{98.2~\%}		&  \multicolumn{3}{c}{-~\%}		 \\ 					
  Cosmic muon rejection	 *				& \multicolumn{3}{c}{98.2~\%}		&  \multicolumn{3}{c}{-~\%}		 \\ 
 $ \geq$ 4 jets ($\pT>30$~\GeV)				& \multicolumn{3}{c}{95.4~\%}		&  \multicolumn{3}{c}{94.6~\%}		 \\ 
  1st jet $\pT > 90$~\GeV				& \multicolumn{3}{c}{95.4~\%}		&  \multicolumn{3}{c}{94.6~\%}		 \\ 			
 \etmiss$>$ 150~\GeV					& \multicolumn{3}{c}{88.7~\%}		&  \multicolumn{3}{c}{88.0~\%}		 \\ 
  Electron veto						& \multicolumn{3}{c}{88.7~\%}		&  \multicolumn{3}{c}{86.1~\%}		 \\ 					
  Muon veto						& \multicolumn{3}{c}{88.2~\%}		&  \multicolumn{3}{c}{85.7~\%}		 \\ 
 $\Delta\phi^{4j}_{min} > 0.5$				& \multicolumn{3}{c}{58.5~\%}		&  \multicolumn{3}{c}{59.9~\%}		 \\ 
 \etmiss$/m^{4j}_{\text{eff}}> 0.2$			& \multicolumn{3}{c}{46.2~\%}		&  \multicolumn{3}{c}{48.7~\%}		 \\ \hline
 Signal region cuts					&     \multicolumn{2}{c|}{SR-0l-4j-A}	&     \multicolumn{2}{c|}{SR-0l-4j-B}	 &     \multicolumn{2}{c}{SR-0l-4j-C} \\
 Source							&     ATLAS		&  \Checkmate{}		&     ATLAS		&  \Checkmate{}	 	&     ATLAS		&  \Checkmate{}	\\ \hline
 $\geq 4j$, $\pT > 30, 50, 50$				&     46.2~\% 		&	48.7~\%		& 	42.8~\%		& 45.3~\% 	 	&	42.8~\%		& 45.3~\%	 	\\
 $\geq 3 b$, $\pT > 30, 50, 50$			&     20.5~\% 		&	20.0~\%		& 	17.9~\%		& 17.5~\% 	 	&	17.9~\%		& 17.4~\%	 	\\
 \etmiss$>$ 200, 350, 250				&     20.5~\% 		&	19.9~\%		& 	16.2~\%		& 15.7~\% 	 	&	17.4~\%		& 17.1~\%	 	\\
 $m^{incl}_{\text{eff}} >$ 1000, 1100, 1300			&     20.3~\% 		&	19.7~\%		&  $15.9\pm0.1$~\%     & $15.6\pm0.2$~\% 	&   $15.9\pm0.1$~\%    & $16.1\pm0.2$~\%	 	\\  
 \etmiss$/\sqrt{H^{4j}_T}>$ 16, 0, 0		&     $10.8\pm0.1$~\%  &   $10.9\pm0.1$~\%    & 	-	 	& -	 	 	&	- 		& -	 	\\  \hline \hline
  \end{tabularx}
  \caption{The cutflow is given as an absolute efficiency in $\%$ for each step of event selection. Final error displayed on signal regions is due to finite Monte Carlo statistics. *\textit{No cosmic muon rejection or event cleaning is performed by 
  \Checkmate{}, instead a flat efficiency factor is included.} \label{tab:atlas_2013_061_CF1} }
\end{table*}

\begin{table*}[h] 
\begin{tabularx}{1.0\textwidth}{ l|>{\centering\arraybackslash}X>{\centering\arraybackslash}X|>{\centering\arraybackslash}X>{\centering\arraybackslash}X|>{\centering\arraybackslash}X>{\centering\arraybackslash}X} \hline \hline
 Process						&\multicolumn{6}{c}{$pp\to\tilde{g}\tilde{g}, \tilde{g} \to t \overline{t} \tilde{\chi}^0_1$}	 	   	 \\
 Point 							&\multicolumn{6}{c}{$m(\tilde{g})=1300$~\GeV, $m(\tilde{\chi}^0_1)=100$~\GeV}	        \\ 
  Channel						&\multicolumn{6}{c}{0 Lepton, 7 Jets}	\\  
\hline
  Source						& \multicolumn{3}{c}{ATLAS}		&  \multicolumn{3}{c}{\Checkmate{}}						 \\ \hline
  No selection						& \multicolumn{3}{c}{100~\%}		&  \multicolumn{3}{c}{-~\%}		 \\ 
  Jet and Event cleaning *				& \multicolumn{3}{c}{98.4~\%}		&  \multicolumn{3}{c}{-~\%}		 \\ 					
  Cosmic muon rejection	 *				& \multicolumn{3}{c}{97.2~\%}		&  \multicolumn{3}{c}{-~\%}		 \\ 
 $ \geq$ 4 jets ($\pT>30$~\GeV)				& \multicolumn{3}{c}{96.9~\%}		&  \multicolumn{3}{c}{97.6~\%}		 \\ 
  1st jet $\pT > 90$~\GeV				& \multicolumn{3}{c}{96.9~\%}		&  \multicolumn{3}{c}{97.5~\%}		 \\ 			
 \etmiss$>$ 150~\GeV					& \multicolumn{3}{c}{88.3~\%}		&  \multicolumn{3}{c}{87.7~\%}		 \\ 
  Electron veto						& \multicolumn{3}{c}{59.7~\%}		&  \multicolumn{3}{c}{60.2~\%}		 \\ 					
  Muon veto						& \multicolumn{3}{c}{41.8~\%}		&  \multicolumn{3}{c}{40.7~\%}		 \\ 
 $\Delta\phi^{4j}_{min} > 0.5$				& \multicolumn{3}{c}{30.0~\%}		&  \multicolumn{3}{c}{29.6~\%}		 \\ 
 \etmiss$/m^{4j}_{\text{eff}}> 0.2$			& \multicolumn{3}{c}{25.9~\%}		&  \multicolumn{3}{c}{25.4~\%}		 \\ 
 $ \geq$ 7 jets ($\pT>30$~\GeV)				& \multicolumn{3}{c}{24.6~\%}		&  \multicolumn{3}{c}{24.1~\%}		 \\ 
 $ \geq$ 3 $b$-jets ($\pT>30$~\GeV)			& \multicolumn{3}{c}{11.5~\%}		&  \multicolumn{3}{c}{11.4~\%}		 \\  \hline 
 Signal region cuts					&     \multicolumn{2}{c|}{SR-0l-7j-A}	&     \multicolumn{2}{c|}{SR-0l-7j-B}	 &     \multicolumn{2}{c}{SR-0l-7j-C} \\
 Source							&     ATLAS		&  \Checkmate{}		&     ATLAS		&  \Checkmate{}	 &     ATLAS		&  \Checkmate{}	\\ \hline
 \etmiss$>$ 200, 350, 250				&     11.3~\% 		&	11.3~\%		& 	9.2~\%		& 8.9~\% 	 &	10.8~\%		& 10.8~\%	 	\\
 $m^{incl}_{\text{eff}} >$ 1000, 1000, 1500			& $11.3\pm0.1$~\%	& $11.2\pm0.1$~\%	& 	$9.2\pm0.1$~\%	& $8.9\pm0.1$~\%&	$9.5\pm0.1$~\%	& $9.2\pm0.1$~\%	 	\\  \hline	\hline
  \end{tabularx}
    \caption{The cutflow is given as an absolute efficiency in $\%$ for each step of event selection.  Final error displayed on signal regions is due to finite Monte Carlo statistics. *\textit{No cosmic muon rejection or event cleaning is performed by 
    \Checkmate{}, instead a flat efficiency factor is included.} \label{tab:atlas_2013_061_CF2} }
\end{table*}

\begin{table*}[h] 
\begin{tabularx}{1.0\textwidth}{l|>{\centering\arraybackslash}X>{\centering\arraybackslash}X|>{\centering\arraybackslash}X>{\centering\arraybackslash}X|>{\centering\arraybackslash}X>{\centering\arraybackslash}X} \hline \hline
 Process						&\multicolumn{6}{c}{$pp\to\tilde{g}\tilde{g}, \tilde{g} \to t \overline{t} \tilde{\chi}^0_1$}	 	   	 \\
 Point 							&\multicolumn{6}{c}{$m(\tilde{g})=1300$~\GeV, $m(\tilde{\chi}^0_1)=100$~\GeV}	        \\ 
\hline
  Channel						&\multicolumn{6}{c}{1 Lepton}	\\  
  Source						& \multicolumn{3}{c}{ATLAS}		&  \multicolumn{3}{c}{\Checkmate{}}						 \\ \hline
  No selection						& \multicolumn{3}{c}{100~\%}		&  \multicolumn{3}{c}{-~\%}		 \\ 
  Jet and Event cleaning *				& \multicolumn{3}{c}{98.4~\%}		&  \multicolumn{3}{c}{-~\%}		 \\ 					
  Cosmic muon rejection	 *				& \multicolumn{3}{c}{97.2~\%}		&  \multicolumn{3}{c}{-~\%}		 \\ 
 $ \geq$ 4 jets ($\pT>30$~\GeV)				& \multicolumn{3}{c}{96.9~\%}		&  \multicolumn{3}{c}{97.6~\%}		 \\ 
  1st jet $\pT > 90$~\GeV				& \multicolumn{3}{c}{96.8~\%}		&  \multicolumn{3}{c}{97.5~\%}		 \\ 			
 \etmiss$>$ 150~\GeV					& \multicolumn{3}{c}{88.3~\%}		&  \multicolumn{3}{c}{87.5~\%}		 \\ 
  $\geq$ 1 signal lepton				& \multicolumn{3}{c}{37.0~\%}		&  \multicolumn{3}{c}{40.6~\%}		 \\ 					
  $\geq$ 6 jets $(\pT > 30$~\GeV)			& \multicolumn{3}{c}{33.8~\%}		&  \multicolumn{3}{c}{36.7~\%}		 \\ 
 $ \geq$ 3 $b$-jets ($\pT>30$~\GeV)			& \multicolumn{3}{c}{14.3~\%}		&  \multicolumn{3}{c}{16.1~\%}		 \\  \hline
 Signal region cuts					&     \multicolumn{2}{c|}{SR-1l-6j-A}	&     \multicolumn{2}{c|}{SR-1l-6j-B}	 &     \multicolumn{2}{c}{SR-1l-6j-C} \\
 Source							&     ATLAS		&  \Checkmate{}		&     ATLAS		&  \Checkmate{}	 &     ATLAS		&  \Checkmate{}	\\ \hline
 $m_T >$ 140, 140, 160~\GeV				&	11.3~\%		&	12.0~\%		& 	11.3~\%		& 12.0~\%  	 &	10.7~\%		& 11.3~\%	 	\\
 \etmiss$>$ 175, 225, 275~\GeV			&	10.9~\%		&	11.6~\%		& 	10.0~\%		& 10.7~\%        &	8.8~\%		& 9.3~\%	 	\\
  \etmiss$/\sqrt{H_T}>$ 5~\GeV$^{1/2}$		&	10.8~\%		&	11.3~\%		& 	10.0 ~\%	& 10.6~\% 	 &	8.8~\%		& 9.3~\%	 	\\
 $m_{\text{eff}} >$ 700, 800, 900~\GeV				&  $10.8\pm0.1$~\%	&  $11.3\pm0.2$~\%	& 	$10.0\pm0.1$~\%	& $10.6\pm0.2$~\% & $8.8\pm0.1$~\%	& $9.2\pm0.1$~\%	 	\\  \hline \hline
  \end{tabularx}
    \caption{The cutflow is given as an absolute efficiency in $\%$ for each step of event selection. Final error displayed on signal regions is due to finite Monte Carlo statistics. *\textit{No cosmic muon rejection or event cleaning is performed 
    by \Checkmate{}, instead a flat efficiency factor is included.} \label{tab:atlas_2013_061_CF3} }
\end{table*}

 \clearpage
\subsection{atlas\_conf\_2013\_089}
\noindent Strongly produced SUSY with two leptons (razor), \cite{ATLAS-CONF-2013-089} \\
Energy: 8 \TeV \\
Luminosity: 20.3 fb$^{-1}$ \\
Validation notes:
\begin{itemize}
\item Validation has been performed versus all published cutflows.
\item WARNING: Discrepancies exist in the final signal regions between ATLAS and \Checkmate{}. However, 
we do not believe this shows a systematic failing of \Checkmate{}. First of all, the statistics 
from ATLAS are very small and it is therefore hard to draw a definitive conclusion. Moreover, after 
the trigger has been performed, no cut has an obvious flavour dependence. Consequently it is 
expected (and confirmed by ATLAS) that the $e\mu$ channels should contain the largest number of events.
\item We see a difference in the exclusion for low squark masses and a heavier LSP. We believe this is due to different settings in the Monte Carlo
 parton shower (Pythia 6) that gives a harder initial state radiation distribution. 
\end{itemize}

 \begin{table*}[h] 
\begin{tabularx}{1.0\textwidth}{l|>{\centering\arraybackslash}X>{\centering\arraybackslash}X|>{\centering\arraybackslash}X>{\centering\arraybackslash}X|>{\centering\arraybackslash}X>{\centering\arraybackslash}X} \hline \hline
 Process			&\multicolumn{6}{c}{$\tilde{g}\tilde{g}$ production, $\tilde{g}\to qq \tilde{\chi}^{\pm}_1$, $\tilde{\chi}^{\pm}_1 \to W \tilde{\chi}^0_1$}	 	   	 \\ 
 Point 				&\multicolumn{6}{c}{$m(\tilde{g})=800$~\GeV, $m(\tilde{\chi}^{\pm}_1)=460$~\GeV, $m(\tilde{\chi}^0_1)=60$~\GeV}	     	   \\  \hline
 Source				&\multicolumn{3}{c|}{ATLAS} 						&	\multicolumn{3}{c}{\Checkmate}			 \\  
  No Cuts			&\multicolumn{3}{c|}{59999} 						&	\multicolumn{3}{c}{59999}			 \\\hline
  2 baseline leptons		&\multicolumn{3}{c|}{2073} 						&	\multicolumn{3}{c}{1880}			 \\	 
  Passes trigger		&\multicolumn{3}{c|}{1670} 						&	\multicolumn{3}{c}{1426}			 \\
  $m_{\ell\ell} > 20$~\GeV	&\multicolumn{3}{c|}{1639} 						&	\multicolumn{3}{c}{1408}			 \\		\hline
  Lepton flavour		&\multicolumn{2}{c|}{$ee$} 			&	\multicolumn{2}{c|}{$\mu\mu$}	& \multicolumn{2}{c}{$e\mu$} \\
  Source			&	ATLAS		&	\Checkmate	& 	ATLAS		& \Checkmate	&	ATLAS	&	\Checkmate			\\ \hline
  Lepton separation		&	576		&	420		& 	397		& 397	  	&	666	&	590			\\  
  Signal leptons		&	443		&	377		&  	373		& 397	  	&	549	&	561			\\  
  Trigger + dilepton		&	429		&	377		&  	358		& 397	  	&	517	&	561			\\ 
  $> 2$ jets, $b$-veto		&	341		&	320		& 	297		& 350		&	424	&	480			\\ 
  $Z$-veto 			&	319		&	293		& 	276		& 321	  	&	424	&	480			\\  
  $R>0.35$ 			&	139		&	155		&  	137		& 162	  	&	195	&	252			\\  
  $M'_R>800$~\GeV		&	$53\pm7$	&	$55\pm2$	&  	$50\pm7$	& $60\pm3$  	&  $63\pm8$	&	$92\pm3$	\\ 
  $< 3$ jets, $b$-veto		&	53		&	44		&  	40		& 34	  	&	54	&	56	\\
  $Z$-veto  			&	51		&	41		&  	36		& 31	  	&	54	&	56			\\	  
  $R>0.5$ 			&	16		&	14		&  	19		& 10	  	&	12	&	21			\\  
  $M'_R>400$~\GeV 		&	$14\pm4$	&	$13\pm1$	&  	$19\pm4$	& $10\pm1$	&  $10\pm3$	&	$19\pm1$			\\  \hline \hline
  \end{tabularx}
\caption{The ATLAS column shows the number of Monte Carlo events after each selection cut. The \Checkmate{} 
cutflow is normalised to number of events before any selection cuts are applied. The final error is from 
Monte Carlo statistics for both ATLAS and \Checkmate. \label{tab:atlas_2013_089_CF1} }
\end{table*}

 \begin{figure}[h]
 \centering \vspace{-0.0cm}
 \includegraphics[width=0.49\textwidth]{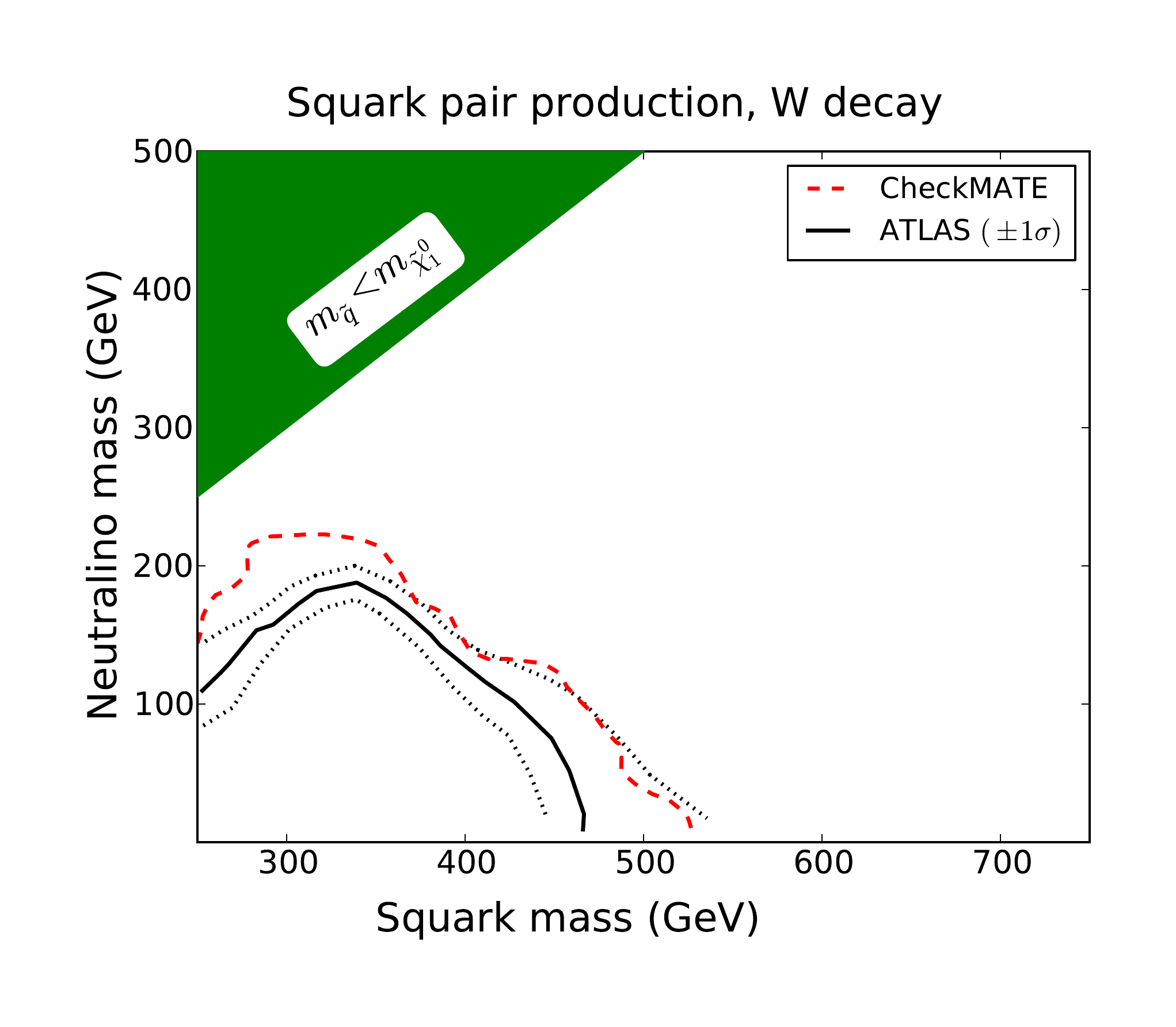}
 \caption{Exclusion curve for a simplified model with squark production followed
 by decay $\tilde{q} \to \tilde{\chi}^{\pm}_1 q, \tilde{\chi}^{\pm}_1 \to W \tilde{\chi}^0_1$. } 
 \label{fig:atlas_2013_089_1}
 \end{figure}
 
 \begin{figure}[h]
 \centering \vspace{-2.4cm}
 \includegraphics[width=0.49\textwidth]{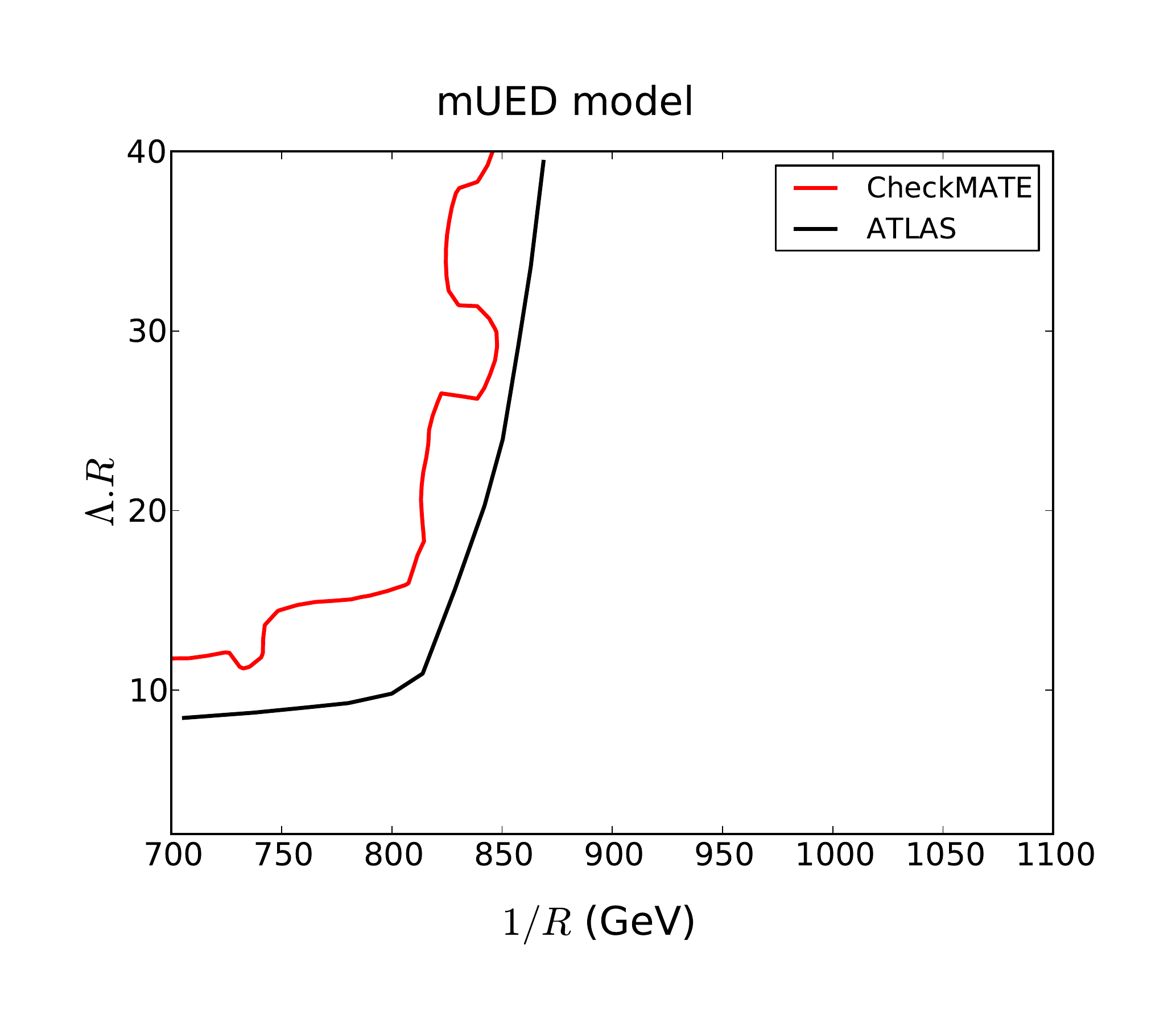} 
 \caption{Exclusion curve for the minimal Universal Extra Dimensions (mUED) model. Jumps in exclusion limit of \Checkmate{} are due to the change of signal region.}
 \label{fig:atlas_2013_089_2}
 \end{figure}

 \clearpage
\subsection{cms\_1303\_2985}
\noindent Hadronic $\alpha_T$ and $b-$jet multiplicity, \cite{Chatrchyan:2013lya} \\
Energy: 8 \TeV \\
Luminosity: 11.7 fb$^{-1}$ \\ 
Validation notes:
\begin{itemize}
  \item No cut flow is provided by CMS, validation is performed with signal 
  region distributions and parameter scans.
  \item The SM background distributions has been taken from the CMS note.
  \item The ATLAS $b$-tagging has been used. 
  \item Jumps in limit in \Checkmate{} parameter scans are due to the single signal 
  region limit setting procedure used (see \Cref{sec:overview}).
\end{itemize}
 
 \begin{figure}[h]
 \centering \vspace{-0.0cm}
 \includegraphics[width=0.49\textwidth]{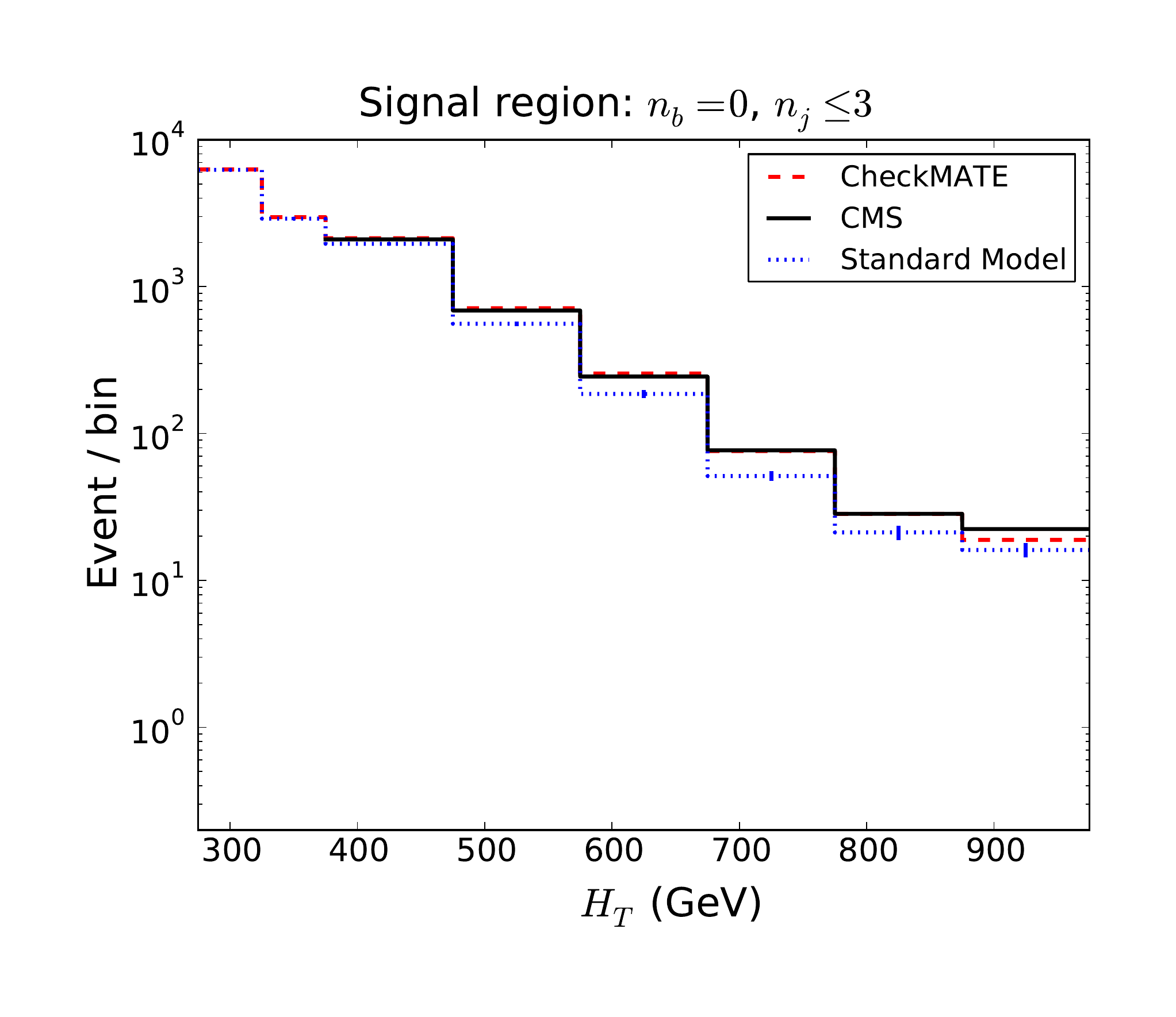}
 \includegraphics[width=0.49\textwidth]{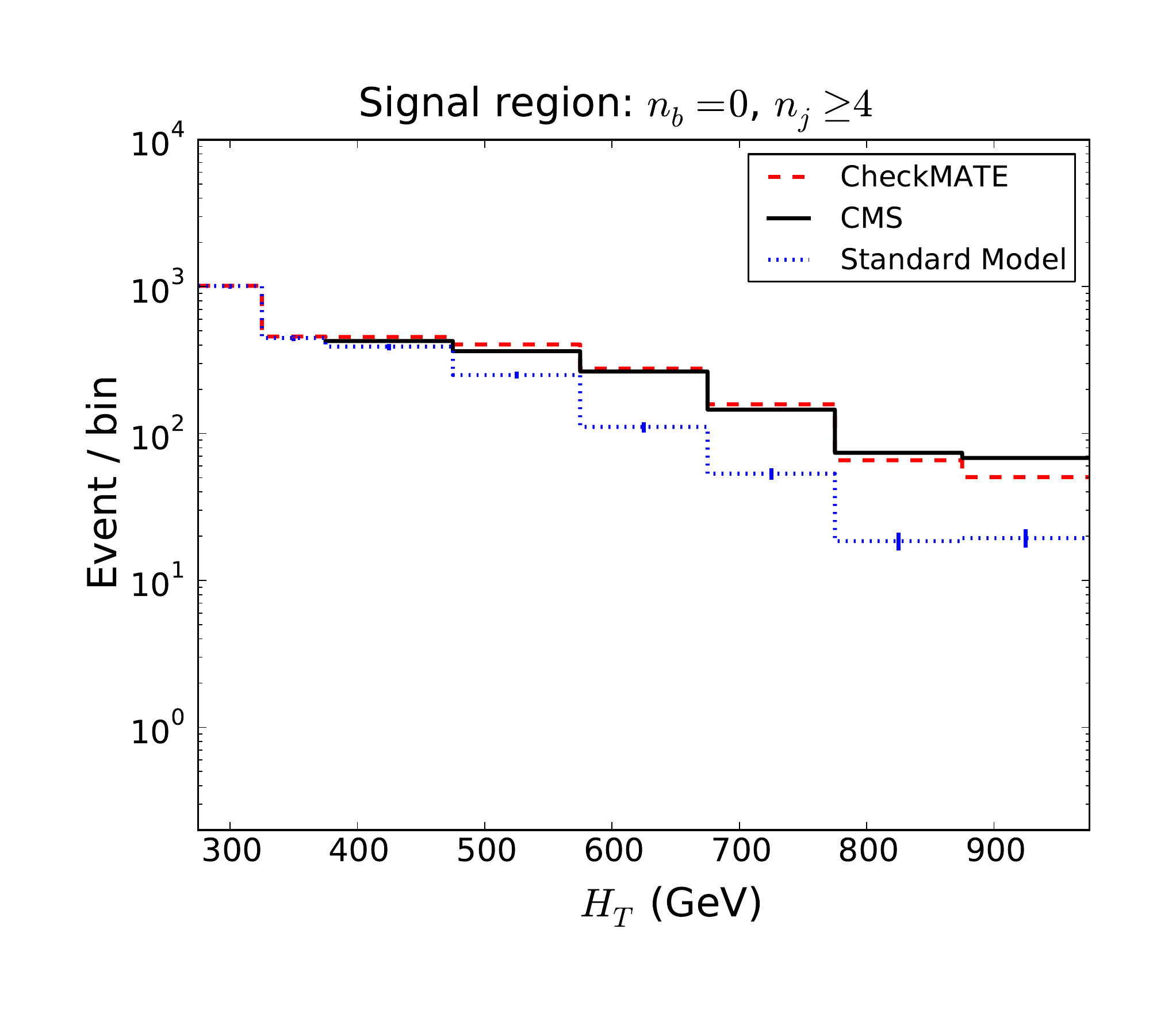}
 \caption{Distributions in $H_T$ for different signal models. 
 Left: $pp\to\tilde{q}\tilde{q}, \tilde{q}\to q {\tilde \chi}^0_1$, 
 ($m_{\tilde{q}}=600$~\GeV, $m_{\tilde{\chi}^0_1}=250$~\GeV). 
 Right: $pp\to\tilde{g}\tilde{g}, \tilde{g}\to q \overline{q} {\tilde\chi}^0_1$, 
 ($m_{\tilde{g}}=700$~\GeV, $m_{\tilde{\chi}^0_1}=300$~\GeV)}. 
 \label{fig:cms_1303_2985_1}
 \end{figure}
 
  \begin{figure}[h]
 \centering \vspace{-0.0cm}
 \includegraphics[width=0.49\textwidth]{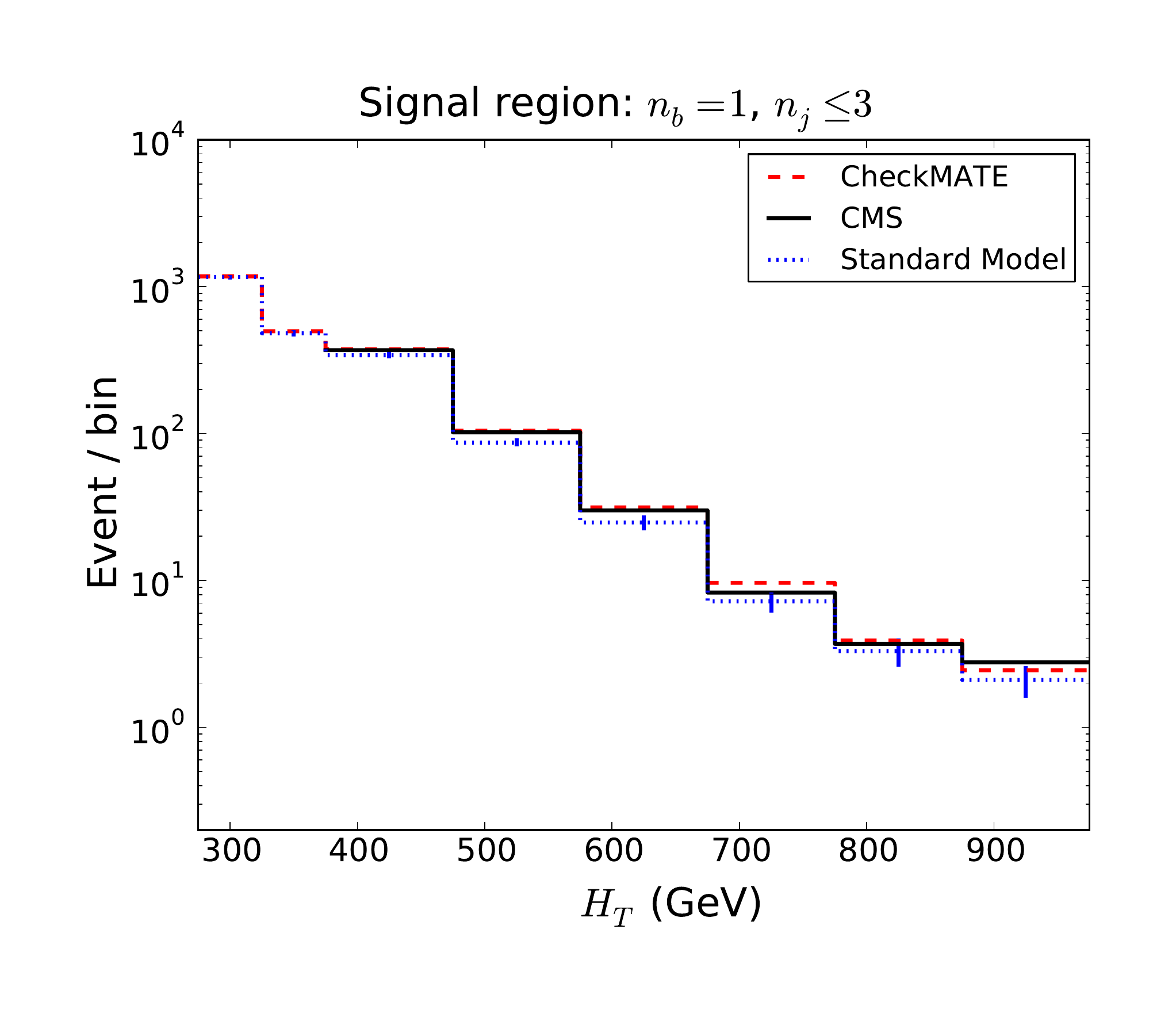}
 \includegraphics[width=0.49\textwidth]{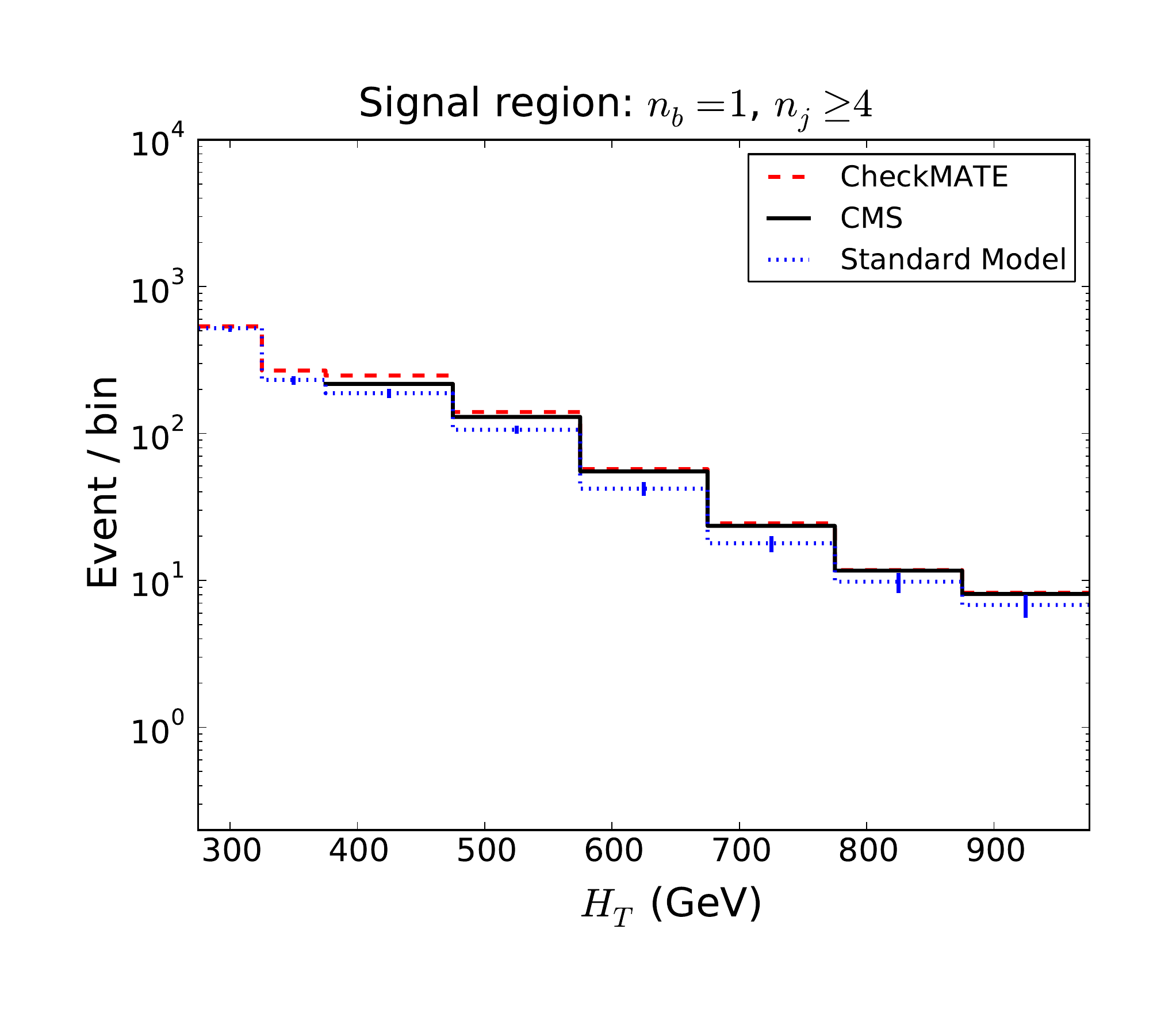}
 \caption{Distributions in $H_T$ for different signal models. 
 Left: $pp\to\tilde{b}\tilde{b}, \tilde{b}\to b {\tilde\chi}^0_1$, 
 ($m_{\tilde{b}}=500$~\GeV, $m_{\tilde{\chi}^0_1}=150$~\GeV). 
 Right: $pp\to\tilde{t}\tilde{t}, \tilde{t}\to t {\tilde\chi}^0_1$, 
 ($m_{\tilde{t}}=400$~\GeV, $m_{\tilde{\chi}^0_1}=0$~\GeV)}. 
 \label{fig:cms_1303_2985_2}
 \end{figure}
 
   \begin{figure}[h]
 \centering \vspace{-0.0cm}
 \includegraphics[width=0.49\textwidth]{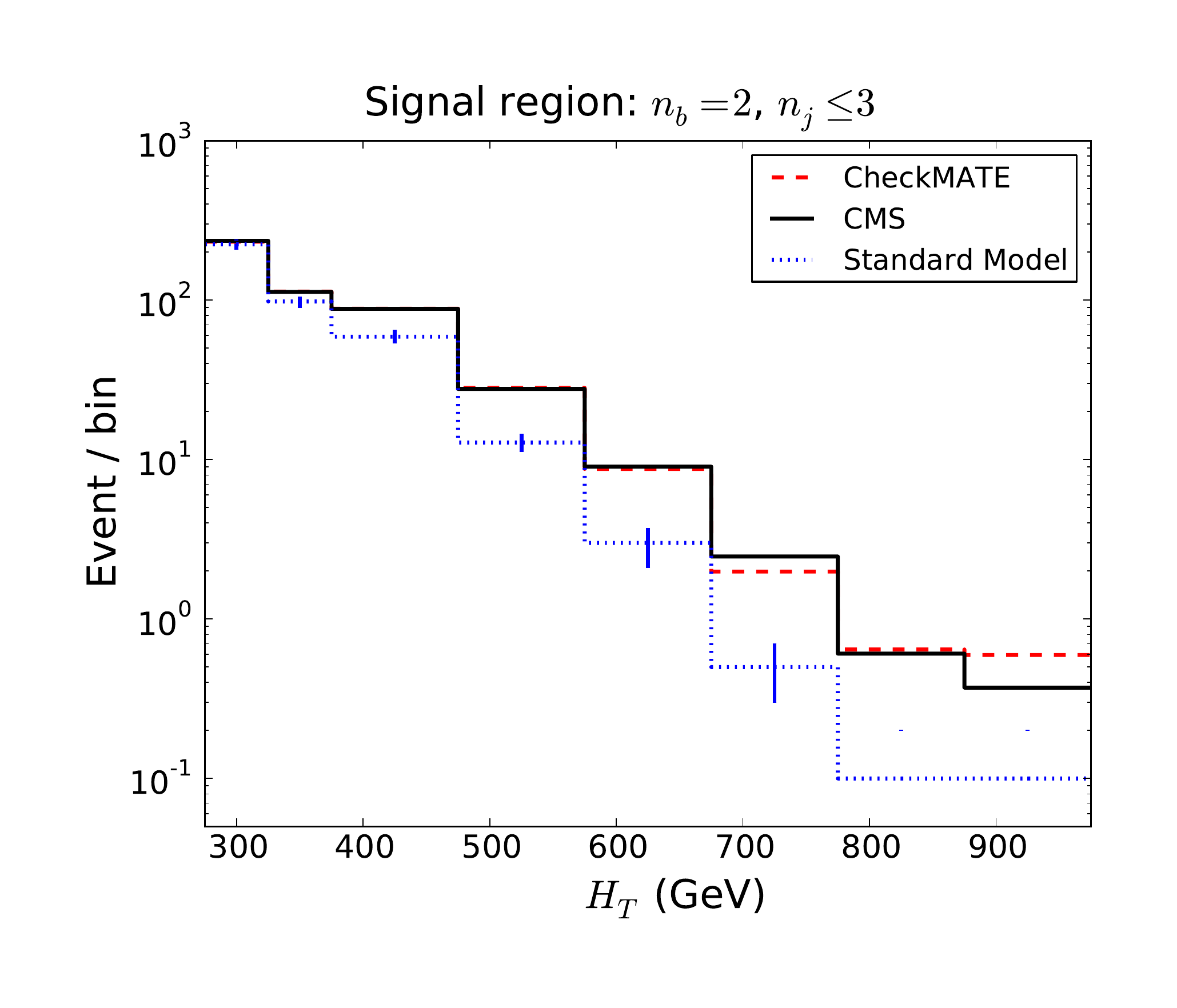}
 \includegraphics[width=0.49\textwidth]{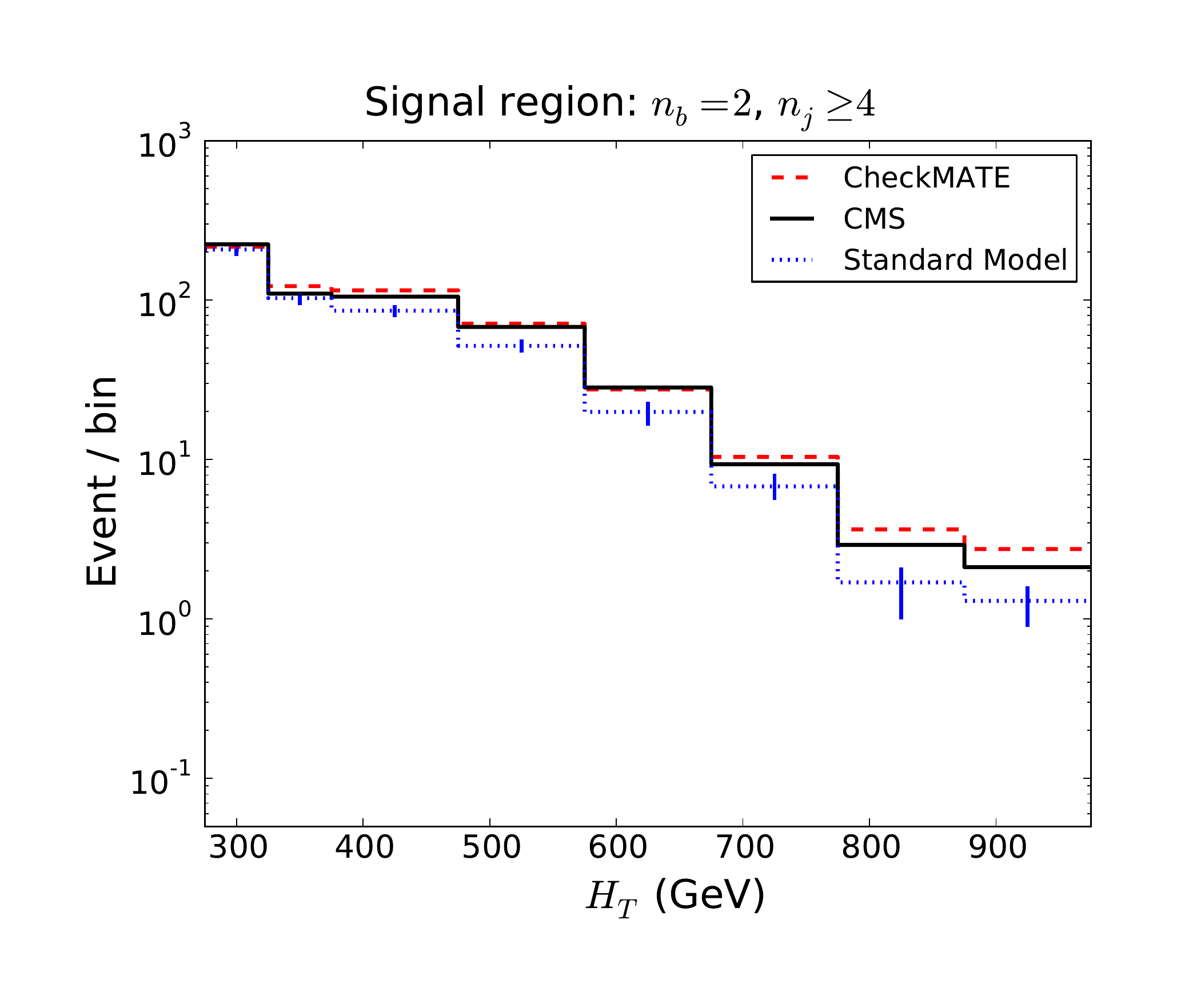}
 \caption{Distributions in $H_T$ for different signal models. 
 Left: $pp\to\tilde{b}\tilde{b}, \tilde{b}\to b {\tilde\chi}^0_1$, 
 ($m_{\tilde{b}}=500$~\GeV, $m_{\tilde{\chi}^0_1}=150$~\GeV). 
 Right: $pp\to\tilde{t}\tilde{t}, \tilde{t}\to t {\tilde\chi}^0_1$, 
 ($m_{\tilde{t}}=400$~\GeV, $m_{\tilde{\chi}^0_1}=0$~\GeV)}. 
 \label{fig:cms_1303_2985_3}
 \end{figure}
 
    \begin{figure}[h]
 \centering \vspace{-0.0cm}
 \includegraphics[width=0.49\textwidth]{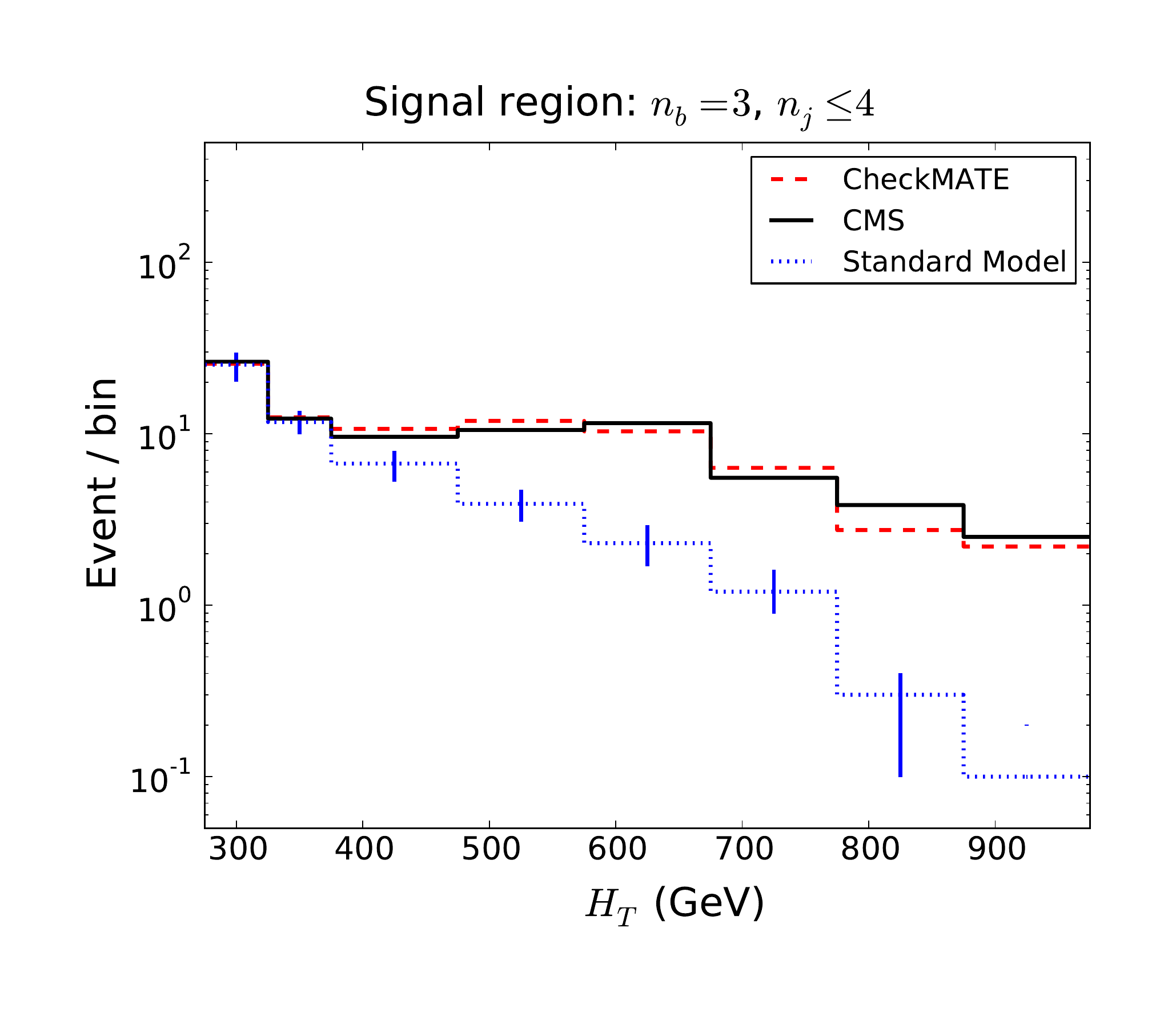}
 \includegraphics[width=0.49\textwidth]{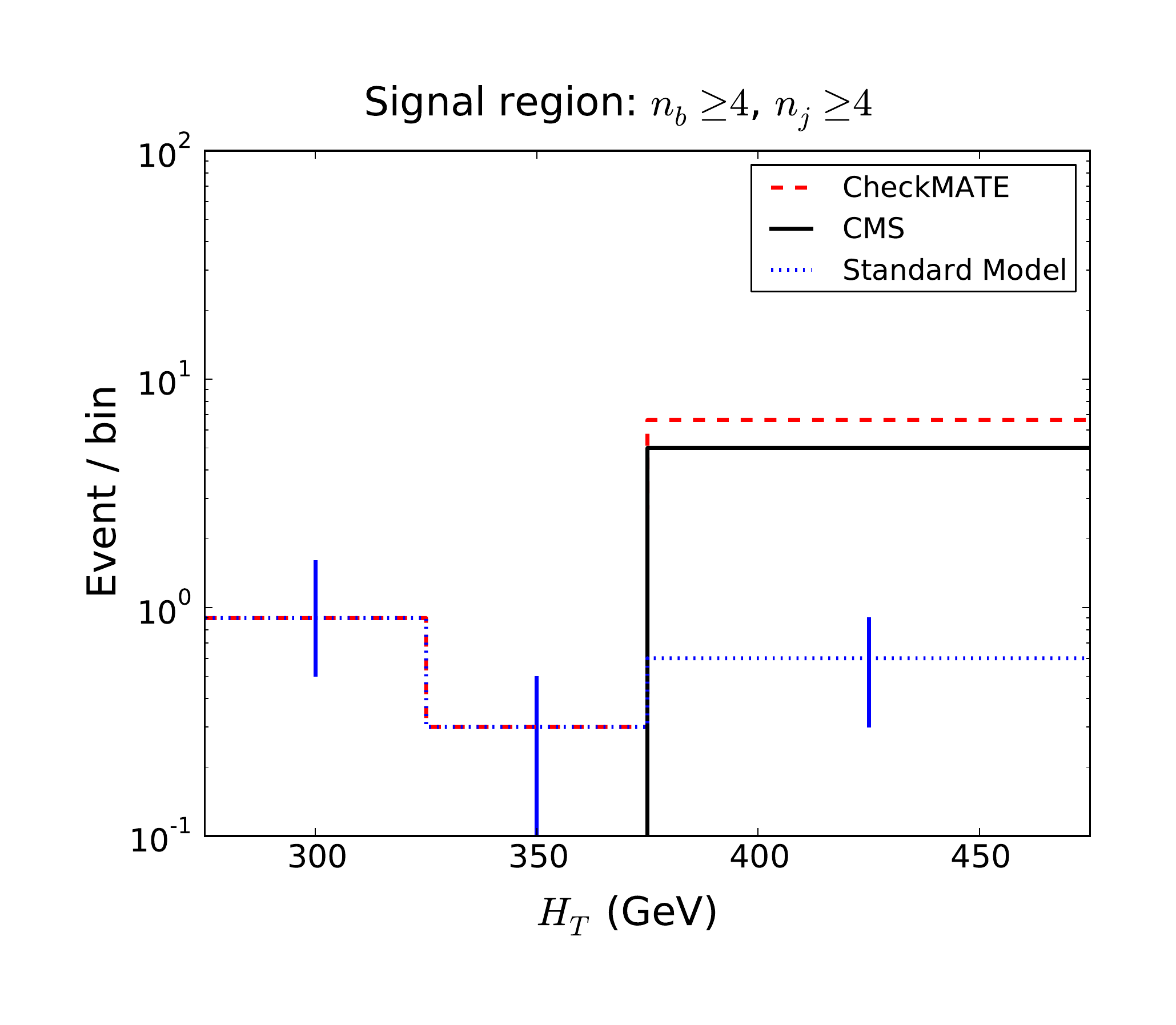}
 \caption{Distributions in $H_T$ for different signal models. 
 Left: $pp\to\tilde{g}\tilde{g}, \tilde{g}\to b \overline{b} {\tilde\chi}^0_1$, 
 ($m_{\tilde{g}}=900$~\GeV, $m_{\tilde{\chi}^0_1}=500$~\GeV). 
 Right: $pp\to\tilde{g}\tilde{g}, \tilde{g}\to t \overline{t} {\tilde\chi}^0_1$, 
 ($m_{\tilde{g}}=850$~\GeV, $m_{\tilde{\chi}^0_1}=250$~\GeV)}. 
 \label{fig:cms_1303_2985_4}
 \end{figure}
 
     \begin{figure}[h]
 \centering \vspace{-0.0cm}
 \includegraphics[width=0.49\textwidth]{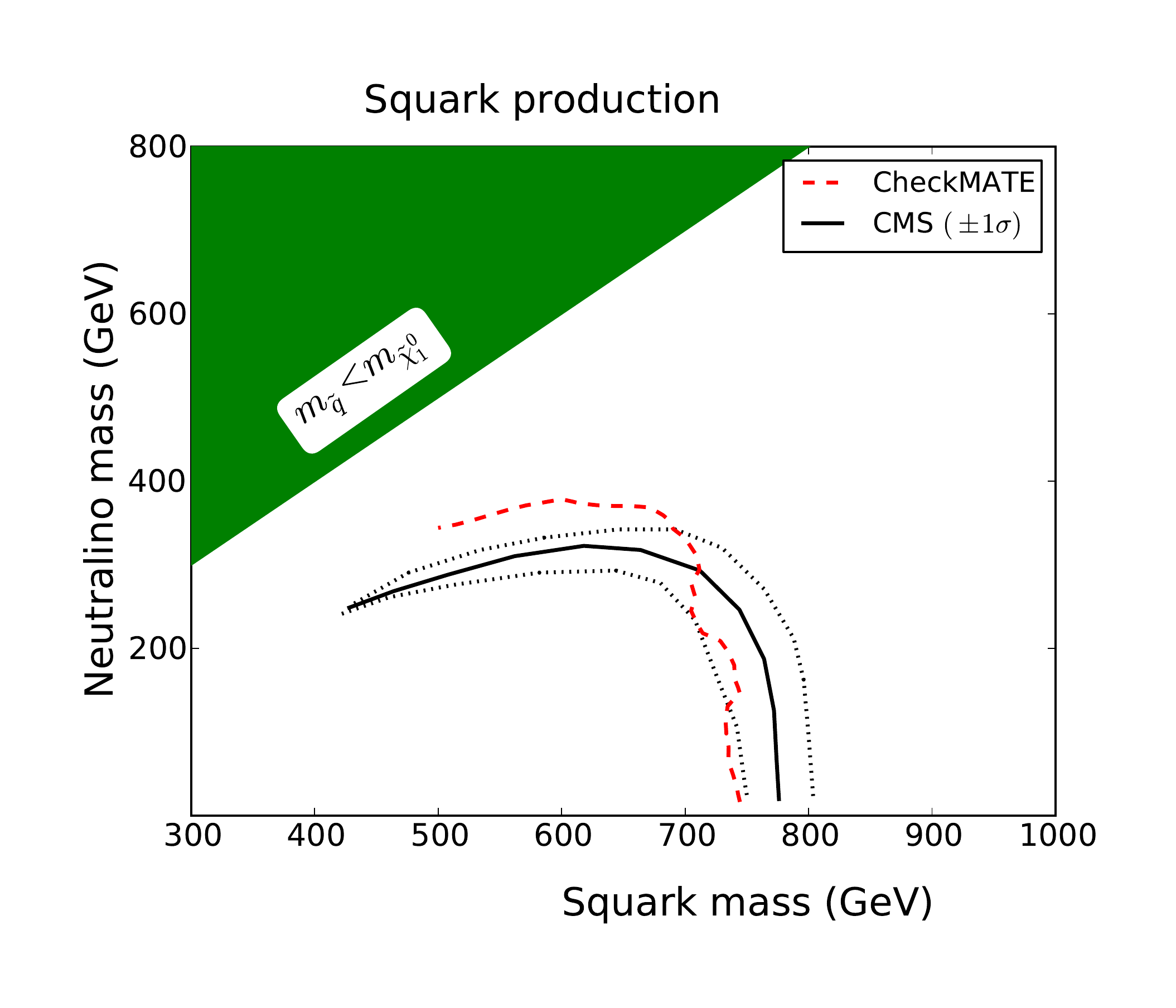}
 \includegraphics[width=0.49\textwidth]{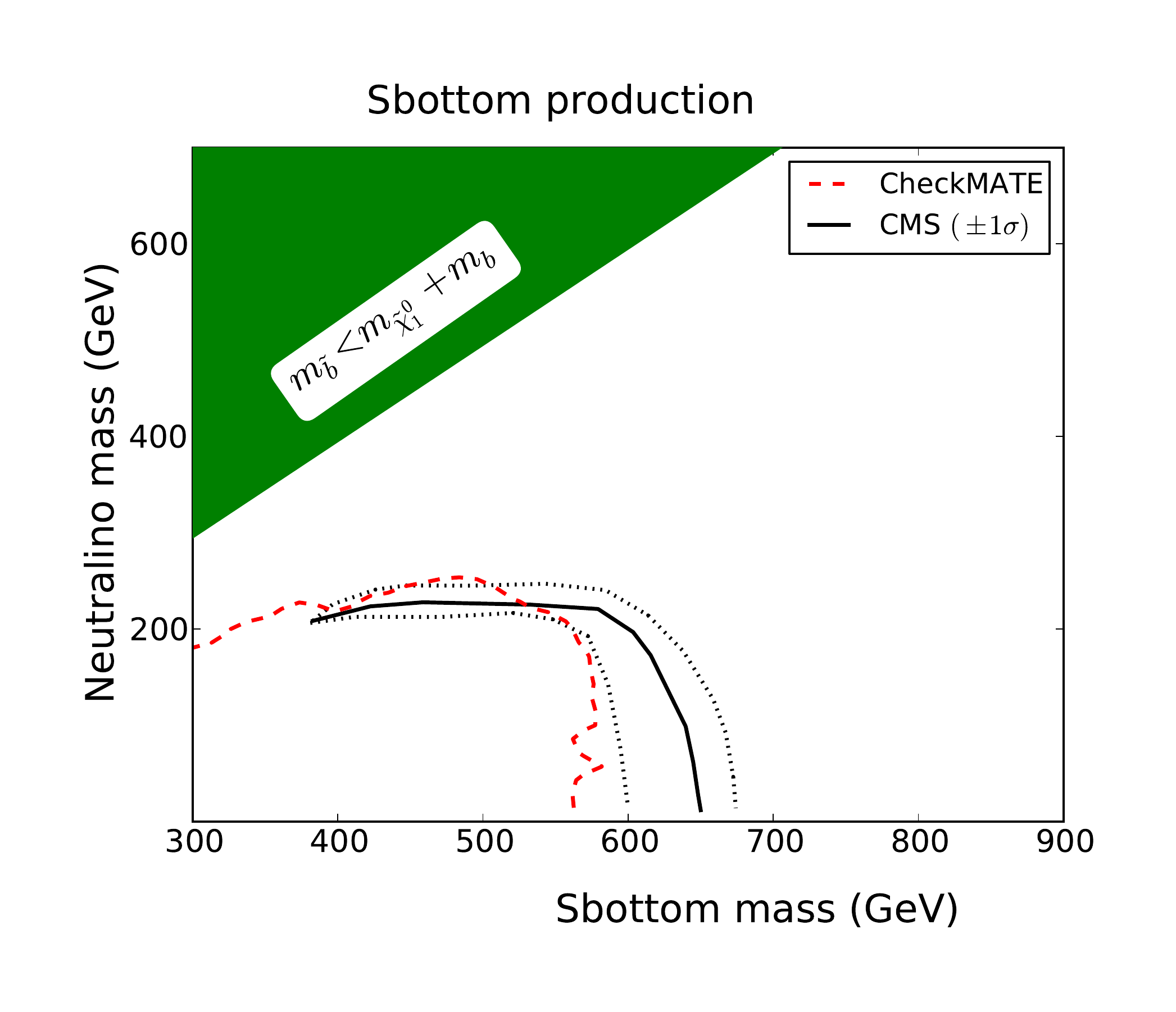}
 \caption{Exclusion curve for a simplified model with only first and 
 second generation squark production (left) and only bottom squark production (right). 
 We see a difference in exclusion for first and second generation squark production with smaller 
 squark masses and a heavier LSP. We believe this is due to different settings in the Monte Carlo
 parton shower (Pythia 6) that gives a harder initial state radiation distribution.} 
 \label{fig:cms_1303_2985_5}
 \end{figure}
 
      \begin{figure}[h]
 \centering \vspace{-0.0cm}
 \includegraphics[width=0.49\textwidth]{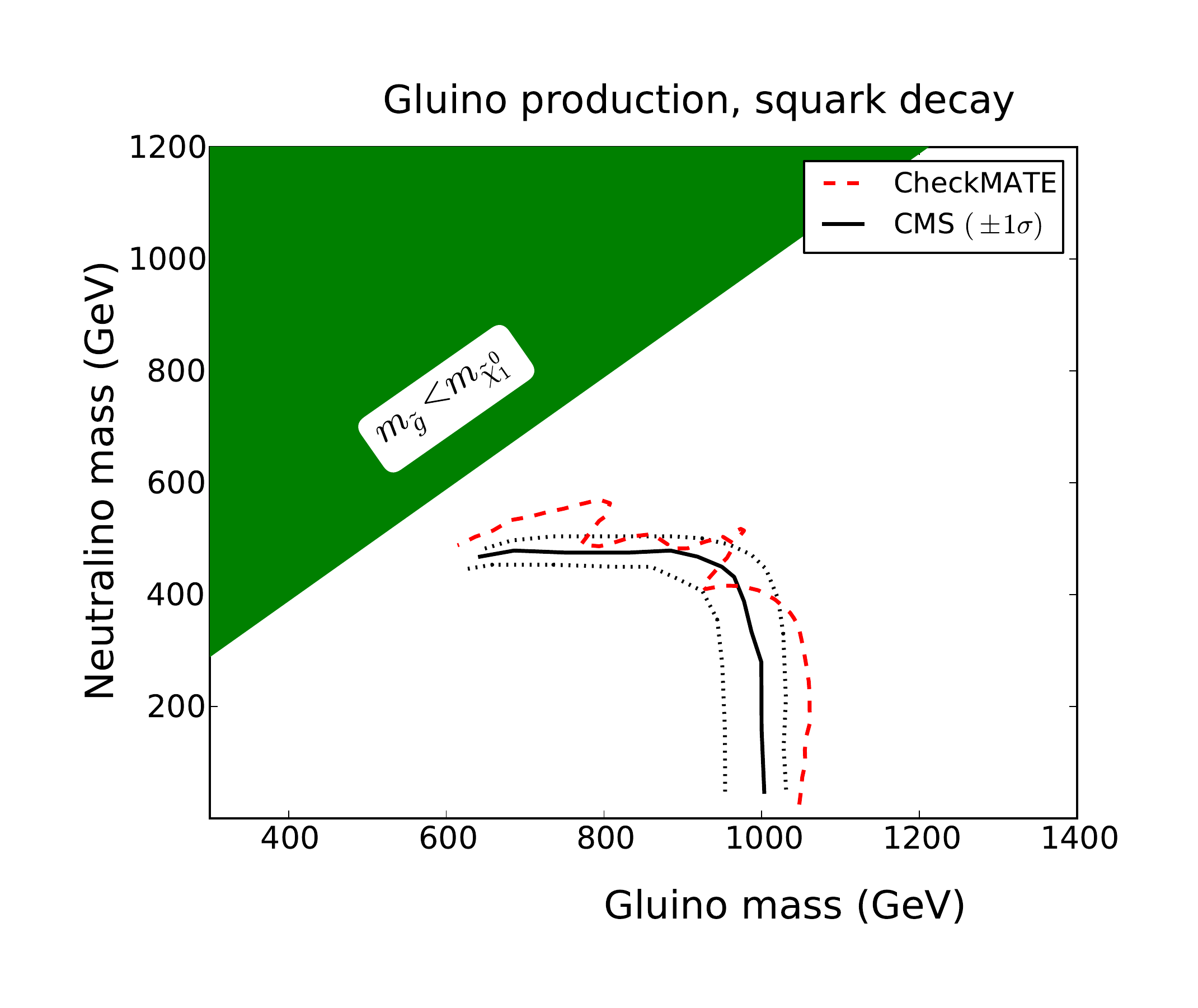}
 \includegraphics[width=0.49\textwidth]{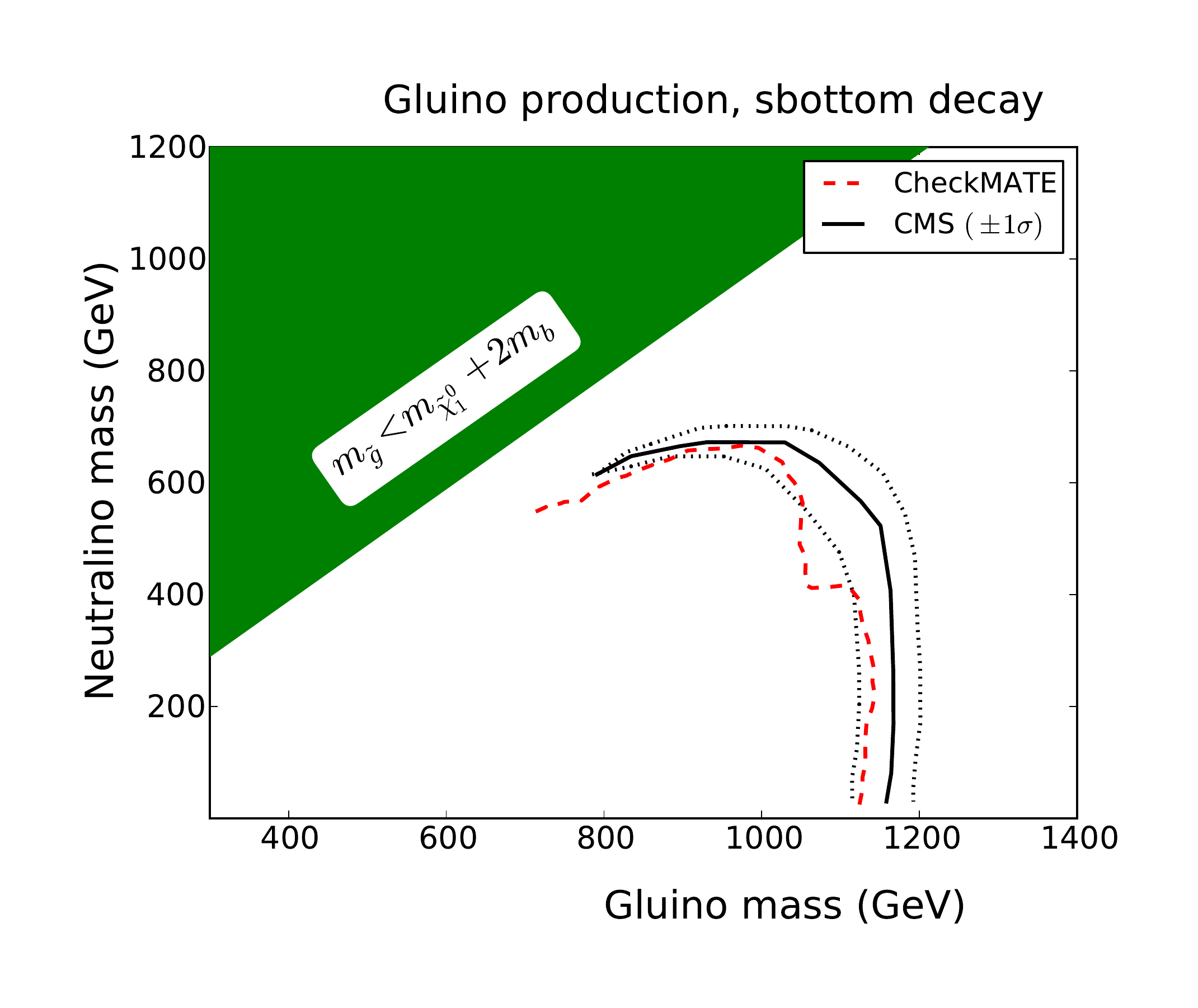}
 \caption{Exclusion curve for a simplified model with gluino production followed
 by decay into a $q \overline{q} \tilde{\chi}^0_1$ final state (left) or
 $b \overline{b} \tilde{\chi}^0_1$ final state (right). } 
 \label{fig:cms_1303_2985_6}
 \end{figure}
 
     \begin{figure}[h]
 \centering \vspace{-0.0cm}
 \includegraphics[width=0.49\textwidth]{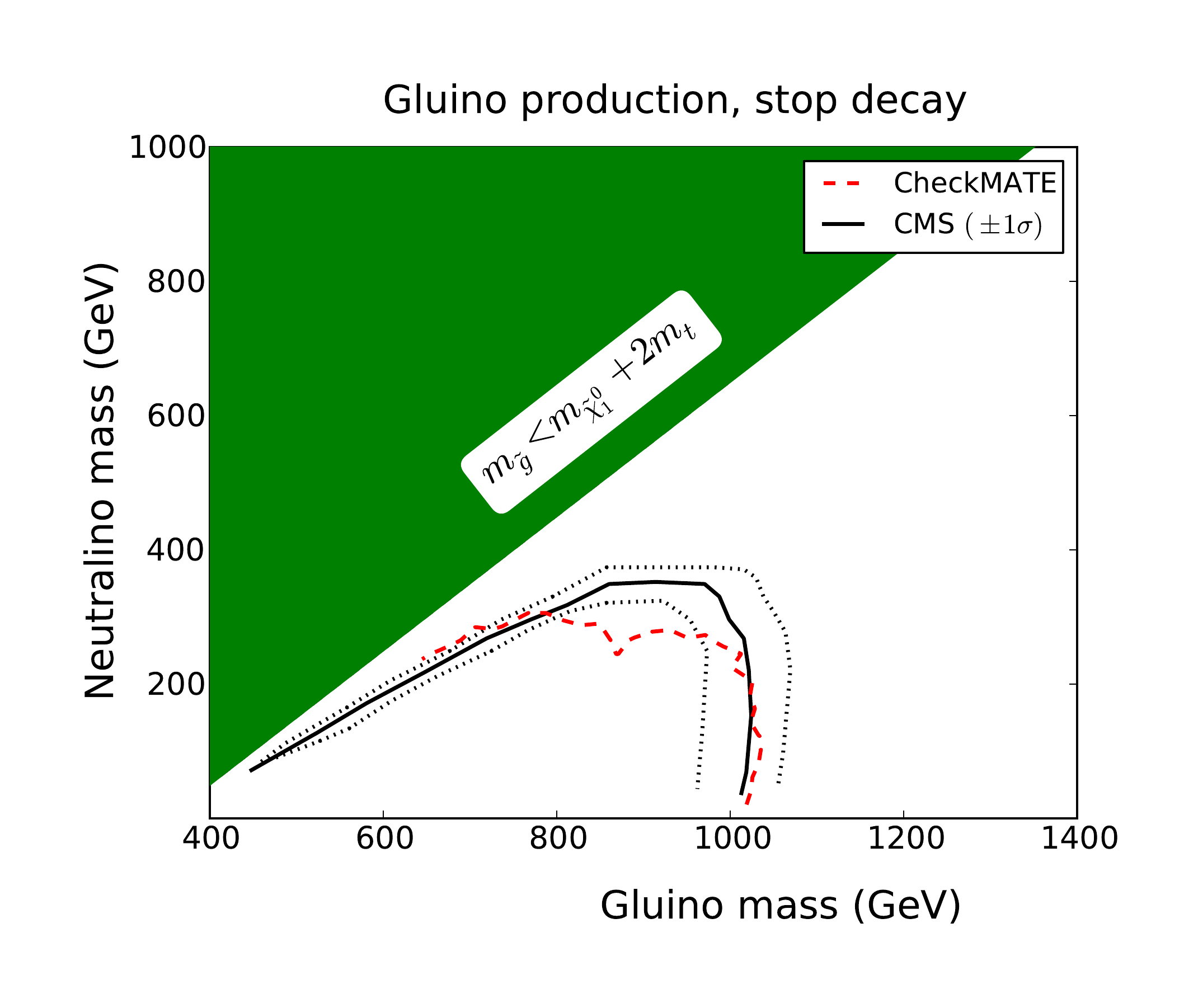}
 \caption{Exclusion curve for a simplified model with gluino production followed
 by decay into a $t \overline{t} \tilde{\chi}^0_1$ final state. } 
 \label{fig:cms_1303_2985_7}
 \end{figure}
\end{appendices} 

\clearpage
\bibliographystyle{ieeetr}
\bibliography{bibtex}

\end{document}